\documentclass{aa}

\usepackage{graphicx}
\usepackage{upgreek}
\usepackage{placeins}
\usepackage{longtable}
\usepackage{xtab,booktabs}
\usepackage{txfonts}
\usepackage{siunitx}
\usepackage[colorlinks,linkcolor=blue,citecolor=blue]{hyperref}
\usepackage{subfig}
\usepackage{mathtools} %cases in eq
\usepackage{makecell} % thead in table
\addto\extrasenglish{
    
}
\addto\extrasenglish{
    
}
\addto\extrasenglish{
    
} 
\addto\extrasenglish{
    
}
\addto\extrasenglish{
    
}

\newcommand*\dif{\mathop{}\!\mathrm{d}}
\newcommand{\insmyr}[1]{\left(\ensuremath{\frac{#1}{\mathrm{M}_\odot\,\mathrm{yr}^{-1}}}\right)}

\newcommand{\inwphz}[1]{\left(\ensuremath{\frac{#1}{\mathrm{W}\,\mathrm{Hz}^{-1}}}\right)}
\newcommand{\hi}{{\sc H\,i}}
\newcommand{\hii}{{\sc H\,ii}}
\newcommand{\nii}{{\sc N\,ii}}

\let\upmu\muup
\let\upalpha\alphaup

\def\aap{A\&A}

\def\apjs{ApJ\ Suppl.}
\def\apjl{ApJ\ Let.}

\begin{document} 

   \title{ViCTORIA project: The LOFAR-view of environmental effects in Virgo Cluster star-forming galaxies}
   \titlerunning{The LOFAR-view of environmental effects in Virgo Cluster star-forming galaxies}
   %\title{ViCTORIA project: The LOFAR-view of ram-pressure stripping in the Virgo Cluster}
   \subtitle{}

 \author{H. W. Edler
          \inst{1}
          \and
          I. D. Roberts\inst{2}
          \and
          A. Boselli\inst{3,4}
          \and
          F. de Gasperin\inst{1,5}
          \and      
          V. Heesen\inst{1}
          \and
          M. Brüggen\inst{1}
          \and
          A. Ignesti\inst{6}
          \and 
          L. Gajović\inst{1}
          %T. W. Shimwell\inst{3,5}
          %\and
          %H. McCall\inst{6}
          %\and
          %T. H. Reiprich\inst{6}
          %\and
          %M. J. Hardcastle\inst{7}
          %\and
          }

   \institute{Hamburger Sternwarte, University of Hamburg,
              Gojenbergsweg 112, D-21029, Hamburg, Germany\\
              \email{henrik.edler@hs.uni-hamburg.de}
              \and
              Leiden Observatory, Leiden University, PO Box 9513, 2300 RA Leiden, The Netherlands
              \and
              Aix Marseille Univ, CNRS, CNES, LAM, Marseille, France
              \and
              INAF - Osservatorio Astronomico di Cagliari, via della Scienza 5, 09047 Selargius , Italy
              \and
              INAF - Istituto di Radioastronomia,
              via P. Gobetti 101, Bologna, Italy
              \and
              INAF - Astronomical Observatory of Padova, 
              vicolo dell'Osservatorio 5, 35122 Padova, Italy 
              %%\and 
              %%Th{\"u}ringer Landessternwarte, Sternwarte 5, D-07778 Tautenburg, Germany 
              %\and
              %ASTRON, Netherlands Institute for Radio Astronomy, Oude Hoogeveensedijk 4, 7991 PD, Dwingeloo, The Netherlands 
              %\and
              %Argelander-Institut für Astronomie (AIfA), Universität Bonn, Auf dem Hügel 71, %53121 Bonn, Germany
              %\and
              %Centre for Astrophysics Research, University of Hertfordshire, College Lane, Hatfield AL10 9AB, UK
              }

   \date{Received October XX, 2023; accepted XXXXXXXX XX, XXXX}

% \abstract{}{}{}{}{} 
% 5 {} token are mandatory
 
  \abstract
   {Environmental effects such as ram-pressure stripping (RPS) shape the evolution of galaxies in dense regions.}
   {We use the nearby Virgo cluster as a laboratory to study environmental effects on the non-thermal components of star-forming galaxies.}
   {We constructed a sample of 17 RPS galaxies in the Virgo cluster and a statistical control sample of 119 nearby galaxies from the Herschel Reference Survey. All objects in these samples are detected in LOFAR 144\,MHz observations and come with H$\upalpha$ and/or far-UV star formation rate (SFR) estimates.}
   {We derived the radio-SFR relations, confirming a clearly super-linear slope of $\approx1.4$. We found that Virgo cluster RPS galaxies have radio luminosities that are a factor of 2-3 larger than galaxies in our control sample. We also investigated the total mass-spectral index relation, where we found a relation for the Virgo cluster RPS galaxies that is shifted to steeper spectral index values by $0.17\pm0.06$. Analyzing the spatially resolved ratio between the observed and the expected radio emission based on the hybrid near-UV + 100\,$\upmu$m SFR surface density, we generally observe excess radio emission all across the disk  with the exception of a few leading-edge radio-deficient regions.}
   {The radio excess and the spectral steepening for the RPS sample could be explained by an increased magnetic field strength if the disk-wide radio enhancement is due to projection effects. For the galaxies that show the strongest radio excesses (NGC\,4330, NGC\,4396, NGC\,4522), a rapid decline of the SFR ($t_\mathrm{quench} \leq 100$\,Myr) could be an alternative explanation. We disfavor shock acceleration of electrons as cause for the radio excess since it cannot easily explain the spectral steepening and radio morphology.}


   \maketitle
%
%-------------------------------------------------------------------

\section{Introduction} \label{sec:intro}
Galaxies that inhabit dense environments such as galaxy clusters show a lower cold gas content \citep[e.g.][]{Catinella2013,Boselli2014molecgasstripping} and a reduced star-forming (SF) activity \citep[e.g.][]{Kennicutt1983,Boselli2014TheEvolution,Boselli2016quenching} compared to those located in poorer environments. It is thought that those differences are caused by environmental processes, i.e.\ perturbations due to interactions with other galaxies or the intra-cluster medium (ICM).  The process which is often thought to be the dominant perturbation affecting galaxies in massive ($>$$10^{14}\,\mathrm{M}_\odot$), low-redshift clusters is ram-pressure stripping \citep[RPS; see][for recent reviews]{Cortese2021,Boselli2022RamEnvironments}. RPS is the removal of the interstellar medium (ISM) of a galaxy moving at high velocity $v$ relative to the ICM. The ram-pressure scales as $P\propto \rho v^2$ where $\rho$ is the ICM density. Consequently, it is most effective in massive clusters where the galaxy velocities and the ICM densities are high. 

RPS impacts the diffuse atomic phase of the ISM due to the advection of the loosely bound neutral hydrogen (H\,{\sc i}) which can give rise to tails that can be traced by 21\,cm line observations \citep[e.g.][]{Chung2007Tails}. This can also  explain the truncated radial \hi{}-profiles \citep[e.g.][]{Chung2009} and the statistical \hi{}-deficiency \citep[e.g.][]{Boselli2006EnvironmentalClusters} of galaxies in clusters.
Subsequent ionization of the stripped \hi{} due to interactions with the ICM creates tails of ionized gas that are most commonly observed in the H$\upalpha$ line \citep[e.g.][]{Gavazzi2001IonizedGasA1367,Yagi2010HalphaComa,Boselli2016Spectacular4569}; as those recently observed during the GAs Stripping Phenomena in galaxies with MUSE \citep[GASP;][]{Poggianti2017GASP} program and the Virgo Environmental Survey Tracing Ionised Gas Emission \citep[VESTIGE;][]{Boselli2018VESTIGE1}.
In some cases, RPS may also affect the dense molecular gas, which is fueling star formation, either indirectly by displacing the atomic gas or directly by displacing the molecular gas from the stellar disk \citep[e.g.][]{Cramer2020,Watts2023}; this will in turn reduce the star formation rate (SFR). 
The time-scale on which the star formation is quenched may depend on a number of parameters such as galaxy mass, orientation and velocity with respect to the ICM as well as the ICM density and dynamical state. 
Observational evidence points to quenching times of 
$\leq 1\,\mathrm{Gyr}$ \citep[e.g.][]{Boselli2006NGC4569,Boselli2016quenching,Ciesla2016, Fossati2018}.

Current RPS events in star-forming galaxies show tails in the atomic or ionized hydrogen distribution, as well as in the radio continuum \citep[e.g.][]{Gavazzi1978A1367,Gavazzi1995peculariA1367}. The radio continuum emission of star-forming galaxies is caused by cosmic-ray electrons (CRe) that were shock-accelerated in supernovae gyrating in weak magnetic fields (i.e. synchrotron radiation). Thus, it is a tracer of the SFR \citep{Condon1992,Gurkan2018LOFAR/H-ATLAS:Relation,Heesen2019CalibratingLOFAR,Heesen2022NearbyRelation}. The CRe are transported by diffusion processes in the galactic magnetic fields, but may also be subject to advection through ram pressure, creating asymmetric or tailed radio continuum profiles. The advance of sensitive radio surveys, such as those undertaken with the LOw-Frequency ARray \citep[LOFAR;][]{vanHaarlem2013}, allowed the identification of $>$$100$ ongoing RPS events in the past few years \citep{Roberts2021LoTSSClusters,Roberts2021II,Roberts2022LoTSSCluster,Roberts2022IIIPerseus,Ignesti2022WalkGalaxies,Ignesti2023,Edler2023}.

Observations in the radio continuum are well suited to identify RPS events and trace galaxy SFR. They also allow us to probe the non-thermal phase of the ISM, i.e. the CRe and the magnetic fields via their synchrotron emission. The radio emission of RPS galaxies often appears to be in excess of what is expected given their SFR inferred from observations at other wavelengths \citep[e.g.][]{Gavazzi1991MultifrequencyPhotometry,Murphy2009,Vollmer2010,Chen2020,Ignesti2022Gasp,Ignesti2022WalkGalaxies}. 
Several explanations for the radio-excess of cluster SF-galaxies have been discussed in the literature:
\citet{Gavazzi1999OnGalaxies} proposed that the ram-pressure leads to a compression of the magnetic field and consequentially, increased synchrotron luminosity. 
\citet{Volk1994} and \citet{Murphy2009} favored a scenario where the higher radio luminosity is not simply caused by compression, but instead due to shocks driven into the ISM by collisions with fragments of cold gas present in the ICM. Diffusive shock acceleration can then accelerate CRe in the ISM, which should result in a flatter radio spectral index.
A third explanation of the radio-excess was brought up more recently in \citet{Ignesti2022WalkGalaxies} and \citet{Ignesti2022Gasp} based on the analysis of so-called jellyfish galaxies which are the most extreme examples of galaxies undergoing strong RPS events. The strong radio-excess compared to the H$\upalpha$-emission was interpreted as the consequence of a rapid quenching of the star-forming activity due to RPS. While the H$\upalpha$-emission is a nearly instantaneous SF-tracer, with a typical delay of only a few Myr, the radio-emitting CRe have a typical lifetime $>100$\,Myr. Thus, if the star formation is quenched on time-scales shorter than the CRe lifetime, we will observe an apparent excess of radio emission simply due to the different time-scales probed by H$\upalpha$ and the radio observations. As a consequence, the spectral index of those objects should be rather steep owing to spectral aging. %Furthermore, the radio-excess should be significantly less prominent compared to UV and IR-inferred SFRs, since those also trace longer time scales $\sim 100$\,Myr \citep{Kennicutt2012StarGalaxies}.

This work is part of the VIrgo Cluster multi-Telescope Observations in Radio of Interacting galaxies and AGN (ViCTORIA) project, a broadband radio imaging campaign of the Virgo cluster (de Gasperin et al. in prep.).
In \citet{Edler2023}, we recently published a 144\,MHz survey of the Virgo cluster region using the LOFAR high-band antenna system as first data release of ViCTORIA. The survey lead to the radio detection of 112 cluster members, with $19$ objects that show signs of RPS in their radio morphology. We will use this data set to study the RPS phenomenon in the Virgo cluster.

In this work, we assume a flat $\mathrm{{\Lambda}CDM}$ cosmology with $\Omega_\mathrm{m}=0.3$ and $H_0=70\,\mathrm{km\,s^{-1}\,Mpc^{-1}}$. At the distance of M\,87, for which we adopt a value of 16.5\,Mpc \citep{Mei2007,Cantiello2018TheBeyond}, one arcsecond corresponds to 80\,pc. 
This paper is arranged as follows: In \autoref{sec:data}, we report the data used for this work and our samples. In \autoref{sec:statistical}, we present a statistical study of the radio and star-forming properties and in \autoref{sec:local}, we carry out a spatially resolved analysis of the LOFAR maps. Finally, our conclusions are outlined in \autoref{sec:conclusion}.

\section{Data and sample}\label{sec:data}

This paper builds primarily on the LOFAR HBA Virgo Cluster Survey \citep{Edler2023} published as part of the ViCTORIA-project. This 144\,MHz-survey covers a $132\,\mathrm{deg}^2$ field around the Virgo cluster and is to date the deepest blind radio continuum survey of this particular region. 

\subsection{RPS sample}\label{sec:rpssample}
\begin{table*}
\begin{tiny}
    \centering
    \caption{LOFAR sample of ram-pressure stripped galaxies in the Virgo cluster.}\label{tab:rpsgalaxies}
    \begin{tabular}{llllllcllll}
    \hline\hline
        VCC & NGC & IC & Morphology & $\log{{M_\star}}$ & \hi-def. & $\Delta{v}_\mathrm{rad, M87}$ & Literature & Comment \\ 
        & & & & [$\log{M_\odot}$] & & [km/s] & & \\
        (1) & (2) & (3)&  (4)   & (5) & (6) & (7) & (8) & (9) \\        \hline
        241 & ~ & 3105  &  SW tail            & 7.89  & 0.41   & $-1439$& B22 & new radio tail,   H$\upalpha$ tail \\ 
        307 & 4254 & ~  &  N tail             & 10.39 & 0.06   &  1124& V05, H07, V12b, B18a, M09 & RPS + tidal interaction  \\ 
        497 & 4302 & ~  &  light N tail       & 10.44 & 0.50   & $-153$  & C07, V13, W12 & new radio tail, \hi{} tail  \\ 
        630 & 4330 & ~  &  S tail             & 9.52  & 0.92   &  321  & C07, M09, V12a, V12b, V13, F18 & \hi{} tail, H$\upalpha$ tail  \\ 
        664 & ~ & 3258  &  W tail             & 8.20  & 0.57   & $-1713$  & ~ &  new candidate \\ 
        %836 & 4388 & ~ & asym & ~ & 0.9      8 & V07, V10, V  & 18, M09, W12 & ~ & ~ \\ 
        865 & 4396 & ~  &  NW tail            & 9.25  & 0.20   & $-1398$  & C07, V07, V10, M09 & \hi{} tail  \\ 
        873 & 4402 & ~  &  NW tail            & 10.04 & 0.83   & $-1049$  & CR05, V07, M09, V10, V12b & \hi{} tail, stripped dust  \\ 
        979 & 4424 & ~  &  small S tail       & 10.17 & 0.98   & $-842$  & C07, V13, B18b & \makecell[tl]{\hi{} tail, H$\upalpha$ tail/outflow, \\radio deficient}  \\ 
        %1043 & 4438 & ~  & asym & ~ & 1      .20 & W07,V07,   & V08, V09, V10 & Tidal interaction & ~ \\ 
        1401 & 4501 & ~ &  NE tail            & 10.98 & 0.58   & 999  & W07, V07, V10, V12b & ~  \\ 
        1450 & ~ & 3476 &  long W tail        & 9.02  & 0.66   & $-1112$ & B21 & \makecell[tl]{new radio tail, H$\upalpha$ tail \\ M87 side-lobes } \\ 
        1516 & 4522 & ~ & long NW tail        & 9.38  & 0.68   & 1037 & V04, M09, V12b & ~  \\ 
        1532 & ~ & 800  & NE tail             & 9.07  & 1.00   & 1051 & ~ &  new candidate \\ 
        1615 & 4548 & ~ & N-S asymmetry       & 10.74 & 0.94   & $-802$ & V99, W07 & old stripping event & ~ \\ 
        1690 & 4569 & ~ &  SW tail, outflows & 10.66 & 1.05   & $-150$3 & M09, V12b, B16  & H$\upalpha$ tail \\ 
        1868 & 4607 & ~ &  asymmetry          & 9.6 0 & 1.20   & 973 & ~ & new candidate \\ 
        1932 & 4634 & ~ &  SW tail            & 9.57  & 0.45   & $-1163$ & S18 &  \makecell[tl]{new candidate, star-forming \\ object in the tail (S18)}  \\ 
        1987 & 4654 & ~ & asymmetry           & 10.14 &$-0.05$ &$-240$ & V03, V07, C07, W07, V10 & \makecell[tl]{\hi{} tail, RPS and  tidal \\interaction} \\ 
    \hline
    \end{tabular}
    \tablefoot{ Columns 1 to 3 list the identifiers of the galaxies in the VCC, NGC and IC. Column 4 comments on the LOFAR morphology of the sources and columns 5 to 7 show the stellar mass \citep[from ][]{Boselli2015HGalaxies}, \hi{}-deficiency (from \citet{Boselli2014coldgas} and \citet{Koppen2018RamApproach}) and the radial velocity relative to M\,87 (taken from HyperLeda\footnote{\url{https://leda.univ-lyon1.fr/}}). In columns 8 and 9, we list the relevant literature and additional comments.\\
    \textbf{References.} V99 -- \citet{Vollmer1999}; -- \citet{Vollmer2003NGCStripping}; \citet{Vollmer2004N4522} -- V04; CR05 -- \citet{Crowl2005Dense4402}; V05 -- \citet{Vollmer2005NGCCluster}; C07 -- \citet{Chung2007Tails}; H07 -- \citet{Haynes2007NGCSurvey}; W07 -- \citet{Wezgowiec2007}; V07 -- \citet{Vollmer2007TheGalaxies}; M09 -- \citet{Murphy2009}; V10 -- \citet{Vollmer2010}, V12a -- \citet{Vollmer2012Ram4330}; V12b -- \citet{Vollmer201212Gals}, W12 -- \citet{Wezgowiec2012}; V13 -- \citet{Vollmer2013}; B16 -- \citet{Boselli2016Spectacular4569}; B18a -- \citet{Boselli2018VESTIGE3}; B18b -- \citet{Boselli2018VESTIGE4}; S18 -- \citet{Stein2018}; B21 -- \citet{Boselli2021Astrophysics3476}; B22 -- \citet{Boselli2022RamEnvironments}.}
\end{tiny}
\end{table*}
\begin{figure*}
    \centering
    \includegraphics[width=0.3\linewidth]{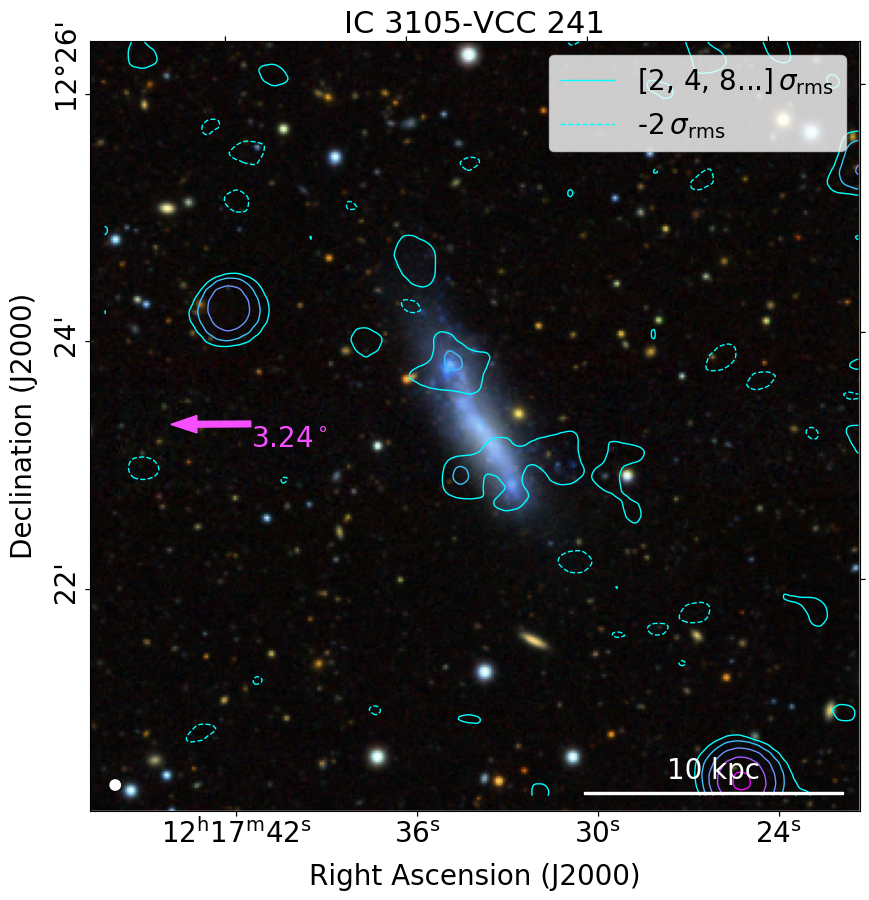}
    \includegraphics[width=0.3\linewidth]{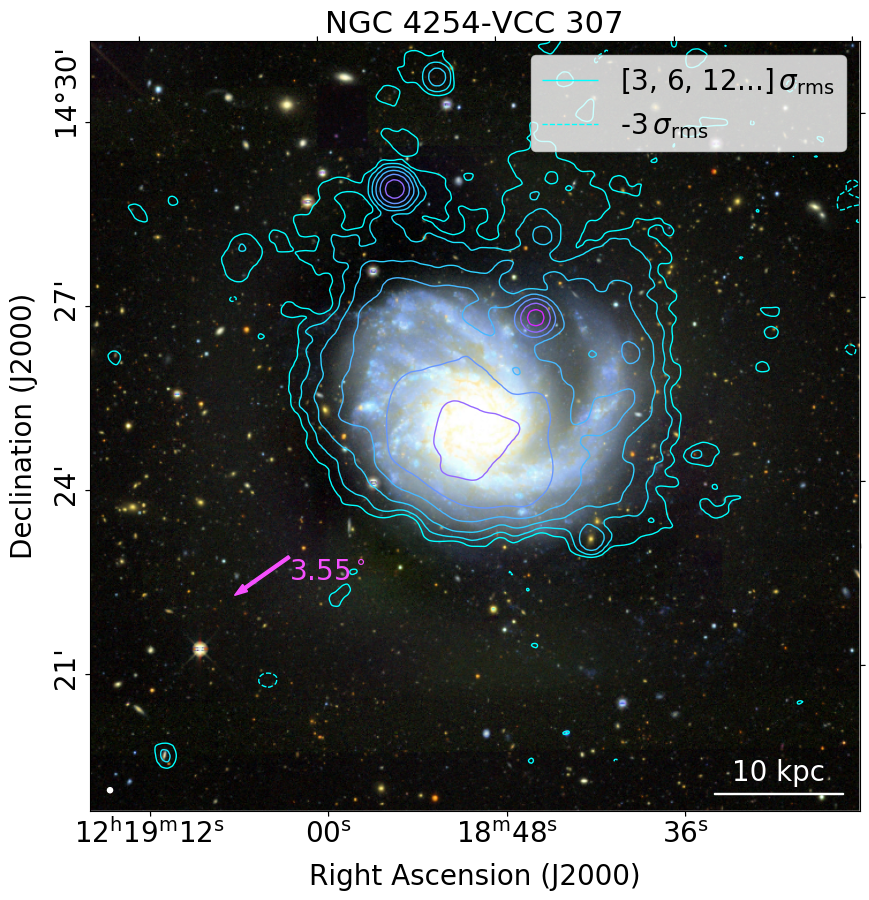}
    \includegraphics[width=0.3\linewidth]{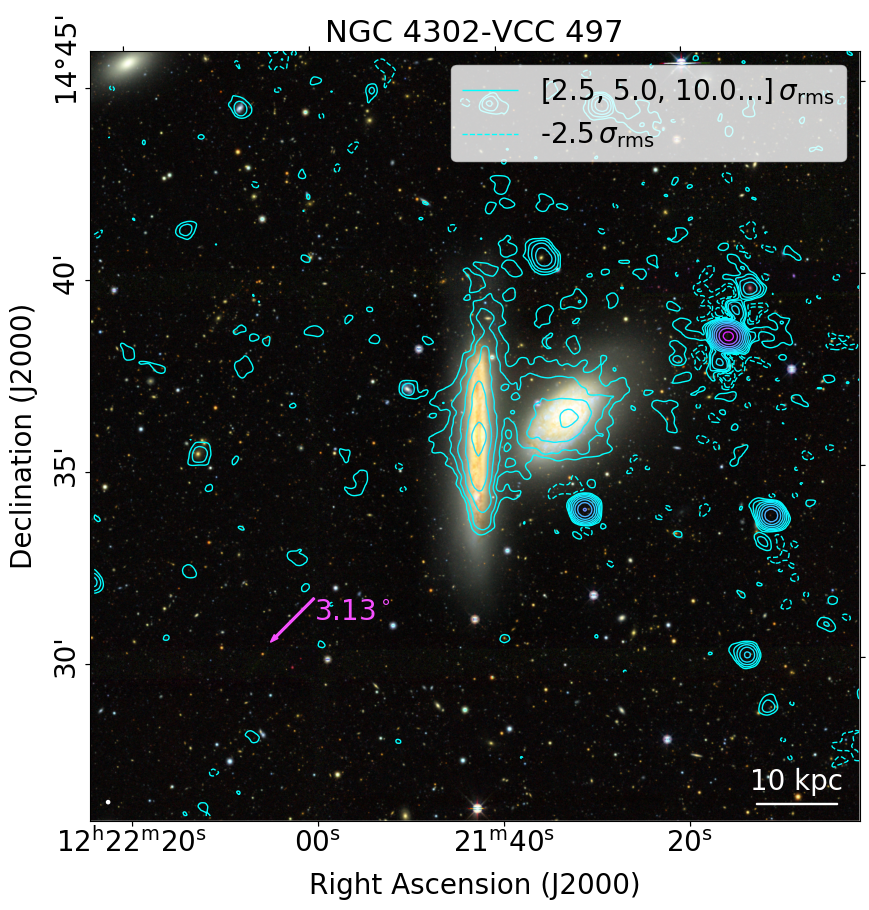}
    \includegraphics[width=0.3\linewidth]{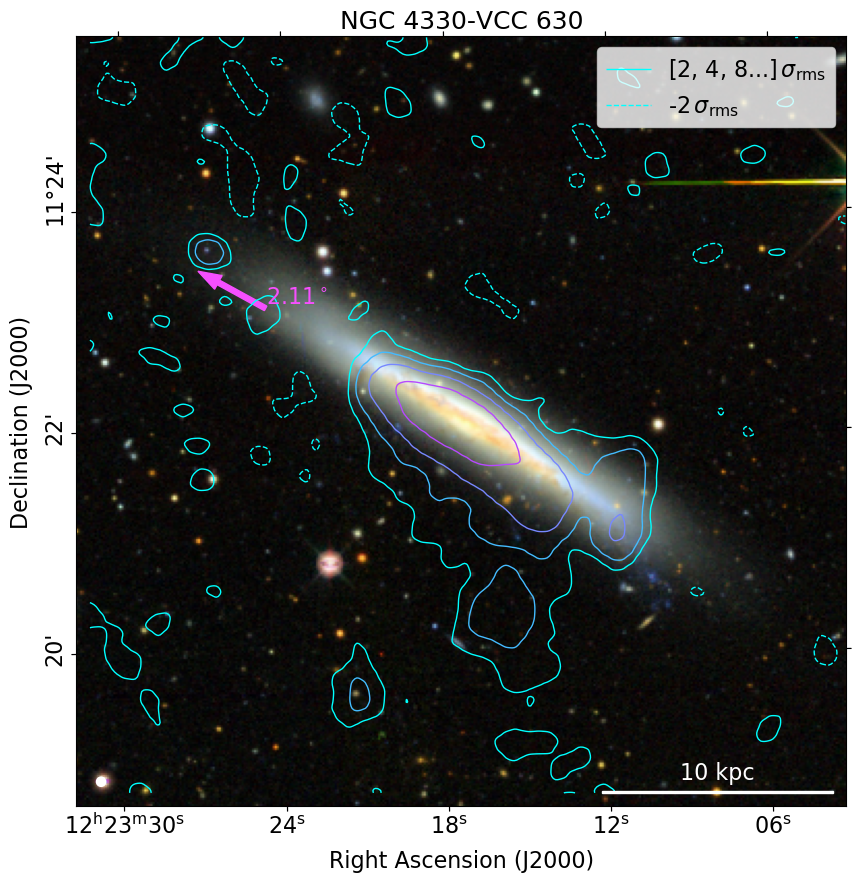}
    \includegraphics[width=0.3\linewidth]{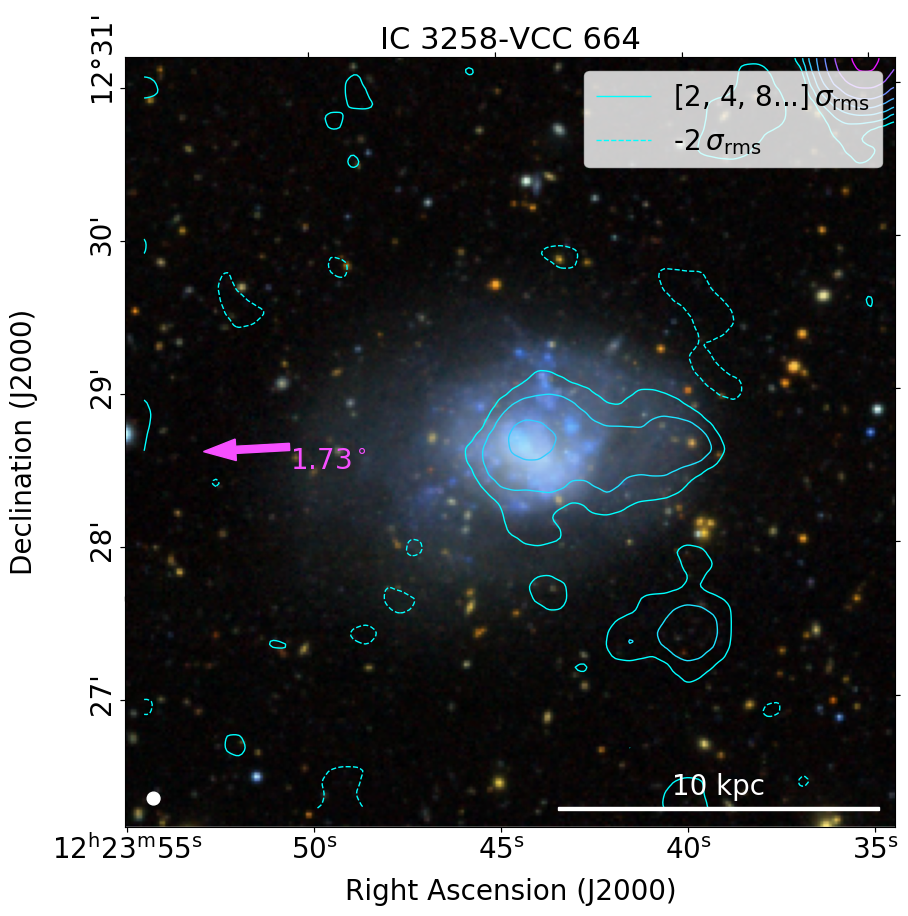}
    \includegraphics[width=0.3\linewidth]{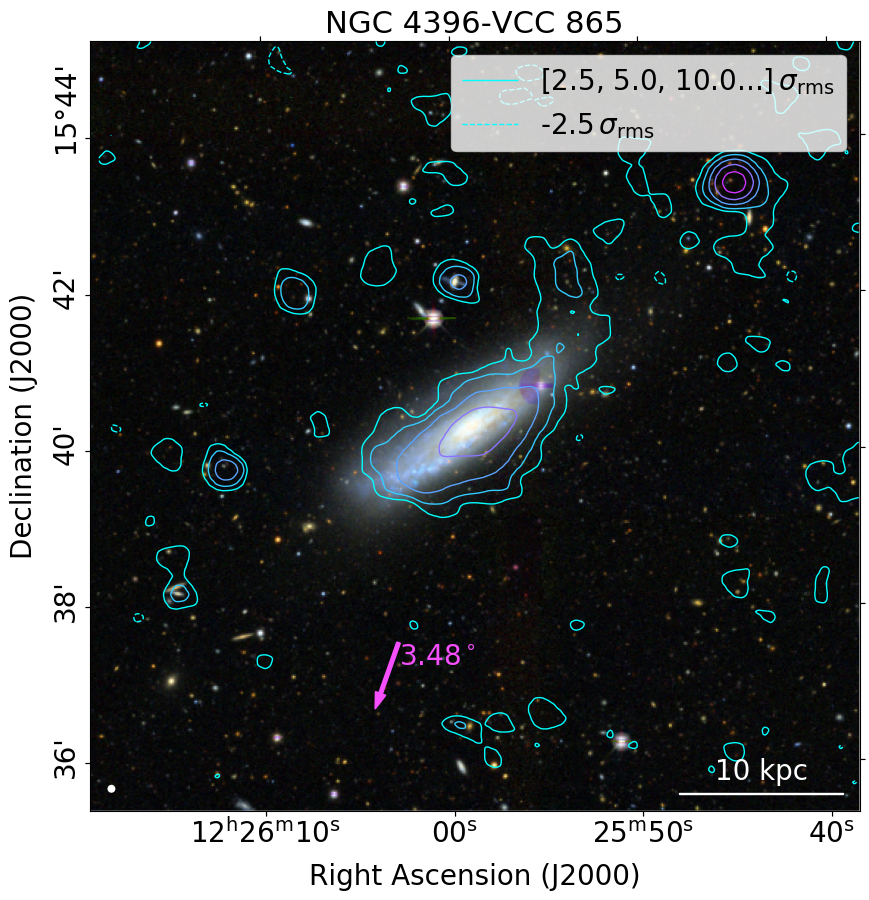}
    \includegraphics[width=0.3\linewidth]{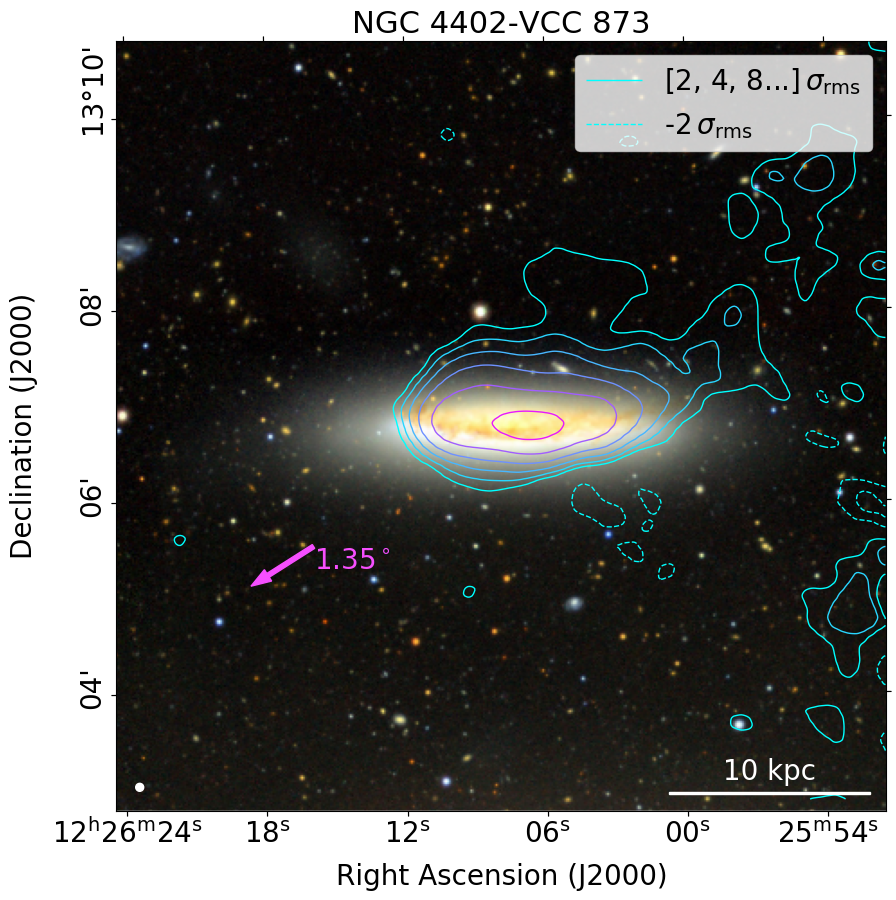}
    \includegraphics[width=0.3\linewidth]{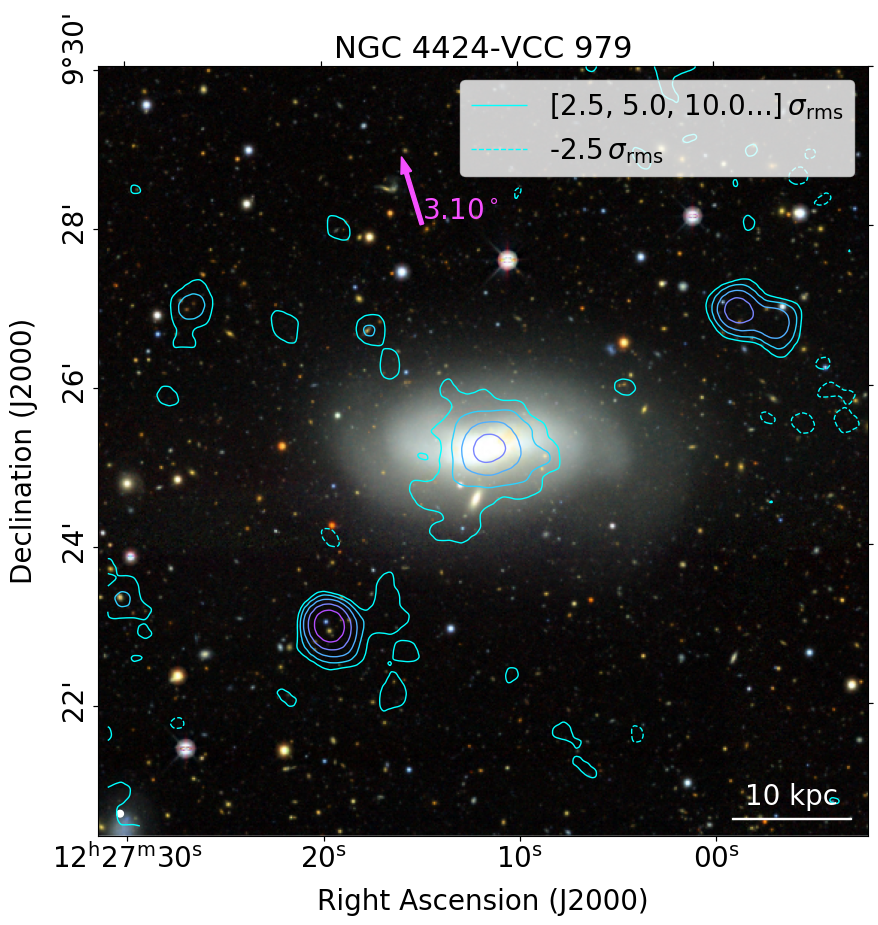}
    \includegraphics[width=0.3\linewidth]{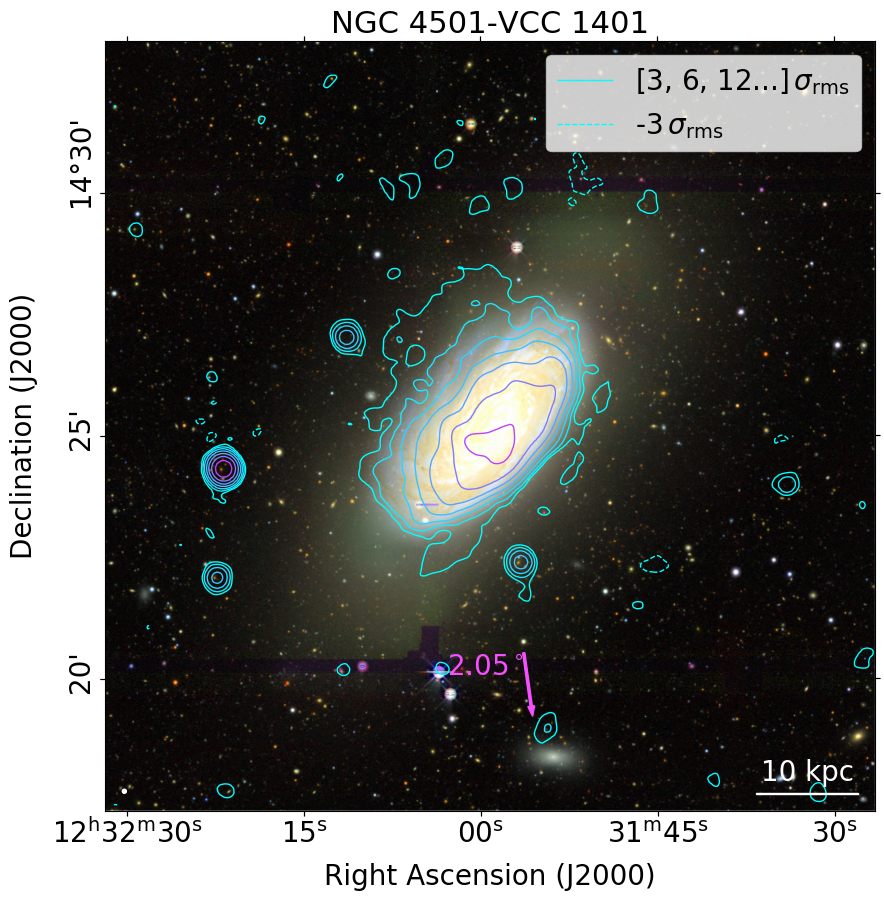}
    \includegraphics[width=0.3\linewidth]{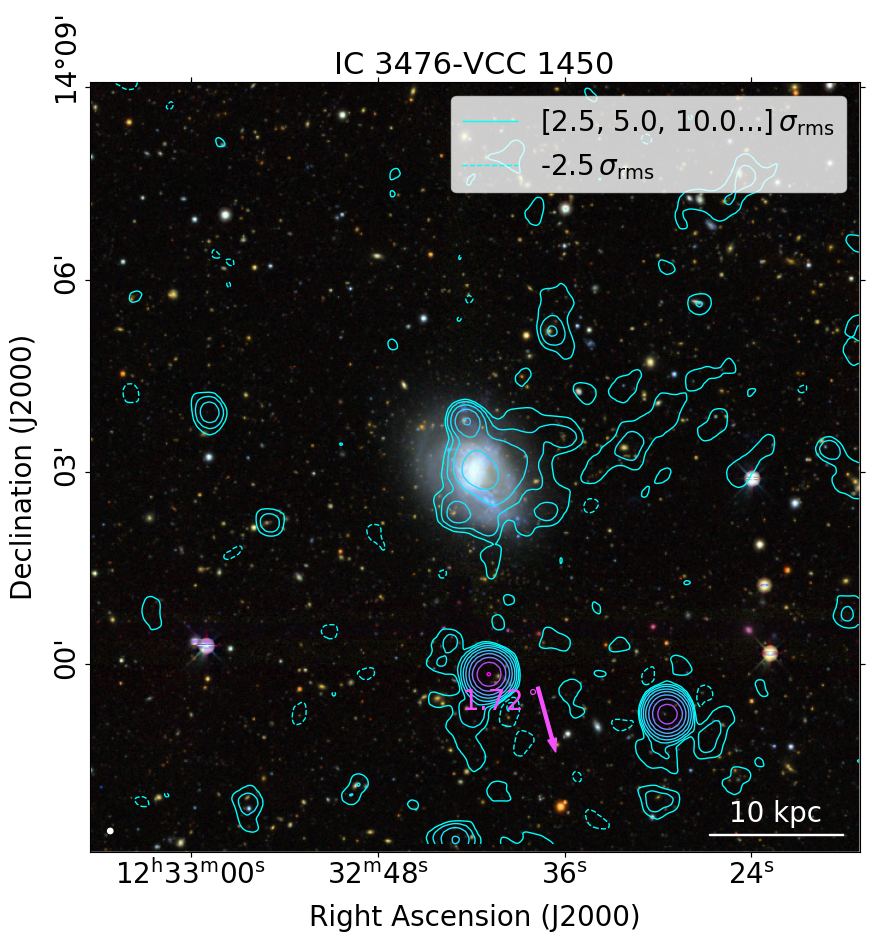}
    \includegraphics[width=0.3\linewidth]{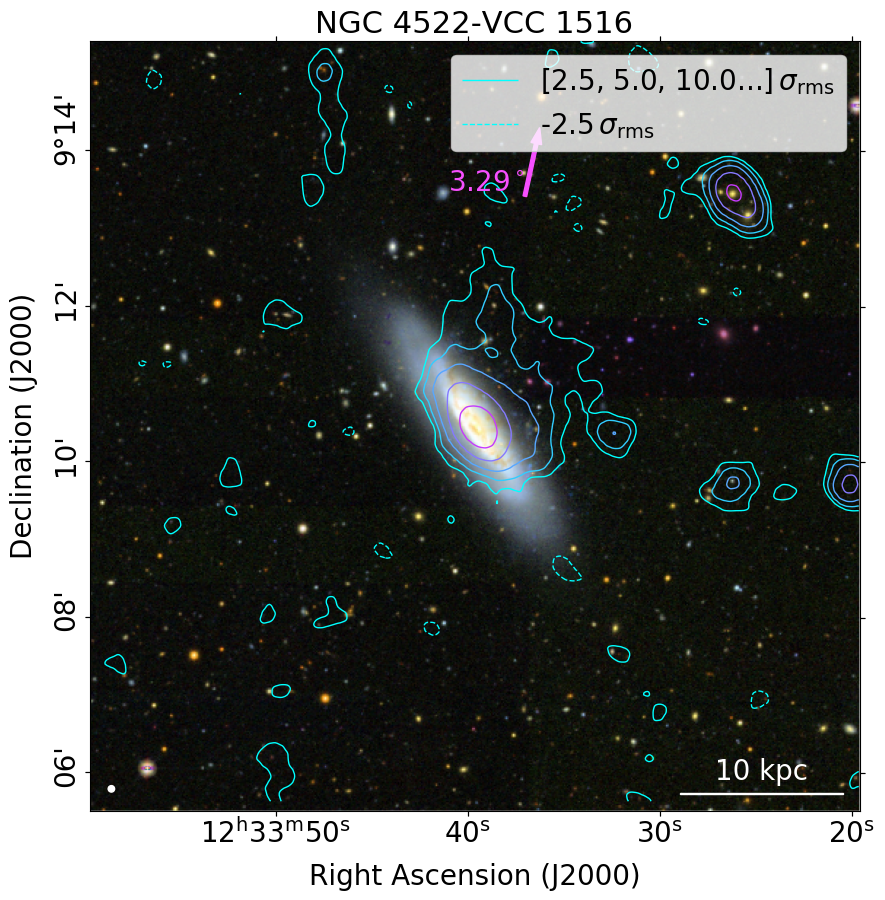}
    \includegraphics[width=0.3\linewidth]{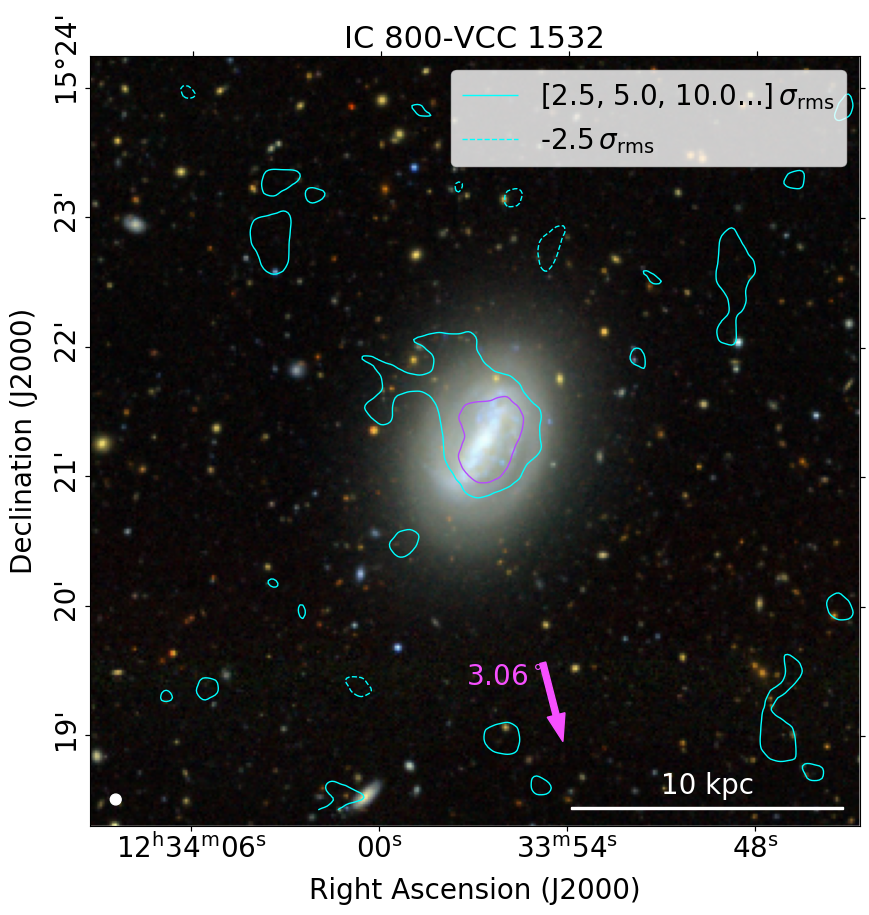}
    \caption{Radio continuum emission in galaxies with a radio-morphology indicative of RPS. We show LOFAR radio continuum intensity contours at 144\,MHz from the $20''$-resolution maps overlaid on top of optical images from the DESI Legacy Survey. Contour-levels increase in increments of powers of two and pink arrows indicate the direction and angular distance (in units of degree) to the center of the Virgo cluster.}
    \label{fig:contours}
\end{figure*}
\begin{figure*}[t]
\ContinuedFloat
    \centering
    \includegraphics[width=0.3\linewidth]{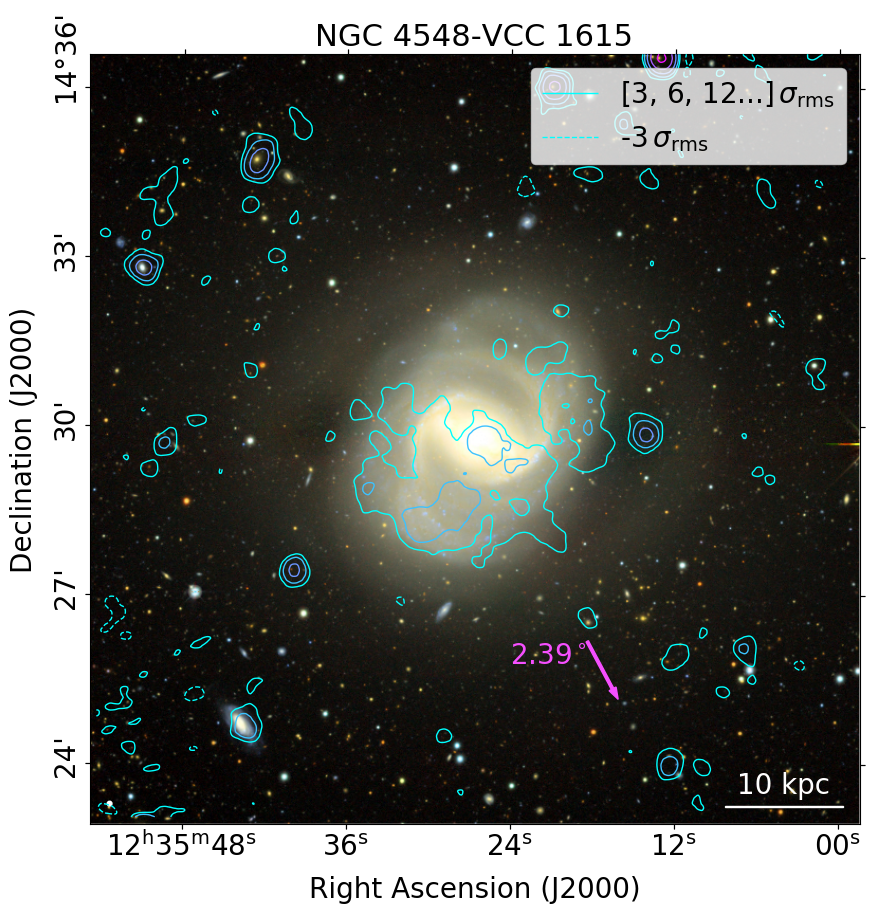}
    \includegraphics[width=0.3\linewidth]{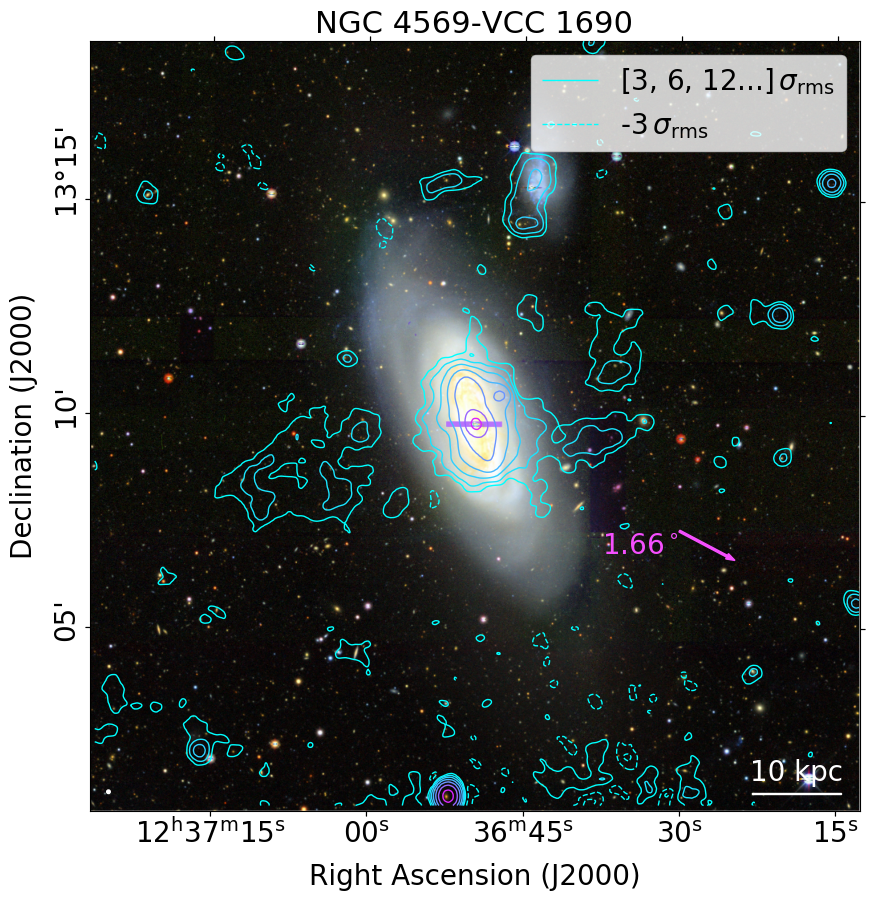}
    \includegraphics[width=0.3\linewidth]{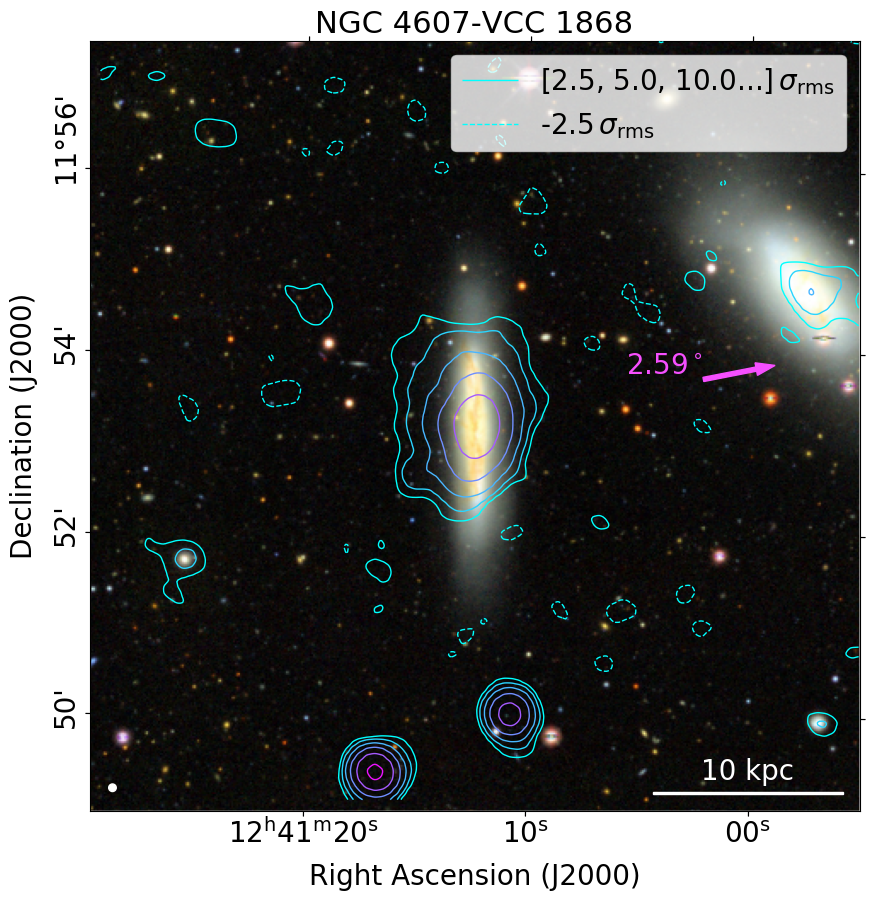}
    \includegraphics[width=0.3\linewidth]{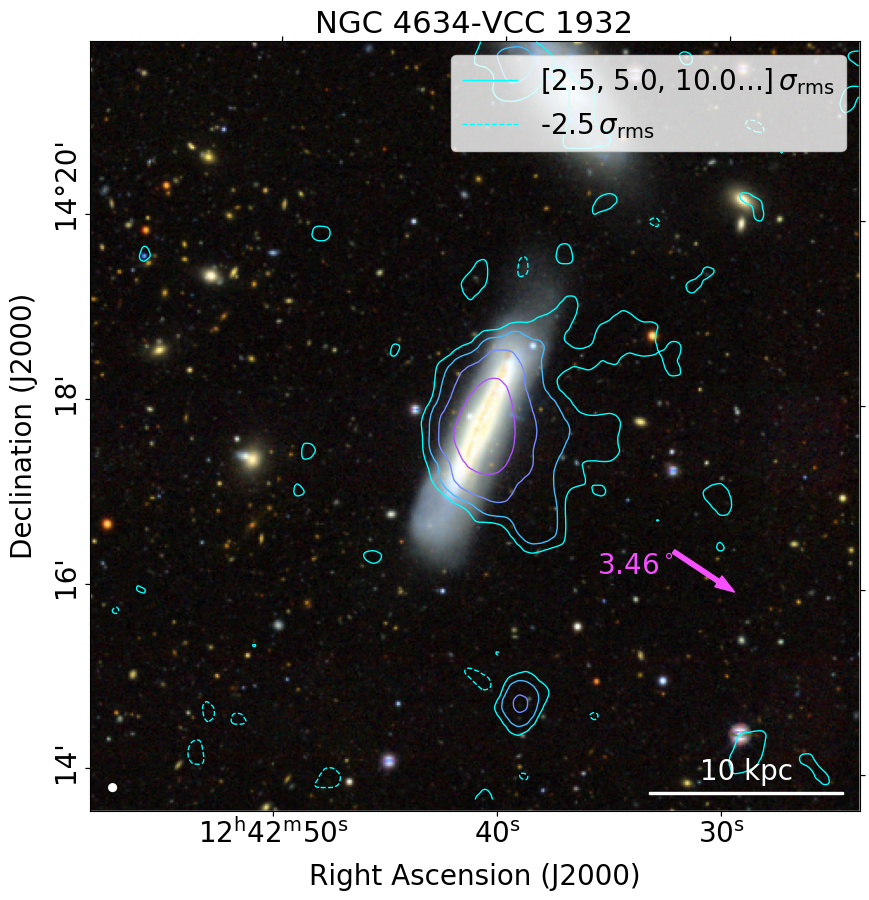}
    \includegraphics[width=0.3\linewidth]{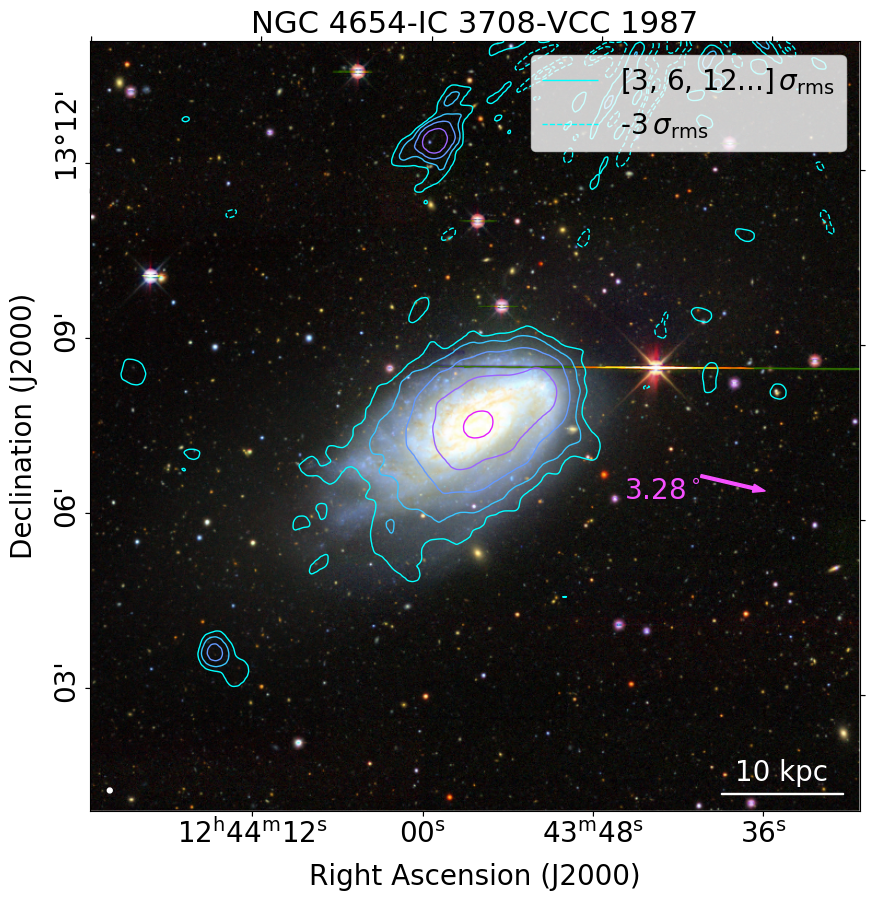}
    \caption{Continued.}
\end{figure*}
For our LOFAR-study of star-forming galaxies in the Virgo cluster, we composed a sample of objects detected in \citet{Edler2023} that show signs of RPS in the radio-continuum. As criteria, we require an asymmetric radio morphology in combination with an undisturbed optical appearance. The latter is to exclude cases where the dominant perturbing mechanism is tidal interaction. If the available literature clearly shows that an object suffers both from tidal interactions and RPS, we include it in our sample. This is the case for NGC\,4254 \citep{Vollmer2005NGCCluster}, NGC\,4438 \citep{Vollmer2005NGC4438} and NGC\,4654 \citep{Vollmer2003NGCStripping}. 
The Virgo cluster has been the target of several multi-frequency surveys and dedicated studies of individual objects. Out of the 19 galaxies with evident tails in radio continuum, 15 were already identified as suffering a RPS event in these detailed studies, as summarized in \citet{Boselli2022RamEnvironments}. They show tails of stripped atomic or ionized gas, have truncated gaseous and star-forming discs, and have been identified as suffering a RPS event from tuned models and hydrodynamic simulations.   
Four of the objects are for the first time reported to be RPS candidates based on our LOFAR data: IC\,800 shows a radio tail to the northeast, in agreement with being on a radial orbit towards the cluster center, and is strongly \hi{} deficient. IC\,3258 is a dwarf galaxy with a radio tail that is also in agreement with a highly radial orbit. NGC\,4607 shows a mild asymmetry with an extension of the radio contours towards the northeast and a strong gradient in the opposite direction, it is also highly deficient in neutral hydrogen. Lastly, NGC\,4634 shows a radio tail towards the west and is slightly deficient in \hi{}, a star-forming object is coincident with the radio tail \citep{Stein2018}. Preliminary images from the ViCTORIA MeerKAT Survey confirm the morphological peculiarities found in LOFAR for all four objects.
For four other galaxies in our RPS sample (IC\,3105, IC\,3476, NGC\,4302 and NGC\,4424), the impact of RPS was previously observed at other wavelengths \citep{Chung2007Tails,Boselli2021Astrophysics3476,Boselli2022RamEnvironments} but not in the radio continuum. 

Since our analysis focuses on the star-forming properties, radio emission due to active galactic nucleus (AGN) activity is a strong contamination. Thus, we investigate the possible contribution of AGN-emission for the galaxies that show signs of ram-pressure stripping in the radio continuum.
Three of those objects, NGC\,4388, NGC\,4438 and NGC\,4501 are Seyfert-galaxies. The first two have a radio-morphology that is clearly dominated by nuclear emission \citep{Hummel1991}. However, in NGC\,4501, the nuclear point-source accounts for only 1\% of the total radio emission. The other galaxies mostly host \hii{}-nuclei in the BPT \citep[Baldwin, Phillips \& Terlevich;][]{Baldwin1981} and WHAN \citep[$W_\mathrm{H\upalpha}$ versus N\,{\sc ii}/H$\upalpha$;][]{CidFernandes2011} diagrams according to \citet{Gavazzi2018}. Exceptions are NGC\,4302, NGC\,4548 and NGC\,4569 and which have a nuclear WHAN classification as AGN and an inconclusive nuclear BPT classification. NGC\,4607 is also classified as AGN in the WHAN diagram and as LINER \citep[Low Ionization Nuclear Emission-line Region;][]{Heckman1980} or transition object in the BPT, but has an integrated classification as an \hii{}-galaxy. Of those, only NGC\,4569 shows a compact nuclear source in the LOFAR maps. While a minor contribution of an AGN cannot be ruled out for this object, this central source is dominantly fueled by nuclear star formation giving rise to prominent outflows \citep{Boselli2016Spectacular4569}. Thus, the only AGN we remove from our sample are NGC\,4388 and NGC\,4438.

We are left with a sample of 17 RPS galaxies (see \autoref{tab:rpsgalaxies}). Their $20\arcsec$-resolution LOFAR contours on top of the optical images of the Dark Energy Spectroscopic Instrument (DESI) Legacy Survey data release 10 \citep{Dey2019} are displayed in \autoref{fig:contours}. They range from galaxies with clear and prominent 144\,MHz tails, such as NGC\,4330, NGC\,4396, NGC\,4522 or IGC\,3476 to objects with only mild asymmetry like NGC\,4548 and NGC\,4607.
With the availability of high-sensitivity images of the nearby Virgo cluster, we start probing the regime of dwarf galaxies with radio continuum tails by detecting lower mass systems, down to a stellar mass of $M_\star = 7.8\times10^7 \mathrm{M_\odot}$. Previous studies of more distant clusters and groups are limited to objects with masses $M_\star \geq 10^9 \mathrm{M_\odot}$ \citep{Roberts2021LoTSSClusters,Roberts2021II,Roberts2022LoTSSCluster,Ignesti2022WalkGalaxies}. 

A number of galaxies that were reported in the literature to suffer from RPS \citep{Boselli2022RamEnvironments} are not part of our LOFAR RPS sample. Of those, IC\,3412, IC\,3418, NGC\,4506 and UGC\,7636 are non-detections in LOFAR at $4\,\sigma$. Others, mostly galaxies which have tails in \hi{}, are radio-detected but show a symmetric radio morphology. Those objects are NGC\,4294, NGC\,4299, NGC\,4469, NGC\,4470 NGC\,4491 and NGC\,4523 \citep{Chung2007Tails,Boselli2023a}. For these objects, we speculate that they experience rather mild ram-pressure, which mostly affects the outskirts of the ISM, such that a tail in radio continuum is below our sensitivity threshold. Alternatively, the peculiar \hi{} morphology of these objects could be due to other process than RPS. In the case of NGC\,4470, the non-detection of a continuum tail is due to local dynamic range limitations in the LOFAR map, indeed preliminary images of our ViCTORIA MeerKAT survey reveal a prominent tail at 1.3\,GHz. For NGC\,4523, a continuum counterpart to the \hi{} tail which we reported in \citet{Boselli2023a} is also observed in the preliminary MeerKAT maps, although at low significance. 

\subsection{LOFAR-HRS sample}

To assess how the non-thermal properties of galaxies in the Virgo cluster are shaped by their local surroundings, we compile a comparison sample of nearby galaxies with high-quality star formation tracer data available. For this, the Herschel Reference Survey \citep[HRS;][]{Boselli2010TheSurvey}, which also contains all but two of the galaxies in our RPS sample, is well suited. This survey consists of a statistically complete, $K$-band limited sample of 323 nearby (15--25\,Mpc distance) galaxies. Around a quarter of those galaxies reside in the Virgo sub-clusters around either M\,87 or M\,49, another quarter is located in other sub-structures of the Virgo cluster or its outskirts. The remaining objects are in less dense environments such as groups and pairs or are isolated galaxies. This allows us to compare the properties of galaxies in the Virgo cluster with those that are inhabiting poorer surroundings.
%The HRS sample was observed in the mid-to-far infrared (IR) with Spitzer-MIPS, in the far-infrared with Herschel-PACS and -SPIRE \citep{Ciesla2012SubmillimetreSurvey,Cortese2014PACSGalaxies} in H$\mathrm{\alpha}$ \citep{Boselli2015HGalaxies} and in the ultra-violet (UV) with GALEX \citep{Cortese2012TheGalaxies}, spectral lines \citep{Boselli2013IntegratedGalaxies}.
% Distances from \citet{Cortese2012TheGalaxies}.

We construct a sample of the 144\,MHz LOFAR-detected star-forming galaxies in the HRS. For this, we use data of the LOFAR HBA Virgo Cluster Survey \citep{Edler2023}, which covers the majority of the galaxies in the Virgo Cluster. In addition, all the HRS galaxies at declinations of $\delta>30\si{\degree}$ are covered by the second data release of the LOFAR Two-metre Sky Survey \citep[LoTSS-DR2;][]{Shimwell2022TheRelease}. Further galaxies were observed by more recent, previously unpublished observations of LoTSS, processed by the LOFAR Surveys Key Science Project\footnote{\url{https://lofar-surveys.org/}}. All LOFAR HBA observations were taken with identical observational settings and processed using the \texttt{ddf-pipeline} algorithm \citep{Tasse2021TheImaging}. For the Virgo cluster observations, additional pre-processing was necessary due to the close proximity to the extremely luminous radio galaxy M\,87 (Virgo\,A), those steps are described in \citet{Edler2023}.

\begin{figure}
    \centering
    \includegraphics[width=\linewidth]{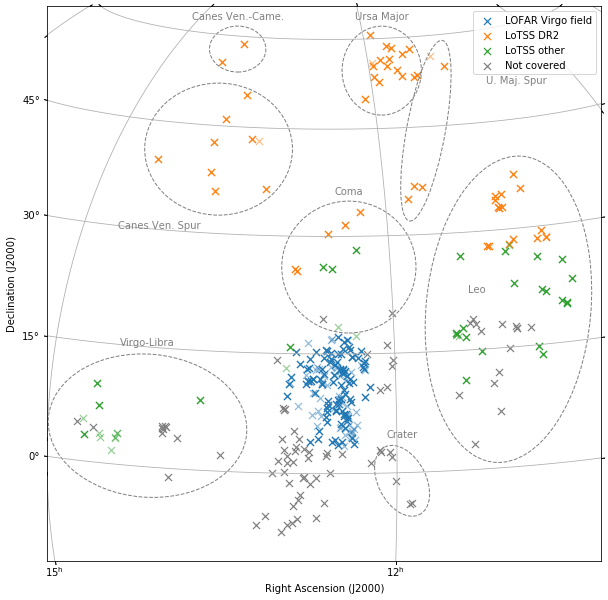}
    \caption{LOFAR-coverage of the HRS. The galaxies are either in the LOFAR Virgo Cluster Survey (blue), in LoTSS-DR2 (orange) or covered in further LoTSS observations (green). Grey crosses are not covered by LoTSS, and colored but partially transparent points are covered but not detected. Dashed regions mark structures of galaxies other than the Virgo cluster.}
    \label{fig:coverage}
\end{figure}

For the HRS galaxies in the footprint of the LOFAR HBA Virgo Cluster Survey, we use the 144\,MHz measurements reported in \citet{Edler2023} based on the  $20\arcsec$-resolution LOFAR-maps. In this paper, all HRS galaxies within the survey area were considered. We extend this sample also to the HRS galaxies covered by LoTSS-DR2 as well as to those which are within $1.97^\circ$ angular distance of unpublished LoTSS-observations processed before August 2023. We manually inspect all HRS galaxies in the LoTSS $20\arcsec$-maps and, if they are visible in the radio images, we measure the flux density in a region encompassing the $3\sigma$-contours as it was done in \citet{Edler2022} for the Virgo objects. For  observations outside of LoTSS-DR2, we use the pointing closest to the galaxy of interest and we take into account a field-dependent factor to align the flux density scale with the one of the 6\,C-survey \citep{Hales1988TheFormula} and NRAO VLA Sky Survey  \citep[NVSS;][]{Condon1998TheSurvey}. This factor was calculated as described in \citet{Shimwell2022TheRelease}. 
We consider all galaxies with an integrated signal-to-noise ratio above four as radio-detected. While we use the 10\% systematic uncertainty of the flux density scale for LoTSS-DR2 as reported in \citep{Shimwell2022TheRelease}, a larger uncertainty of 20\% is assumed for the other LOFAR maps due to the reduced overlap and statistics, the higher uncertainty of the primary beam model for low declinations, and the presence of bright sources such as M\,87  (for sources in the Virgo field, \citet{Edler2023}). 
In \autoref{fig:coverage}, we show the sky distribution of the HRS galaxies and their coverage by the different LOFAR projects. 
In total, out of the 261 late-type galaxies in the statistically complete sample of the HRS (excluding ellipticals and lenticulars), 193 are in the footprint of our LOFAR observations.
Of those, 141 galaxies are detected, 76 are in the LOFAR Virgo field, 44 are from LoTSS-DR2, and 21 are in LoTSS-fields outside of the LoTSS-DR2 footprint. 

Since we are interested in radio emission as tracer of the SFR, we need to exclude objects where a significant fraction of the radio emission is due to an active galactic nucleus (AGN). By excluding elliptical and lenticular galaxies, we remove most objects with strong AGN contamination. In \citet{Gavazzi2018}, nuclear spectroscopy-based BPT and WHAN classifications for the HRS galaxies were presented. Eleven of the LOFAR-detected objects were classified as either strong or weak AGN in the WHAN diagram and as Seyfert galaxy in the BPT diagram. Visually inspecting the radio maps of those revealed that for six of them the nuclear point-like sources contribute $>$$15\%$ to the flux density. These objects (NGC\,3227, NGC\,4313, NGC\,4419, NGC\,4586) were removed from our sample.
% Seyfert AGN + sAGN
% 3227 ~ 33% nuclear
% 3380 ~ <5% nuclear
% 3982 - <5% nuclear
% 4412 ~ 15% nuclear
% 4419 ~ 24% nuclear
% Seyfert AGN + wAGN
% 3512 <5%
% 4313 strong nuclear
% 4501 ~ M88 <5% nuc
% 4586 strong nuclear
% 4438 strong nuc. -> LINER in Gavazzi18 but Seyfert in others
All galaxies in the RPS sample described in \autoref{sec:rpssample} except for IC\,3105 and IC\,3258, which are fainter than the limiting $K$-band magnitude of 12, are also part of the HRS. Those two objects will also be part of our analysis but are excluded from any fitting since they do not meet the selection criteria of the HRS. In the following, we will refer to the objects in our LOFAR-HRS sample minus the objects in our RPS sample for simplicity as the non-RPS sample. 

% WHAT to put in table? NGC / IC / UGC ? Ra /DEC , Mstar, L144, HI-def, dist.?, SI?
In \autoref{sec:appendix0}, we display the 144\,MHz measurements of the 137 LOFAR-detected star-forming galaxies used in this work.
The spectral luminosity $L_{144}$ at 144\,MHz is calculated from the measured flux densities $S$ according to $L_{144} = 4 \uppi d^2 S$; since our sample only consists of nearby galaxies at $v < 1750\,\mathrm{km\,s^{-1}}$, we neglect $k$-correction. 
We employed distances $d$ following the HRS \citep{Boselli2010TheSurvey}, with the difference that we set the distance to objects in the Virgo cluster to 16.5\,Mpc instead of 17\,Mpc to be consistent with what was assumed in the NGVS \citep{Ferrarese2012} and VESTIGE \citep{Boselli2018VESTIGE1}.

\subsubsection{Star formation tracers}
The integrated radio luminosities serve as a tracer of the SFR of the individual galaxies in the sample. A key advantage of radio-inferred SFR is that it is not affected by dust-attenuation \citep{Condon1992,Murphy2011CalibratingNGC6946}. Thus, no extinction-correction is required.
% intrinsic scatter of 015 dex Tabatabaei2017
While at low radio frequencies, the radio emission is almost free from the Bremsstrahlung-contribution of thermal electrons, the synchrotron lifetime of CRe in a magnetic field $B$ \citep{Beck2005}:
\begin{equation}
t_\mathrm{syn} = 88\,\mathrm{Myr} \left(\frac{B}{10\,\upmu\mathrm{G}}\right)^{-3/2} \left(\frac{\nu}{144\,\mathrm{MHz}}\right)^{-1/2}
\end{equation} is longer compared to CRe probed at higher frequencies. So an underlying assumption of SFRs derived from low frequency observations is that the SFR is constant on timescales of $\approx$$100$\,Myr.

To compare the radio luminosity to further tracers of the star-forming activity, we consider SFRs based on far-UV (FUV) and H$\upalpha$. The SFRs obtained from H$\upalpha$ and FUV were reported in \citet{Boselli2015HGalaxies} for the HRS star-forming galaxies based on the Salpeter initial mass function (IMF) and the calibration of \citet{Kennicutt1998}. We converted the SFR to a Chabrier IMF \citep{Chabrier2003} by applying a factor of 0.63 to the SFRs \citep{Madau2014}.     
The SFRs also need to be corrected for dust attenuation. For the UV-based SFR, \citet{Boselli2015HGalaxies} employed a correction based on the 24\,$\upmu$m emission. For the H$\upalpha$-inferred SFR, two approaches were compared in \citet{Boselli2015HGalaxies} -- a correction based on the Balmer-decrement using spectroscopic data \citep{Boselli2013IntegratedGalaxies} and a method relying on the \SI{24}{\micro\metre} dust emission. The authors found that the correction with the Balmer decrement $C(\mathrm{H}\beta)$ as defined in \citet{Lequeux1981} is only accurate if the fractional uncertainty is $\sigma[C(\mathrm{H}\beta)] < 0.1$; on the other hand, the correction using the \SI{24}{\micro\metre} emission can be biased for systems with a particularly low specific SFR due to the contribution of the old stellar population to the dust heating \citep{Cortese2008,Boselli2015HGalaxies}. Thus, we use the values corrected with the Balmer-decrement if $\sigma[C(\mathrm{H}\beta)]<0.1$, and with the \SI{24}{\micro\metre} emission otherwise. 

No uncertainty estimates are available for the SFRs published in \citet{Boselli2015HGalaxies}. In the following, we assume a systematic uncertainty of $15\%$ for the GALEX UV measurements \citep{GilDePaz2007} and the H$\upalpha$ photometry with the San Pedro Martir telescopes \citep{Boselli2015HGalaxies,Boselli2023SFMS}. We neglect the photometric uncertainty of the Spitzer $24\,\upmu\mathrm{m}$ measurements used for a dust correction as they have a uncertainty of only $2\%$ \citep{Engelbracht2007}.  We note that those estimates are only a rough first-order approximation of the true uncertainties, which also would require us to take into account the complex and hardly quantifiable dependencies on the dust and \nii{}-line corrections and the SFR conversion \citep[for discussions of those, see e.g.][]{Boselli2015HGalaxies,Boselli2016quenching,Boselli2023SFMS}.
We ensure that the SFRs are based on the same distances as the radio luminosities by re-scaling the SFRs by $(d/d_\mathrm{HRS})^2$.
 
Another common SFR-tracer is the infrared-emission which traces the dust heated by the young stellar population. As already mentioned, in systems with low specific SFR, older stellar populations also contribute to the dust heating. Low specific SFR systems are systematically more common in our sample which includes relatively quenched galaxies in the Virgo cluster \citep{Boselli2016quenching}. For this reason, we do not consider SFRs based purely on the infrared emission.

\subsubsection{Sample properties}

\begin{figure}
    \centering
    \includegraphics[width=\linewidth]{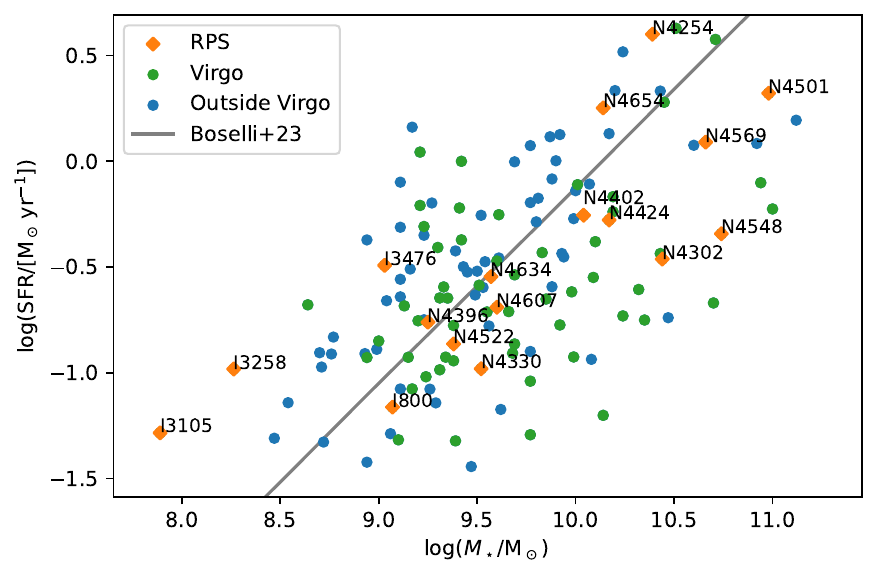}
    \caption{SFR (mean of H$\upalpha$ and UV-based values) as function of stellar mass for galaxies in the RPS sample (orange diamonds), galaxies in the Virgo cluster region (within $10^\circ$ from M\,87, green circles) and galaxies outside of the cluster (blue circles). For comparison, we show thee star-forming main sequence for Virgo cluster galaxies with normal \hi{}-content \citep[][grey line]{Boselli2023SFMS}.}% for the RPS sample (orange diamonds) and the remaining galaxies (blue circles). The black dotted line marks the Virgo cluster virial radius of $3.3^\circ$ \citep{Simionescu2017}.}
    \label{fig:stellarmasssfr}
\end{figure}

%\begin{figure}
%    \centering
%    \includegraphics[width=\linewidth]{hidef_dist.pdf}
%    \caption{\hi{}-deficiency against angular distance to M\,87 for the RPS sample (orange diamonds) and the remaining galaxies (blue circles). Galaxies with values $\approx0$ are not deficient in H\,{\sc i} whereas galaxies with values $>0.5$ are highly deficient. The black dotted line marks the Virgo cluster virial angular radius of $3\fdg 3$ \citep{Simionescu2017}.}
%    \label{fig:hidef}  
%\end{figure}

The stellar masses of our sample were obtained from \citet{Cortese2012TheGalaxies} who used the Chabrier IMF and span a large regime, ranging from $3.0\times 10^{8}\,\rm M_\sun$ to $1.3\times 10^{11}\,\rm M_\sun$. The two additional RPS galaxies outside of the HRS, IC\,3105 and IC\,3258, are of even lower mass with $M_\star = 7.8 \times 10^{7}\,\rm M_\sun$ and $M_\star = 8.3 \times 10^{7}\,\rm M_\sun$, respectively. In \autoref{fig:stellarmasssfr}, the mean of the H$\upalpha$ and UV-based SFR is shown as a function of the stellar mass. We also display the star-forming main sequence relation for Virgo cluster galaxies with normal \hi{}-content \citep{Boselli2023SFMS}. As RPS generally reduces the SFR, RPS galaxies are expected to mostly lie below this relation. In some cases, RPS galaxies can show a high specific SFR due to a temporary enhancement of SFR \citep{Bothun1986BlueDiskComa,Vulcani2018,Roberts2020enhancedSF}. Since our sample is limited to 144\,MHz detected objects, we are biased towards high-SFR objects, in particular at the low-mass end of the distribution.
% In \autoref{fig:hidef}, we display the \hi{}-deficiency against the angular distance from M\,87. The \autoref{fig:hidef} is a logarithmic measure of the  \hi{}-content of a galaxy compared to the expected content for that angular size and Hubble-type \citep{Haynes1984TheGalaxies}. Since galaxies are deprived of their neutral hydrogen through environmental processes, cluster galaxies are systematically \hi{}-deficient, this makes \hi{}-deficiency a proxy of present or past environmental interaction. As can be seen in \autoref{fig:hidef}, the Virgo cluster galaxies are more likely to lack neutral hydrogen with a median \hi{}-deficiency of 0.48 for objects within and 0.08 for objects outside of $2r_\mathrm{vir} = 6\fdg 6$ angular distance from the cluster center. The median deficiency of the RPS sample is 0.66.

\section{Statistical analysis}\label{sec:statistical}
\subsection{Radio-SFR-relation}

\begin{figure}
    \centering
    \includegraphics[width=1.0\linewidth]{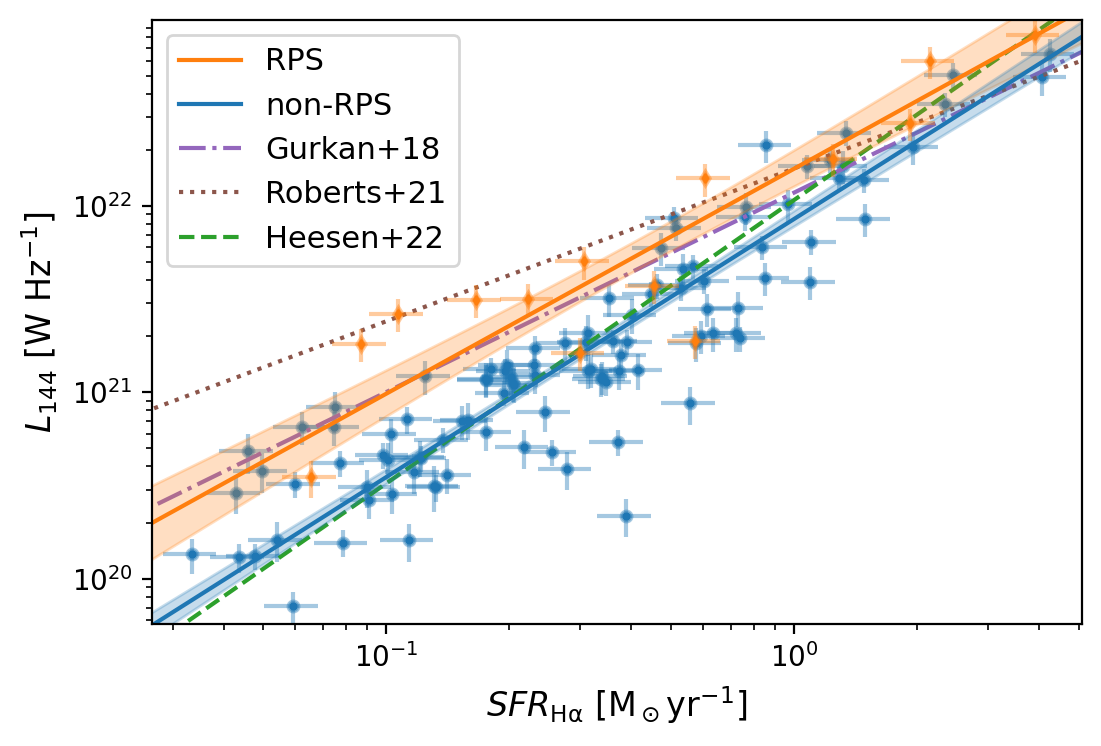}
    \includegraphics[width=1.0\linewidth]{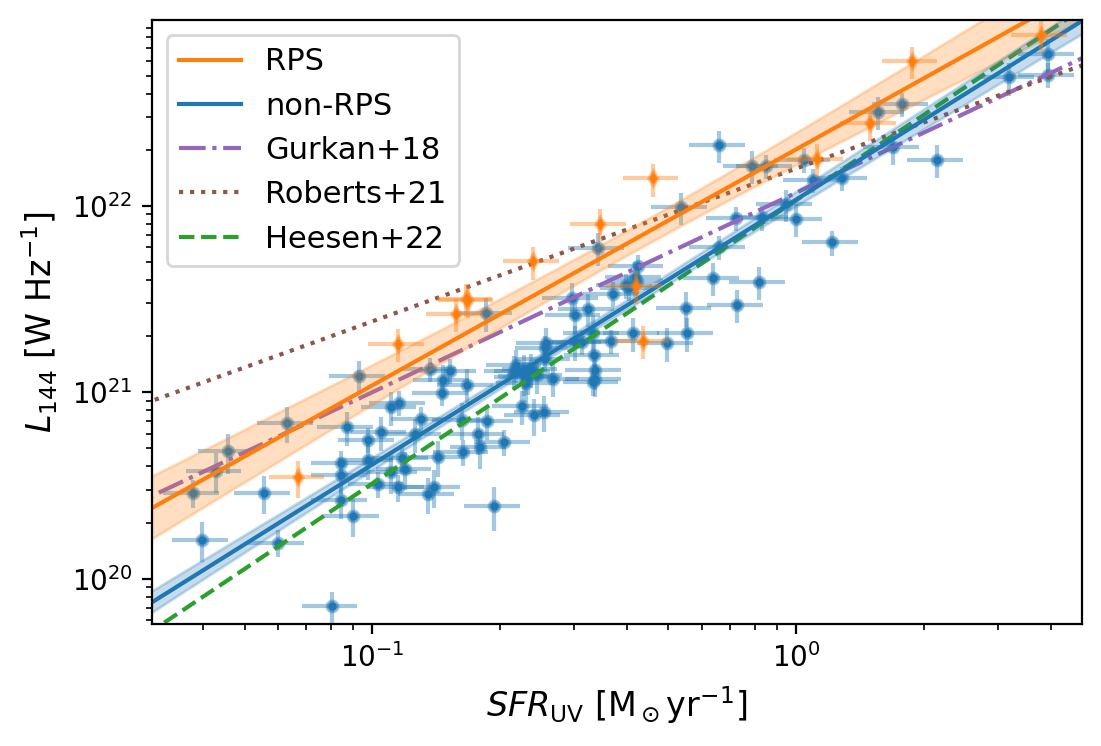}
    \caption{Radio-SFR relations. Top panel: H$\upalpha$-inferred SFRs against radio luminosity for galaxies in the RPS sample (orange data points) and other galaxies in the LOFAR HRS sample (blue data points). The orange and blue lines represent the best-fitting power-law fits and the corresponding $1\,\sigma$ uncertainty bands for those samples. The green, purple and brown lines are fits taken from the literature. Bottom panel: Same as above but for the UV-based SFRs.}
    \label{fig:SFRvsl144}
\end{figure}

The availability of radio and other SFR-tracers for our data allows us to investigate the radio-SFR-relation for our samples. Particularly at low radio-frequencies, this radio-SFR-relation is known to deviate from the linear scenario \citep{Heesen2022NearbyRelation}. Thus, we fit a power-law relation of the form:
\begin{equation}
    \inwphz{L_{144}} = N_0 \insmyr{\mathrm{SFR}}^\beta.
\label{eq:fitpl}
\end{equation}
Fitting is performed in log-log space, where the expression assumes a linear form. We use the orthogonal method of the \emph{bivariate errors and intrinsic scatter} (BCES) regression algorithm \citep{Akritas1996LinearScatter,Nemmen2012ASystems} for the minimization. 

We fitted \autoref{eq:fitpl} for the H$\upalpha$ and UV inferred SFRs and for the HRS galaxies in the RPS sample and those not in the RPS sample independently.
The scatter $\sigma_\mathrm{raw}$ of the data points around the fit is calculated as:
\begin{equation}
    \sigma_\mathrm{raw}^2 =\left(n-2\right)^{-1}\sum_i {\left(\log{{L_{144,i}}} - \beta\log{\mathrm{SFR_i}} - \log{N_0}\right)^2},
\end{equation}
where $n$ is the sample size.
The best-fitting parameters are reported in \autoref{tab:fit} and we display the fit results together with the data points in \autoref{fig:SFRvsl144}. 
\begin{table*}[]
\centering
\caption{Best-fitting parameters for different parametrizations of the radio-SFR relation.}%Best-fitting parameters $N_0$, $\beta$ for the model of \autoref{eq:fitpl} and different SFR-tracers. The scatter of the distribution around the fit is given by $\sigma_\mathrm{raw}$ and the sample size by $n$.}
\label{tab:fit}
\begin{tabular}{ccccccc}
\hline
Tracer    & Sample & $N_0 \,[10^{21}]$ & $\beta$       & $\gamma$ & $\sigma_\mathrm{raw}$ & $n$ \\\hline
H$\upalpha$ & RPS    & $15.8\pm2.1$ & $1.21\pm0.11$ & --            & 0.28 & 14 \\
H$\upalpha$ & others & $8.5\pm0.67$ & $1.39\pm0.06$ & --            & 0.28 & 103 \\
UV        & RPS    & $20.0\pm2.6$ & $1.27\pm0.11$ & --            & 0.23 & 14 \\
UV        & others & $10.8\pm0.8$ & $1.42\pm0.06$ & --            & 0.24 & 102 \\

H$\upalpha$ & RPS    & $13.7\pm2.5$ & --            & $0.08\pm0.13$ & 0.29 & 14 \\
H$\upalpha$ & others & $7.2\pm0.6$  & --            & $0.23\pm0.05$ & 0.27 & 103 \\
UV        & RPS    & $16.0\pm2.5$ & --            & $0.08\pm0.12$ & 0.26 & 14 \\
UV        & others & $7.8\pm0.5$  & --            & $0.21\pm0.04$ & 0.25 & 102 \\

H$\upalpha$ & RPS    & $14.1\pm3.6$  & $1.05\pm0.25$ & $0.05\pm0.21$ & 0.29 & 14  \\
H$\upalpha$ & others & $8.3\pm0.7$  & $1.22\pm0.06$ & $0.12\pm0.06$ & 0.26 & 103 \\
UV        & RPS    & $20.3\pm5.2$ & $1.26\pm0.23$ & $0.09\pm0.19$ & 0.25 & 14 \\
UV        & others & $9.9\pm0.8$  & $1.28\pm0.06$ & $0.08\pm0.05$ & 0.24 & 102 \\
\hline
%IR & $4.1\pm0.2$ & $1.28\pm0.04$  & 0.25 & 132
\end{tabular}
\end{table*}

For both SFR tracers, the relation for the radio continuum luminosity derived for the RPS sample is a factor of 2--3 above the one found for the non-RPS galaxies. This is expected given the reports of enhanced radio-to-SFR ratios for RPS galaxies in the literature \citep{Murphy2009, Vollmer2010, Vollmer2013, Roberts2021LoTSSClusters, Ignesti2022WalkGalaxies}. In \autoref{fig:SFRvsl144}, we also display the relations derived from LOFAR observations of SF-galaxies in other works: \citet{Heesen2022NearbyRelation} analyzed a sample of 45 nearby galaxies, \citet{Gurkan2018LOFAR/H-ATLAS:Relation} studied a sample of > 2000 SF-galaxies across a wide range of redshifts and \citet{Roberts2021LoTSSClusters} published a relation for a sample of 95 jellyfish galaxies in low-redshift clusters. 

\subsection{Mass dependency and radio excess} \label{sec:mass+radioexcess}
\begin{figure}
    \centering
    \includegraphics[width=1.0\linewidth]{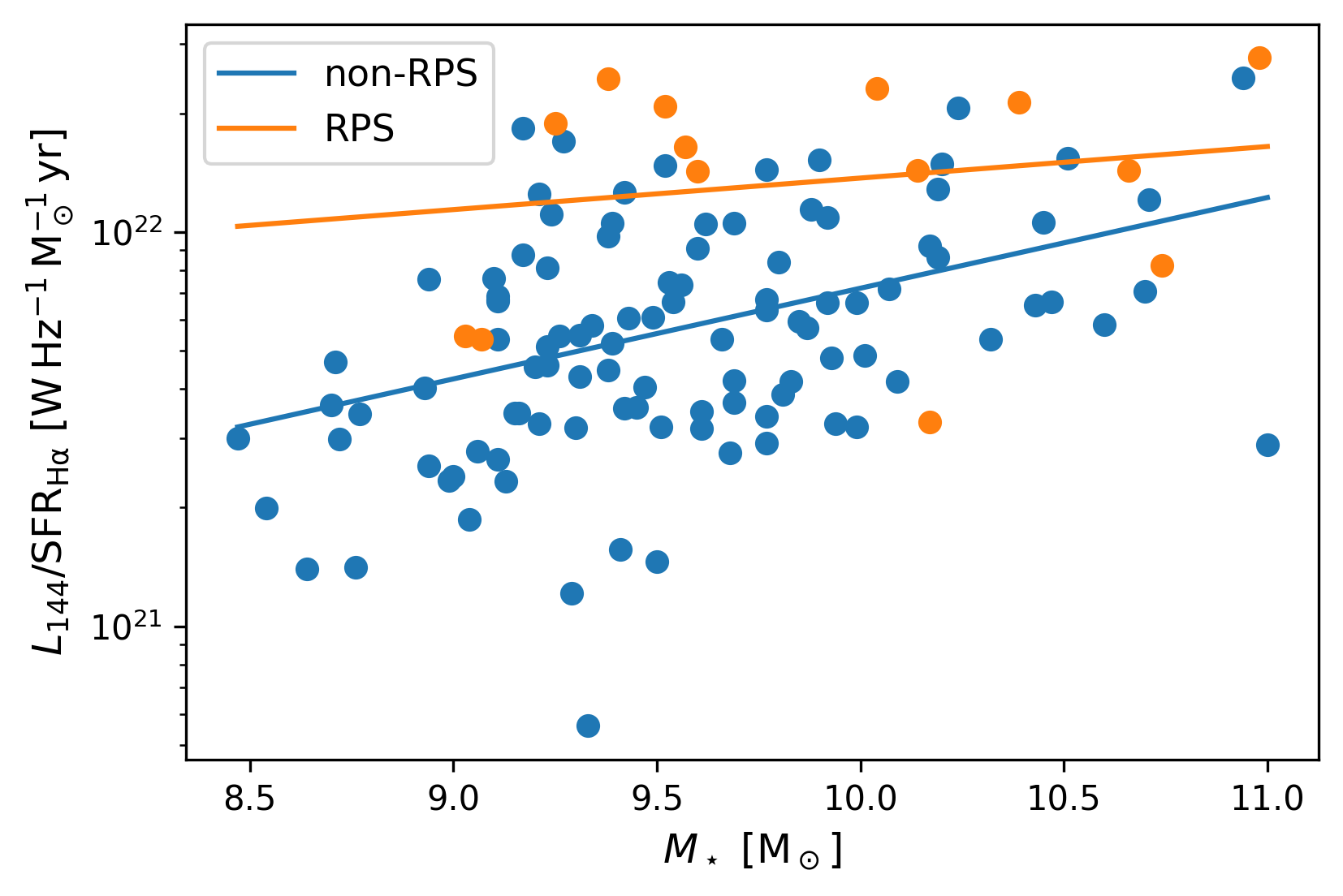}
    \includegraphics[width=1.0\linewidth]{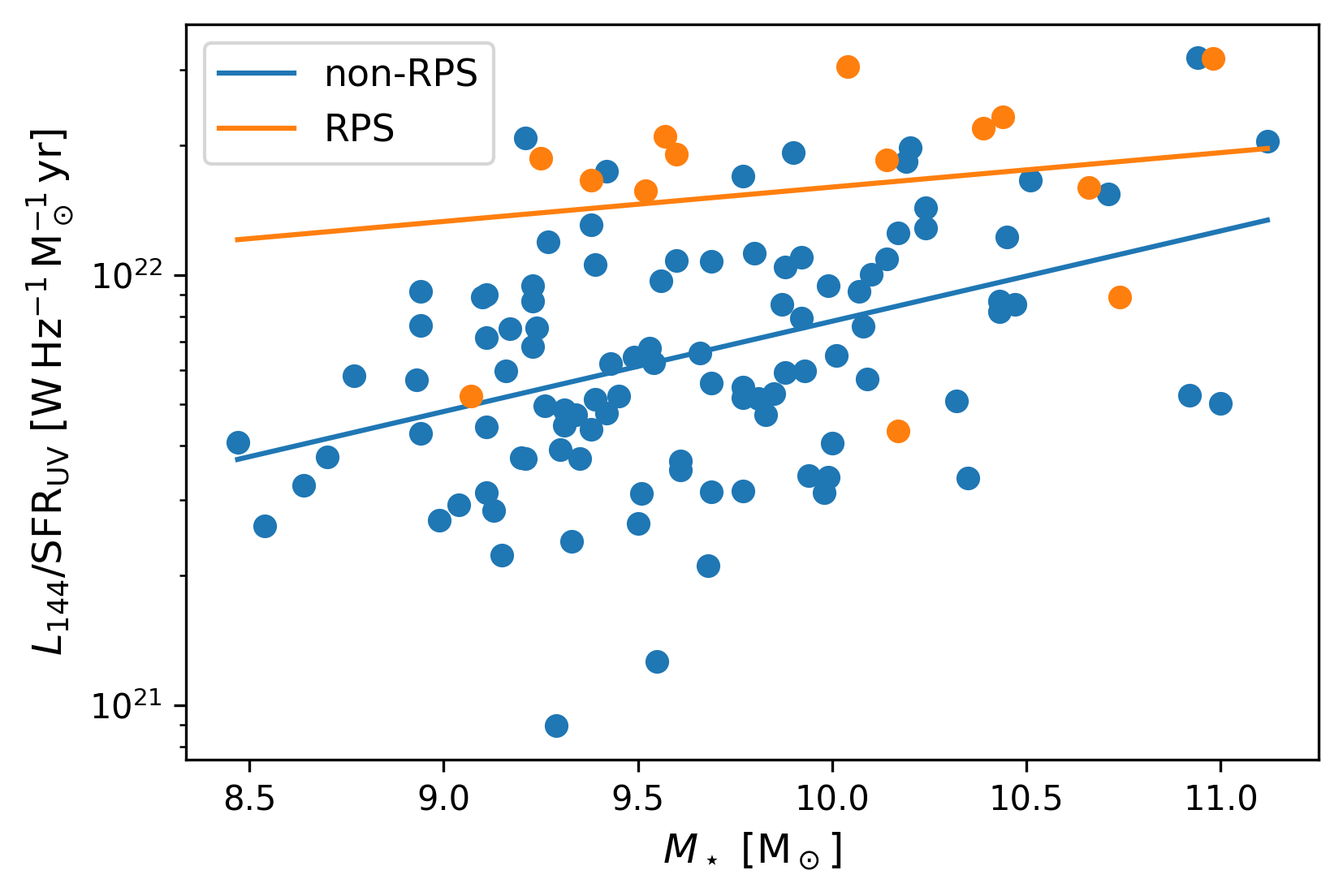}
    \caption{Top panel: Ratio of $L_{144}$ to $\mathrm{SFR}_\mathrm{H\upalpha}$ as a function of $M_\star$. The orange data points mark the galaxies in the RPS sample, the blue ones the galaxies outside the RPS sample. The orange and blue lines are the corresponding fits. Bottom panel: Same but for the UV-inferred SFR.}
    \label{fig:l144sfrvsmass}
\end{figure}
The resulting power-law fits for our samples are super-linear with $\beta \approx 1.4$. 
This super-linearity is thought to originate from a mass-dependence in the calorimetric efficiency $\eta=\eta(M)$, which is the ratio of CRe energy that is radiated withing galaxy. In larger galaxies with a higher mass $M$, CRe have longer escape times $t_\mathrm{esc}$ in relation to their lifetime $t_{\rm syn}$, thus, they lose a greater fraction of their energy inside the galaxy before escaping into regions with low magnetic field strengths. 

We use the following parametrization to enforce a linear radio-SFR relation while taking into account a mass-dependent calorimetric efficiency \citep[see e.g.][]{Heesen2022NearbyRelation}:
\begin{equation}\label{eq:linwithmass}
    L_{144} = N_0 \insmyr{\mathrm{SFR}} \left(\frac{M_\star}{10^{10}\mathrm{M_\odot}}\right)^\gamma.
\end{equation}
Here, $\gamma$ describes the dependency on the stellar mass $M_\star$.
We repeat the fitting for the RPS and non-RPS galaxies for both SFR-tracers. The best-fitting model-parameters are reported in \autoref{tab:fit}, and the data points and fits are shown in \autoref{fig:l144sfrvsmass}. For the galaxies that do not show a RPS-morphology, we can reproduce the systematic increase of $L_{144}/\mathrm{SFR}$ as a function of mass, finding a positive mass-exponent of $\gamma\approx.2$. 

Again, the population of RPS galaxies has a clear radio-excess with 12/14 galaxies above the relation for non-RPS galaxies. However, for them, the mass-dependency is less clear. This could be either due to an increased scatter in $L_{144}/\mathrm{SFR}$ for this population, indirectly connected to the ram-pressure stripping, or due to the small sample size. 

%For H$\alpha$- and UV-inferred SFRs, we find a significantly positive dependence on the (stellar) masses with an exponent $\gamma > 0.2$. The mass-dependence is much less clear for the IR-based SFRs, where we find $\gamma = 0.05\pm0.04$. The primary cause of this lies in systematics which affect $\mathrm{SFR}_\mathrm{IR}$ - in galaxies with an more evolved stellar population, older stars contribute significantly to the dust-heating, such that the IR-emissions are not a direct proxy of the SFR anymore. This leads to an over-estimation of the SFR for earlier-type galaxies. Since more massive galaxies, in particular in a cluster environment, are oftentimes more evolved, the IR-inferred SFR will be overestimated for galaxies with higher mass. This leads to a lower best-fitting value of $\gamma$.

We further investigate a relation with an additional free parameter, which is a power-law in both mass and SFR \citep[see ][]{Gurkan2018LOFAR/H-ATLAS:Relation}:
\begin{equation}\label{eq:dblpl}
    \inwphz{L_{144}} = N_0 \insmyr{\mathrm{SFR}}^\beta \left(\frac{M_\star}{10^{10}\mathrm{M_\odot}}\right)^\gamma.
\end{equation}
The best-fitting results for the this parametrization are displayed in \autoref{tab:fit}. For the non-RPS galaxies, we find a mass-exponent of $\gamma \approx 0.1$ and an SFR-exponent of $\beta \approx 1.25$. 
Those results can be compared to the ones from \citet{Gurkan2018LOFAR/H-ATLAS:Relation}, who found a flatter relation with SFR ($\beta=0.77$) but a steeper exponent in mass ($\gamma=0.44$). 

To quantify the radio-excess (or deficit) compared to the best fit of \autoref{eq:dblpl} for the non-RPS sample, we calculate the distance of the RPS-galaxies to the best-fitting plane in log-space according to:
\begin{equation}
    d_\mathrm{3D} = \frac{\log{\inwphz{L_{144}}} - \beta \log{\insmyr{\mathrm{SFR}}} - \gamma \log{\left( \frac{M_\star}{10^{10}\mathrm{M}_\odot}\right)}}{\sqrt{\beta^2+\gamma^2+1}}.
\end{equation}
\begin{figure}
    \centering
    \includegraphics[width=1.0\linewidth]{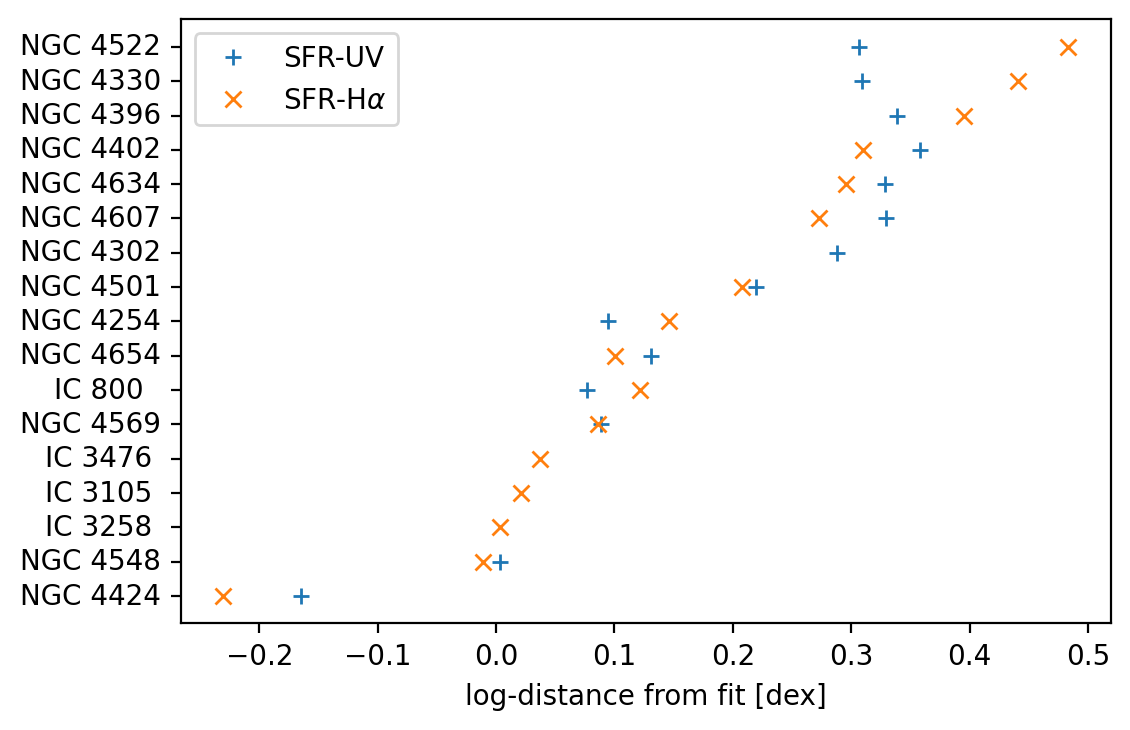}
    \caption{Distance of the RPS galaxies to the best-fitting $L_{144}$-SFR-$M_\star$-plane for the non-RPS galaxies in log-space. The galaxies are sorted by the mean of the $\mathrm{SFR}_\mathrm{UV}$ and $\mathrm{SFR}_\mathrm{H\upalpha}$ based log-distances.  }
    \label{fig:d3d}
\end{figure}
We display those distances in \autoref{fig:d3d}. In general, there is a good agreement between the distances based on the UV and H$\upalpha$-inferred SFR. The only object with a strong radio-deficit is the peculiar galaxy NGC\,4424, which is thought to be a merger remnant and shows outflows of ionized gas \citep{Boselli2018VESTIGE4}. Similarly, the galaxies with a clear RPS morphology in radio show the strongest radio-excesses, those are NGC\,4522, NGC\,4330, NGC\,4396, NGC\,4402 and NGC\,4634. Qualitatively, there appears to be a connection between the asymmetry of the radio emission and the strength of the radio excess. The only exception being IC\,3476, which shows a strong tail but only a mild excess of $\approx$$10\%$. Possibly, the radio emission of this source is underestimated due to masked background sources and side-lobes of M\,87 which unfortunately cover parts of the tail of this source .
% there is no trend between d3ds and inclination
% move next part to discussion? 

\subsection{Spectral properties}
To probe if the spectral properties of the galaxies in the ram-pressure stripped sample differ from the ones of unperturbed galaxies, we determined the integrated spectral index $\alpha$\footnote{Defined as $S\propto nu^\alpha$.} for all objects which have radio-measurements also at $1.4$\,GHz available in the literature. The spectral index $\alpha$ between two frequencies $\nu_1$, $\nu_2$ with flux density measurements $S_1$, $S_2$ is calculated as:
$\alpha = \mathrm{ln}(S_1/S_2)/\mathrm{ln}(\nu_1/\nu_2)$. We estimate the corresponding uncertainties following Gaussian propagation of uncertainty.
The $1.4$\,GHz flux densities or upper limits were collected from  \citet{Murphy2009}, \citet{Chung2009} (re-measured in cases where the maps were provided by the authors), \citet{Vollmer2010} and \citet{Boselli2015HGalaxies}. For the two RPS objects which are not in the HRS, we derived the $4\sigma$ upper limits from the NVSS \citep{Condon1998TheSurvey}.
To allow for comparison with the work of \citet{Heesen2022NearbyRelation}, we estimated the total masses of the galaxies in our sample. For this, we employed the dynamical data derived from H$\upalpha$ \citep{Gomez-Lopez019} or \hi{} \citep{Boselli2014coldgas} observations according to $M_\mathrm{tot} = r_\mathrm{SF} v_\mathrm{rot}^2/G$, where $G$ is the gravitational constant, $v_\mathrm{rot}$ the rotation velocity and $r_\mathrm{SF}$ the size of the star-forming disk, estimated from $D24$ or $D25$ as described in \citet{Gomez-Lopez019}.

\begin{figure}
    \centering
    \includegraphics[width=1.0\linewidth]{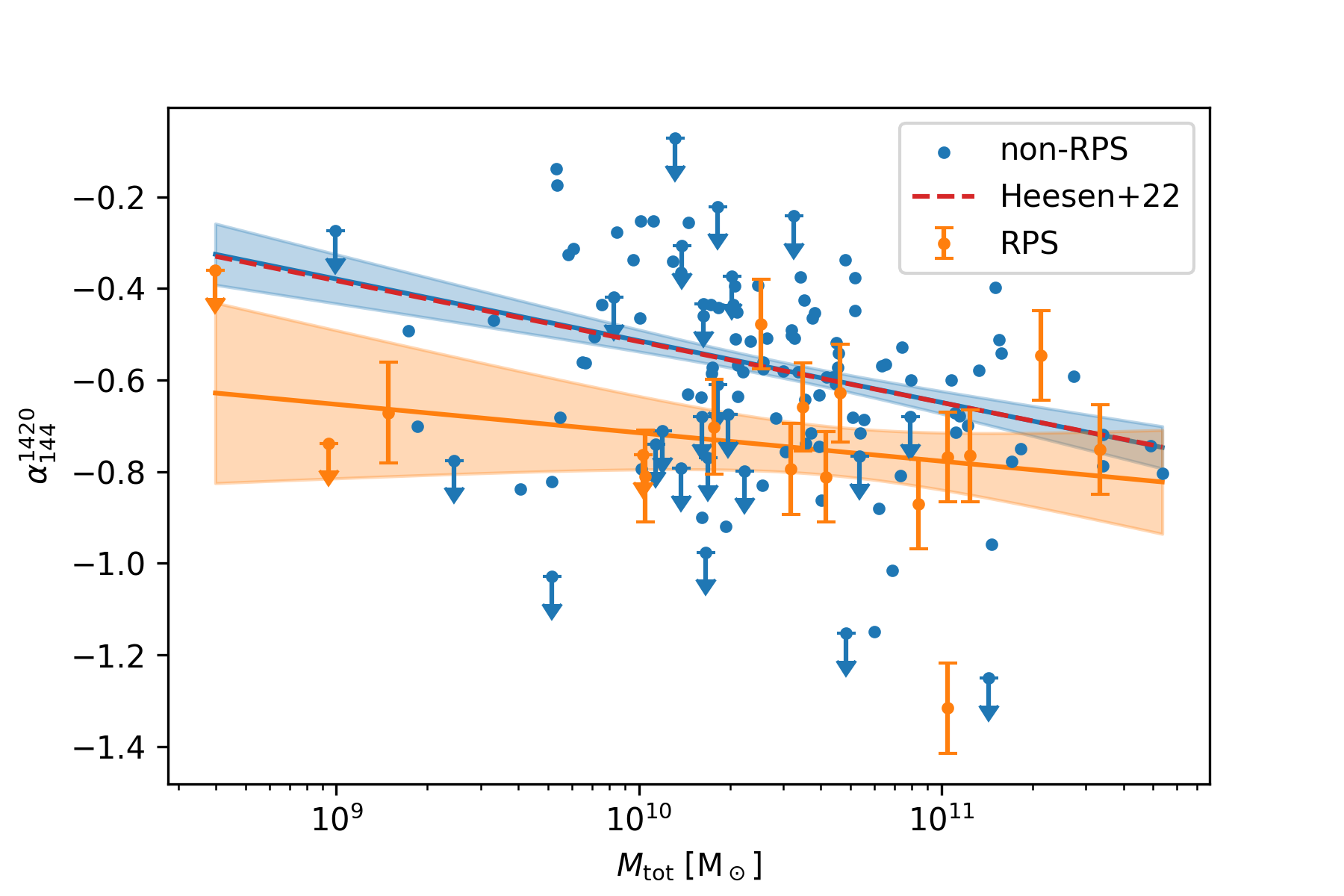}
    \caption{Total mass against spectral index for the RPS sample (orange markers) and the other galaxies in the sample (blue markers). For clarity, uncertainties are only displayed for the RPS sample, they are of comparable size ($\sigma_\alpha \sim 0.1$) for the other galaxies. Downward pointing arrows correspond to $4\,sigma$ upper limits. The orange and blue line show the corresponding best-fitting relations, and the red dashed line shows the fit from \citet{Heesen2022NearbyRelation}. }
    \label{fig:si}
\end{figure}

In \autoref{fig:si}, we display the total mass-spectral index-distribution of the 114 galaxies with available 1.4\,GHz measurements, for the 26 objects without detection, we provide an upper limit. We also show the best-fitting log-linear relation of the form $\alpha = A \log{(M_\mathrm{tot}/M_\odot)} + B$. For the galaxies outside of our RPS sample, we find $A=-0.13 \pm 0.02$ and $B=0.83 \pm 0.20$, in excellent agreement with the parameters derived  by \citet{Heesen2022NearbyRelation} ($A=-0.13$ and $B=0.81$).
However, for the RPS sample, we find evidence that the relation is shifted towards steeper spectral index values with $A=-0.06 \pm 0.05$ and $B=0.09 \pm 0.52$. Evaluated at $M_\mathrm{tot} = 3\times10^{10}\,M_\odot$, the relations differ by $\Delta\alpha=0.17\pm0.06$.
We cross check that this finding is not simply caused by a systematic underestimation in the dynamical estimates of $M_\mathrm{tot}$ due to truncated gas disks in RPS galaxies by repeating the fitting using $M_\star$, where we find $\Delta\alpha=0.13\pm0.07$ at $M_\star = 4\times10^{9}\,M_\odot$. 
A strong outlier from the relation is NGC\,4548 with a particularly steep spectral index of $\alpha= -1.3\pm0.1$ measured from the VIVA \citep{Chung2009} map with low image fidelity. However, the $4\sigma$ upper limit from the NVSS map \citep{Condon1998TheSurvey} agrees with a steep spectrum $\alpha \leq -1.0$.   %A cross-check with a preliminary map from our ViCTORIA-MeerKAT survey yields a steep but less extreme value of $\alpha = -0.9$. Not certain, NVSS / RACS limits also point towards SI < -1.0
Our finding with a best-fitting relation for the RPS galaxies that is shifted to lower values of $\alpha$ holds when repeating the analysis without NGC\,4548 ($\Delta\alpha=0.13\pm0.04$ at $M_\mathrm{tot} = 3\times10^{10}\mathrm{M}_\odot$).

We also investigated if there is a correlation between the spectral index and the log-distance from the best-fitting $L_{144}$-SFR-$M_\star$ plane for the RPS sample. We find a slightly positive Pearson's correlation coefficient of 0.30 using SFR$_\mathrm{UV}$ and 0.377 SFR$_\mathrm{H\alpha}$, however, both are not significant ($p-\mathrm{value} > 0.23$), the slightly positive relation is due to the low luminosity steep spectrum source NGC\,4548.

\section{Spatially resolved radio-SFR analysis}\label{sec:local}

In Sect.\,\ref{sec:mass+radioexcess}, we showed that the radio emission of our Virgo RPS sample is enhanced up to a factor of $3$ compared to the H$\upalpha$ and UV-inferred SFR. The localization of this excess allows to constrain the origin of the surplus radio emission. So could a leading-edge enhancement point towards compression of the magnetic field lines or to ram-pressure driven shocks as additional source of CRe. %If the excess is global, it can point towards a recent quenching of SFR. 
To investigate this, we created maps of the ratio between the observed radio surface brightness $B_{\rm obs}$ and the expected radio surface brightness $B_\mathrm{model}$ based on the SFR-surface density $\Sigma_\mathrm{SFR}$. A similar analysis was carried in \citet{Murphy2009} for six of the galaxies in our sample using 1.4\,GHz VLA observations and {\it Spitzer} far-infrared emission as SFR-tracer. For our analysis, we derived the SFR-surface density from {\it GALEX} NUV \citep{Martin2005} following the calibration of \citet{Leroy2019}.% and converted to a Salpeter IMF using a factor of $1.56$ \citep{Kennicutt2012StarGalaxies}. 
We employed a dust-correction based on Herschel 100\,$\upmu$m measurements obtained from the HRS \citep{Cortese2014PACSGalaxies} and HeVICS \citep{Davies2010} according to \citet{Kennicutt2012StarGalaxies}. Foreground and background sources in the $\Sigma_\mathrm{SFR}$ and LOFAR maps were manually masked and the maps were gridded to the same pixel layout. We convolved the $\Sigma_\mathrm{SFR}$-maps to match the resolution of our LOFAR maps (either $9''\times9''$ or $20''\times20''$) using the \texttt{convolve} method implemented in \texttt{astropy} \citep{Astropy2022}. 
We obtained the model for the radio emission using \autoref{eq:linwithmass} and our best-fitting parameters ($N_0=7.7\times10^{21}$ and $\gamma=0.21$) for the non-RPS sample to convert $\Sigma_\mathrm{SFR}$ to the radio surface brightness $B$:
{\small
\begin{equation}
    \frac{B_\mathrm{model}}{\mathrm{Jy\,beam^{-1}}} = \frac{N_0}{4\pi{\times}10^{-26}} \left(\frac{\Sigma_\mathrm{SFR}}{\mathrm{M_\odot\,yr^{-1}\,kpc^2}}\right) \left(\frac{M_\star}{10^{10}\mathrm{M_\odot}}\right)^\gamma \left(\frac{A_\mathrm{beam}}{\mathrm{kpc^2}}\right)^{\mathrm{-}1} \left(\frac{d}{\mathrm{m}}\right)^{\mathrm{-}2} .
\end{equation}}
Since the CRe have lifetimes of $\sim$$100$\,Myr and are subject to CR-transport, the synchrotron emission of an undisturbed galaxy should be a smoothed version of SFR-surface density \citep[e.g.][]{Heesen2023M51}, this can be expressed in terms of the CRe-transport length $l_\mathrm{CRe}$. The shape of the smoothing kernel used to model the CR-transport depends on the transport process.  We assume as a benchmark the case of pure diffusion modeled by a Gaussian kernel. The width $\sigma$ of the Gaussian kernel used for the smoothing is related to $l_\mathrm{CRe}$ according to $\sigma = (2\mathrm{ln}2)^{-0.5}l_\mathrm{CRe}$. 

Since CRe diffuse away from the star-forming regions into regions with low or no star-forming activity, the resolved $B$-$\Sigma_\mathrm{SFR}$-relation is sub-linear, opposite to the global radio-SFR-relation. Taking into account CRe diffusion with an appropriate choice of $l_\mathrm{CRe}$, the relation can be linearized \citep{Murphy2009,Berkhuisjen2013a,Heesen2023M51}. To linearize this relation, we fit the convolution kernel size $\sigma$. For each $\sigma$, we determine the slope $a$ of the relation by fitting the pixels that are above a signal-to-noise ratio of 5 in both maps. Due to their low surface brightness, we instead require a S/N of 3 for IC\,800 and IC\,3105 and 4 for NGC\,4548. The optimal $\sigma$ is found once the slope reaches unity. 
We carried out the smoothing procedure based on maps at $9''$ resolution (for sources with sufficient surface brightness) and at $20''$ resolution. The resulting transport lengths are listed in \autoref{tab:si_lcre}, there is decent agreement between the values derived at different resolutions. Two notable outliers with a large transport length are NGC\,4254 and NGC\,4548 with $l_\mathrm{CRe} > 5$\,kpc. Those values are larger than  the values $\leq 3.8$ found for field galaxies at 144\,MHz in \citet{Heesen2023NearbyRelation}. This is likely due to the significant contribution of the external perturbations to the CRe-transport, a scenario also suggested by \citet{Ignesti2022Gasp}.

\begin{figure*}
\centering
    \includegraphics[width=0.14\textwidth]{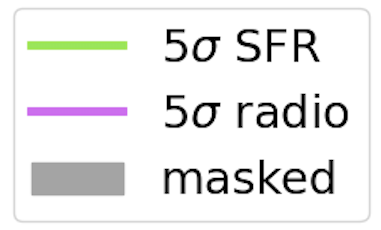}
    \hspace{0.10\textwidth}
    \includegraphics[width=0.5\textwidth]{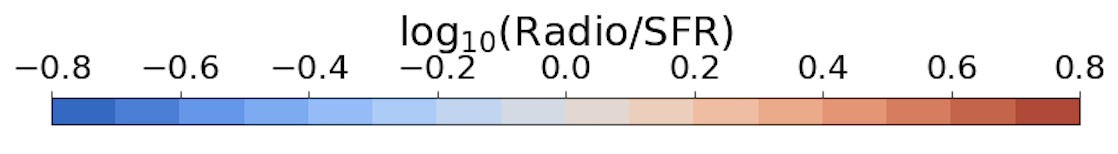}
    \includegraphics[width=0.24\textwidth]{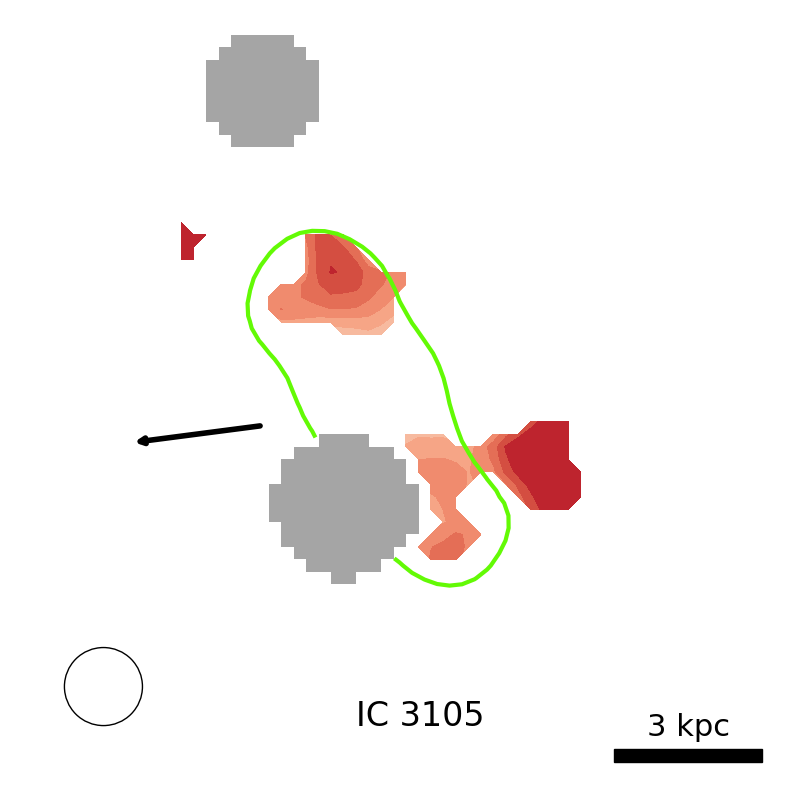}
    \includegraphics[width=0.24\textwidth]{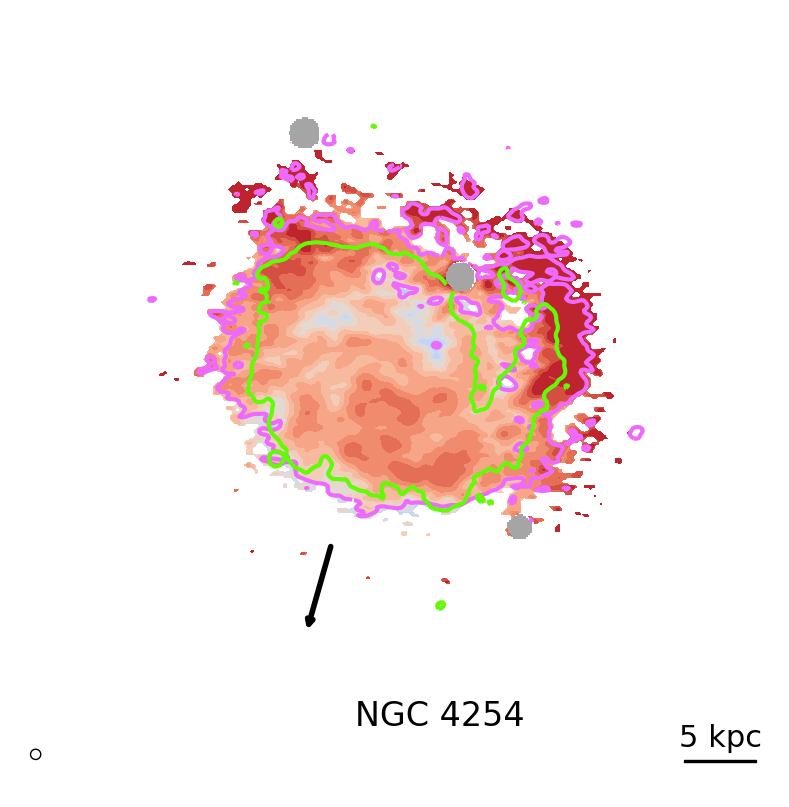}
    \includegraphics[width=0.24\textwidth]{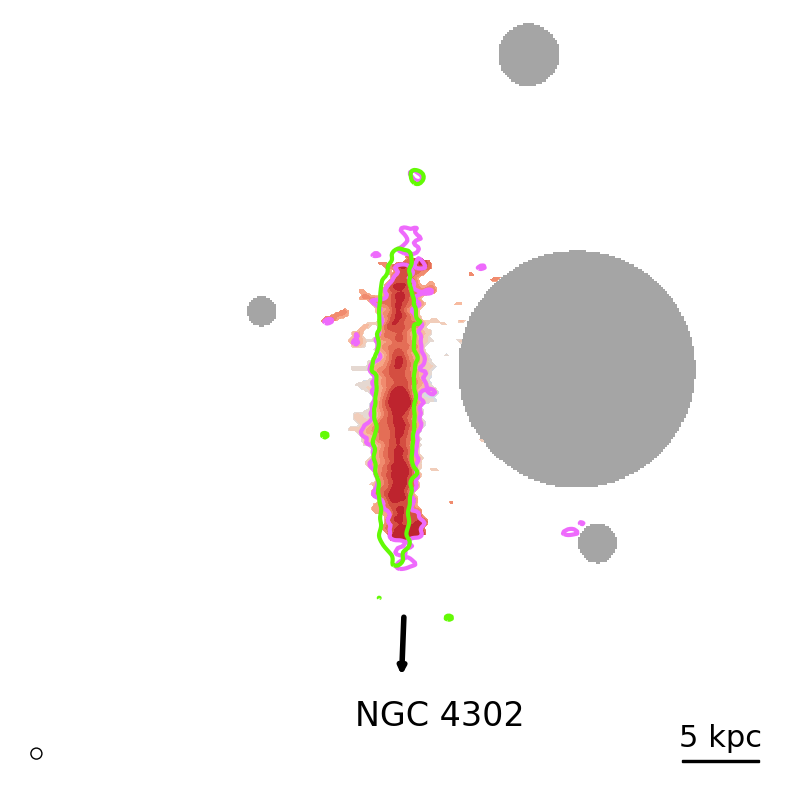}
    \includegraphics[width=0.24\textwidth]{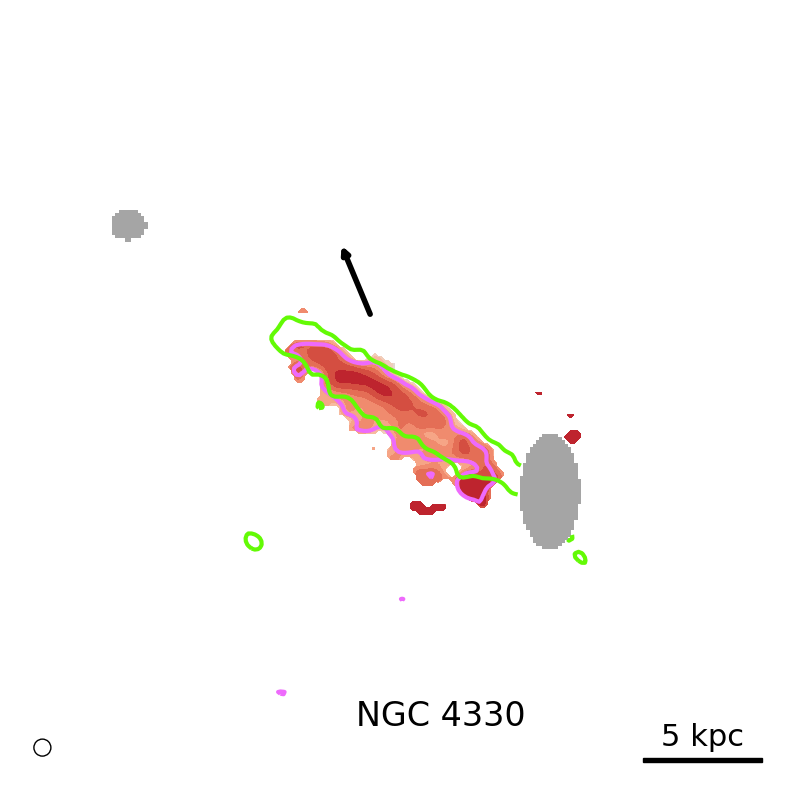}
    \includegraphics[width=0.24\textwidth]{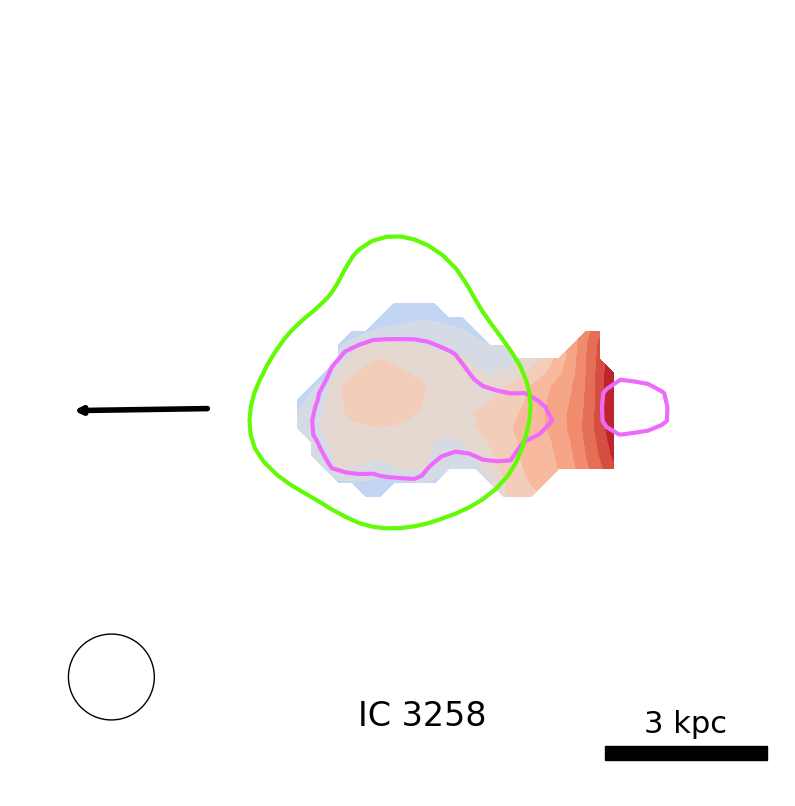}
    \includegraphics[width=0.24\textwidth]{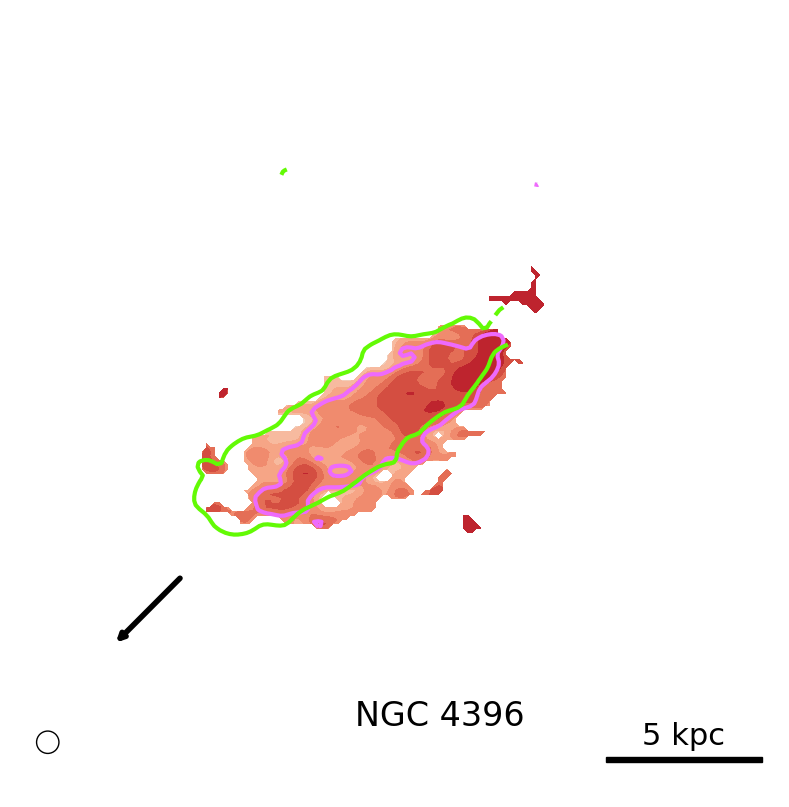}
    \includegraphics[width=0.24\textwidth]{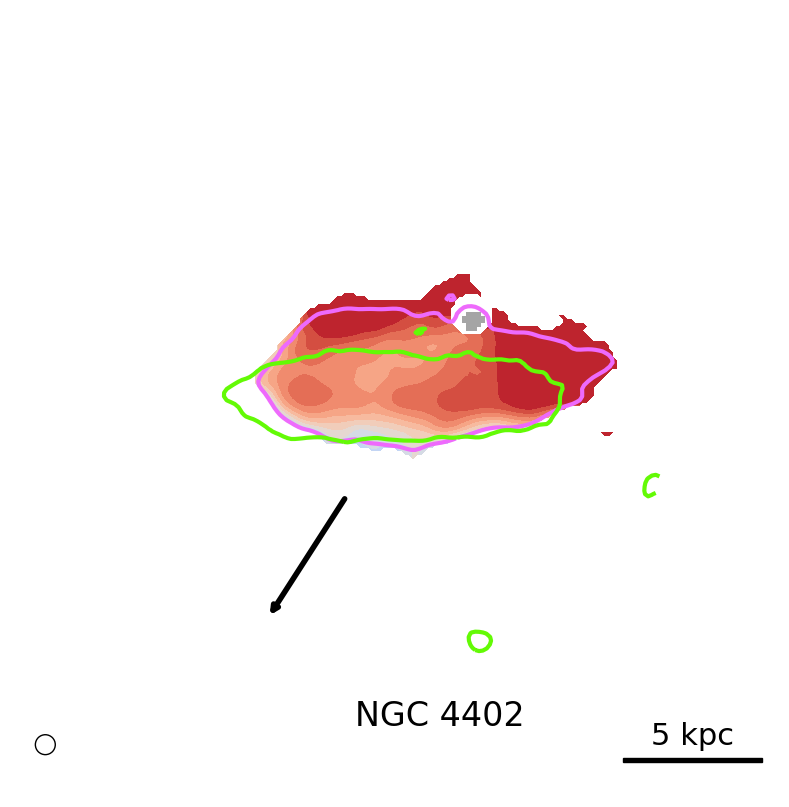}
    \includegraphics[width=0.24\textwidth]{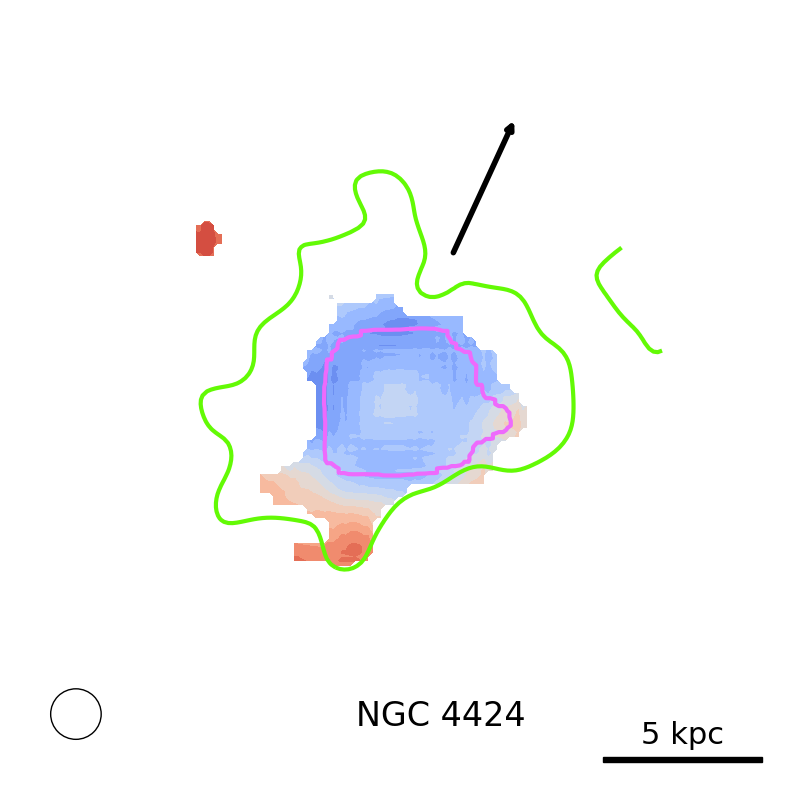}
    \includegraphics[width=0.24\textwidth]{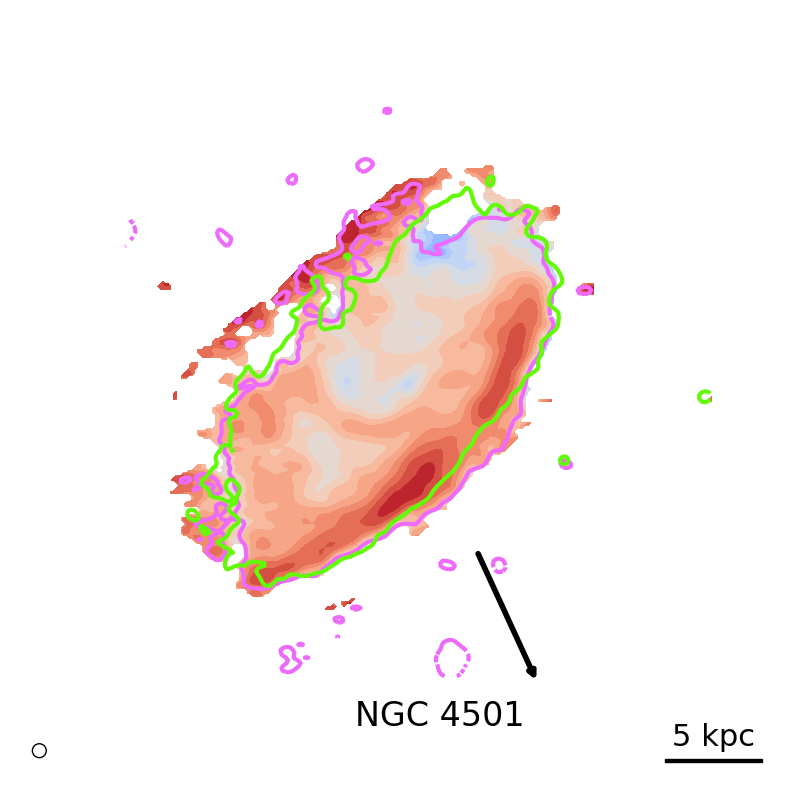}
    \includegraphics[width=0.24\textwidth]{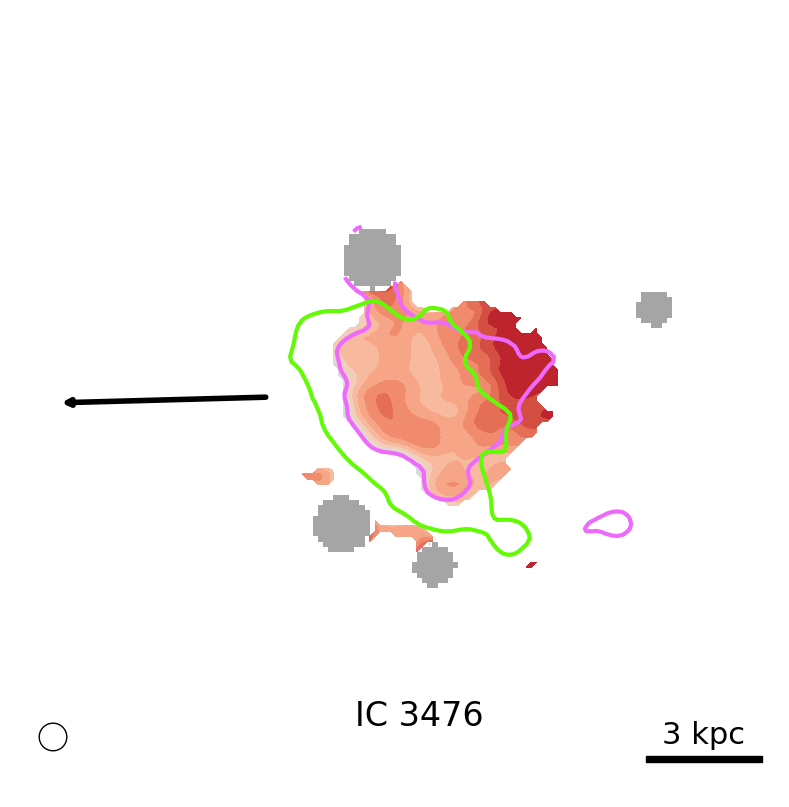}
    \includegraphics[width=0.24\textwidth]{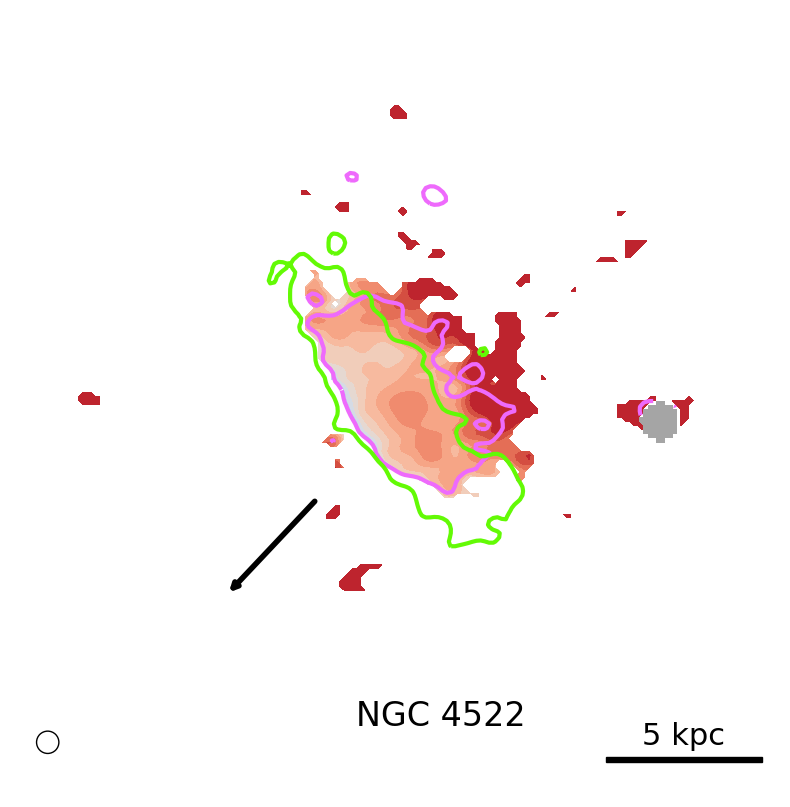}
    \includegraphics[width=0.24\textwidth]{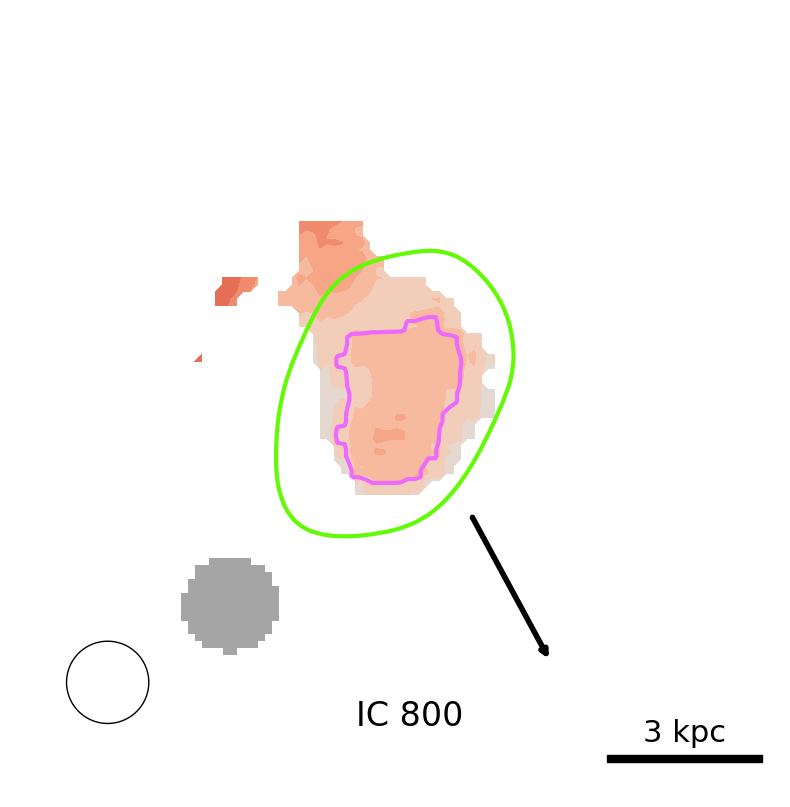}
    \includegraphics[width=0.24\textwidth]{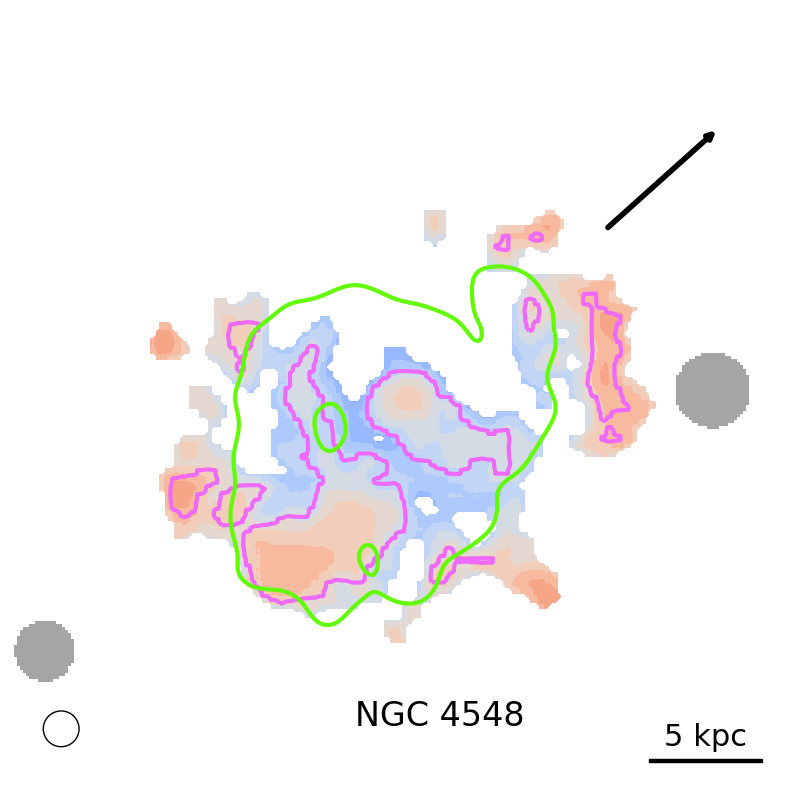}
    \includegraphics[width=0.24\textwidth]{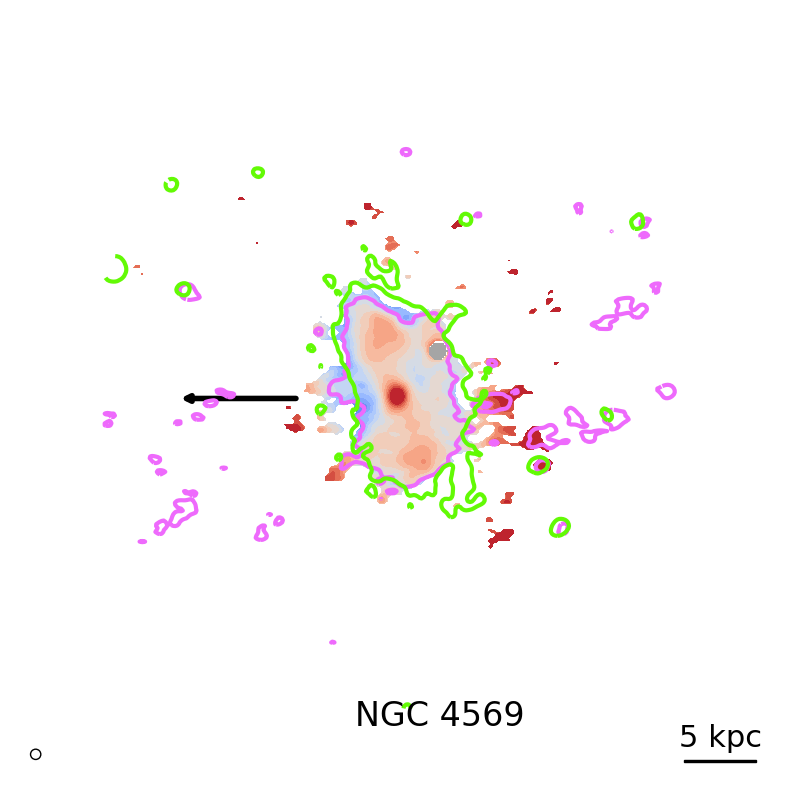}
    \includegraphics[width=0.24\textwidth]{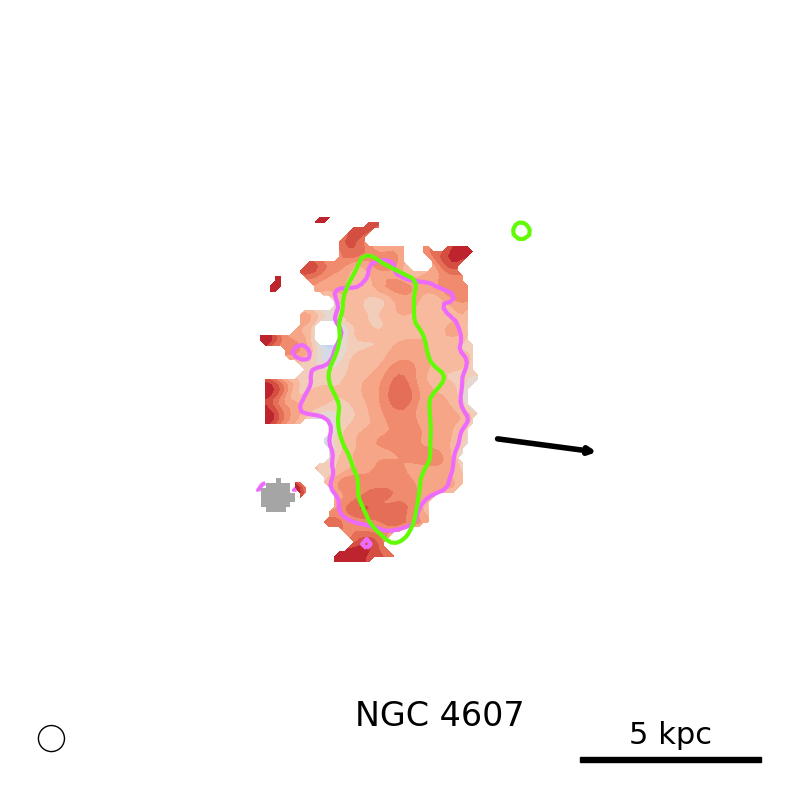}
    \includegraphics[width=0.24\textwidth]{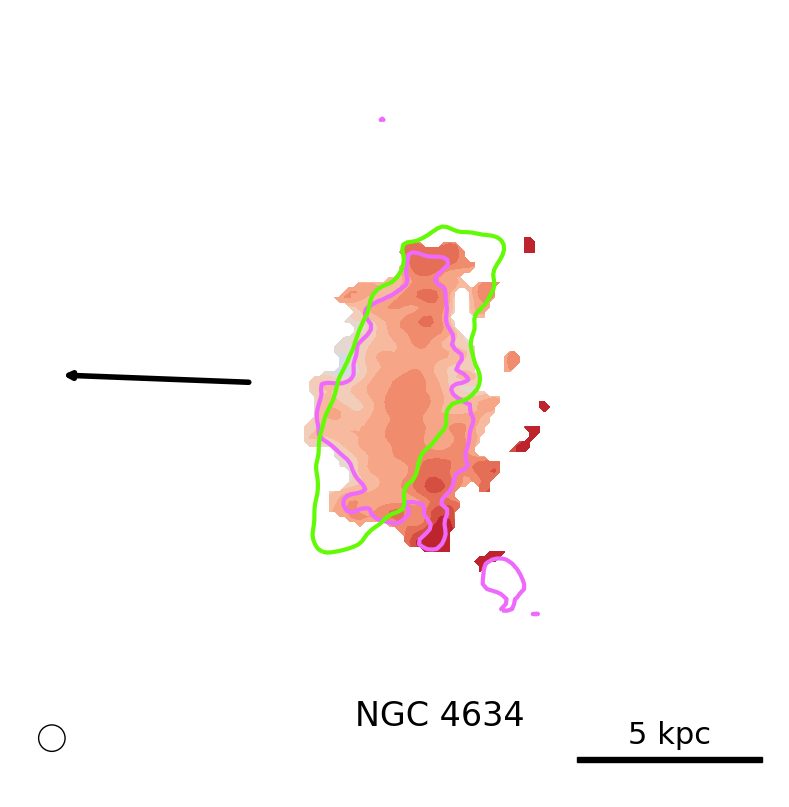}
    \includegraphics[width=0.24\textwidth]{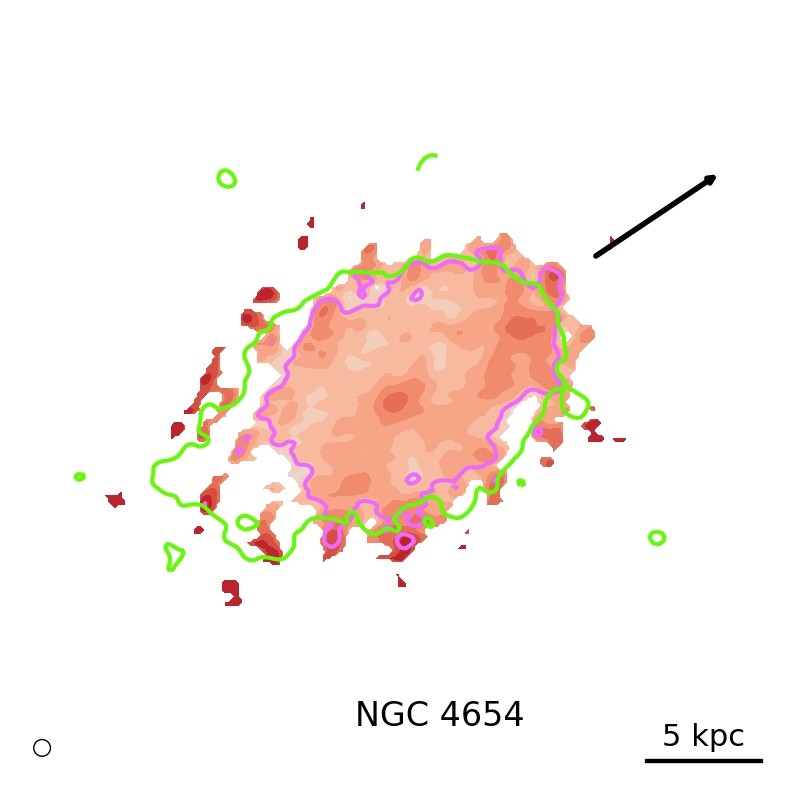}
    \caption{Logarithmic ratio between the observed and modeled radio emission. The color-scale is symmetric and increases in increments of 0.1\,dex, red colors indicate a local radio excess, blue colors a radio deficit and gray areas are masked background sources. Magenta and green lines are the $5\sigma$ contours of $B_\mathrm{obs}$ and $\Sigma_\mathrm{SFR}$, respectively. The black arrow indicates the approximate orbit of the galaxy as inferred from the direction of the tail. The circles in the bottom left correspond to the angular resolution of the radio data, which is $9''$, except for IC\,3105, IC\,3258, NGC\,4424, IC\,800 and NGC\,4654 where we used maps at $20''$ resolution.}\label{fig:ratio}
\end{figure*}

We then convolved the maps of the model surface brightness using a Gaussian kernel with the corresponding width $\sigma$.
In \autoref{fig:ratio}, we display the log-ratio $R=\log_{10}{(B_\mathrm{obs}/B_\mathrm{model})}$ between the observed and the modeled radio emission for all pixels with a signal-to-noise ration above 3 (above 2 in the case of IC\,3105). We also mark the $5\sigma$ contours of the SFR-surface density and radio surface brightness. For sufficiently bright sources, we used the $9''$-resolution maps to derive the ratios, otherwise the $20''$-resolution ones. For the former, we present the $20''$ maps, which are more sensitive to the faint tails, in \autoref{sec:appendix1}.

In general, the radio excess is a global phenomenon and all galaxies but NGC\,4424, NGC\,4548 and IC\,3258 show enhanced radio emission across the disk.
As expected, the excess is strongest at the trailing edge or in the tail, where all objects  show some form of enhanced radio emission or radio contours that extend asymmetrically beyond the SFR contours at the trailing edge. This is most pronounced in NGC\,4330, NGC\,4522 and NGC\,4634, while NGC\,4548 shows only a mild enhancement in parts of the trailing half of the disk. The reason for the excess radio emission in the tails is the advection of CRe from the disk due to RPS to regions with low star-forming activity \citep[e.g.][]{Murphy2009,Ignesti2022WalkGalaxies}.
At the same time, many galaxies show a deficit of radio emission at the leading edge (i.e. $\Sigma_\mathrm{SFR}$ contours that extend beyond the radio contours) - e.g. in NGC\,4330, NGC\,4396, NGC4402\, NGC\,4522 NGC\,4548 and IC\,3476 (see also \citet{Murphy2009}). However, the leading-edge deficit is not omnipresent. Most notably, in NGC\,4501, we instead observe an enhancement of radio emission at the leading edge -- likely connected to a local compression of magnetic fields \citep{Vollmer2008NGC4501}.

%3105 None
%4254 5.19
%4302 4.42
%4330 1.92
%3258 0.24
%4396 2.30
%4402 1.39
%4424 1.85
%4501 1.33
%3476 1.70
%4522 0.97
%800  2.00
%4548 7.47
%4569 2.52
%4607 1.59
%4634 2.01
%4654 2.99

% lres
%3105 2.266
%4254 5.439
%4302 3.925
%4330 1.123
%3258 1.847
%4396 1.274
%4402 1.605
%4424 1.441
%4501 0.875
%3476 1.836
%4522 1.605
%800  2.248
%4548 5.472
%4569 1.950
%4607 1.462
%4634 1.772
%4654 3.000

\begin{table}[!ht]
    \centering
    \caption{CRe transport length and spectral indices of the RPS sample.}\label{tab:si_lcre}
    \begin{tabular}{cccc}\hline\rule{0pt}{1.0\normalbaselineskip}
        Name & $l^{9''}_\mathrm{CRe}$ &$l^{20''}_\mathrm{CRe}$ & $\alpha_{144\,\mathrm{MHz}}^{1.4\,\mathrm{GHz}}$ \\ \rule{0pt}{0.5\normalbaselineskip}
         &  [kpc] & [kpc] &  \\ \hline\hline\rule{0pt}{0.5\normalbaselineskip}
        IC\,3105  &  --    & $2.3$ & $<-0.36^{\,a}$        \\ 
        NGC\,4254 &  $5.2$ & $5.4$ & $-0.77\pm0.10^{\,b}$  \\ 
        NGC\,4302 &  $4.4$ & $3.9$ & $-0.77\pm0.10^{\,c}$  \\ 
        NGC\,4330 &  $1.9$ & $1.1$ & $-0.63\pm0.11^{\,c}$  \\ 
        IC\,3258  &  --    & $1.8$ & $<-0.74^{\,a}    $    \\ 
        NGC\,4396 &  $2.3$ & $1.3$ & $-0.79\pm0.10^{\,c}$  \\ 
        NGC\,4402 &  $1.4$ & $1.6$ & $-0.81\pm0.10^{\,c}$  \\ 
        NGC\,4424 &  $1.9$ & $1.4$ & $-0.67\pm0.11^{\,c}$  \\ 
        NGC\,4501 &  $1.3$ & $0.9$ & $-0.75\pm0.10^{\,c}$  \\ 
        IC\,3476  &  $1.7$ & $1.8$ & $-0.81\pm0.1^{\,a} $  \\ 
        NGC\,4522  &  $1.0$ & $1.6$ & $-0.48\pm0.10^{\,b}$  \\ 
        IC\,800   &  --    & $2.2$ & $<-0.76^{\,a}$        \\ 
        NGC\,4548 &  --    & $5.5$ & $-1.32\pm0.10^{\,d}$  \\ 
        NGC\,4569 &  $2.5$ & $2.0$ & $-0.55\pm0.10^{\,b}$  \\ 
        NGC\,4607 &  $1.6$ & $1.5$ & $-0.70\pm0.10^{\,d}$  \\ 
        NGC\,4634 &  $2.0$ & $1.8$ & $-0.66\pm0.10^{\,e}$  \\ 
        NGC\,4654 &  $3.0$ & $3.0$ & $-0.87\pm0.10^{\,d}$  \\\hline 
    \end{tabular}
    \tablefoot{
    The 1.4\,GHz flux densities for the spectral index estimation are taken from: $^a$NVSS  \citep[$4\sigma$ upper limits][]{Condon1998TheSurvey}; $^b$\citet{Murphy2009}, $^c$\citet{Vollmer2010}; $^d$ measured from the maps provided by \citet{Chung2009} and $^e$\citet{Boselli2015HGalaxies}.}
\end{table}

\section{Discussion}\label{sec:discussion}
%\subsection{Virgo RPS galaxies}
%\subsubsection*{NGC\,4254}
%This galaxy is thought to be affected by ram-pressure stripping and a gravitational perturbation \citep{Vollmer2005NGCCluster,Duc2008,Boselli2018VESTIGE3}. It is among the few Virgo RPS objects that lie above the star-forming main sequence (see \autoref{fig:stellarmasssfr}), star formation is particularly active in the central 4\,kpc. The radio morphology is much broader, which points towards effective CRe transport, this can also be seen in the large transport length ($l_\mathrm{CRe} = 5.44\,kpc$ that we derived for this galaxy.
%\subsubsection*{NGC\,4302}
%
%\subsubsection*{NGC\,4330}
%NGC\,4330 shows a clear tail towards the south, corresponding to a radio-deficient region in the north
%\subsubsection*{NGC\,4396}
%\subsubsection*{NGC\,4402}
%\subsubsection*{NGC\,4424}
% It shows only a mildly asymmetric radio morphology while it has a prominent \hi{}-tail \citep{Chung2007Tails} in the same direction. 
%\subsubsection*{NGC\,4501}
%\subsubsection*{NGC\,4522}
%\subsubsection*{NGC\,4548}
%\subsubsection*{NGC\,4569}
%\subsubsection*{NGC\,4602}
%\subsubsection*{NGC\,4634}
%\subsubsection*{NGC\,4654}
%\subsubsection*{IC\,3105}
%\subsubsection*{IC\,3258}
%\subsubsection*{IC\,3476}
%\subsubsection*{IC\,800}

\subsection{Radio-SFR relations}
In \autoref{fig:SFRvsl144},  our best-fitting radio SFR relations are displayed. As expected, our result for the non-interacting galaxies is in good agreement with \citet{Heesen2022NearbyRelation}, who worked on galaxies which are nearby and are not members of a cluster. 
Comparison of our relation for the non-RPS objects with the work of \citet{Gurkan2018LOFAR/H-ATLAS:Relation} shows that their relation is shifted towards higher radio luminosity, in particular towards the low SFRs. These authors also used LOFAR data, but their sample extends to significantly larger redshifts and has a mean stellar mass that is 2.5 times higher than in our work. We argue that the mass-dependency can mitigate the discrepancy between the various relations, taking into account the mean stellar mass of the samples can account for a discrepancy of $\approx 0.1$\,dex in \autoref{fig:SFRvsl144}.
\citet{Roberts2021LoTSSClusters} on the other hand studied a sample of 95 jellyfish galaxies in low-redshift clusters and found a relation even above our RPS sample. In agreement with the picture that RPS increases the radio-SFR ratio, this relation is closest to the ones for our RPS sample. Indeed, the fact that the relation of \citet{Roberts2021LoTSSClusters} is offset to higher radio luminosities compared to our relations for the RPS sample could be explained by the more efficient RPS for the objects in the \citet{Roberts2021LoTSSClusters} sample, since they are mostly in clusters that are more massive than Virgo. It was also previously reported that Coma and A1367, more massive nearby clusters, show a greater radio excess compared to Virgo \citep[see][and references therein]{Boselli2006EnvironmentalClusters}.

That galaxies suffering from RPS do not follow the radio-SFR relation of normal star-forming galaxies but show a higher radio luminosity is well established now due to observations in nearby galaxy clusters. In our RPS sample, this radio excess is widespread ($\sim 14/17$ objects) and ranges from 0.1 to 0.5\,dex \citep{Murphy2009}, in line with previous radio continuum studies of the cluster \citep[e.g.][]{Niklas1995,Murphy2009}.

\subsection{Spectral index of RPS galaxies}
We additionally report for the first time tentative evidence for a spectral index-mass relation for RPS galaxies that is shifted to steeper spectral index values compared to normal star-forming galaxies (\autoref{fig:si}). While the significance of this result is currently limited by the availability of uniform, high-fidelity continuum data at 1.4\,GHz, it is noteworthy that also other studies reported steep spectral indices for a number of objects suffering from RPS \citep{Chen2020,Muller2021NatAst,Ignesti2022Gasp}. The spectral steepening is indicative of a CRe population that is of higher radiative age. 

Little is known about the spectral indices of RPS galaxies, with studies mostly limited to a few or individual object  \citep{Vollmer2013,Chen2020,Muller2021NatAst,Ignesti2022Gasp,Lal2022,Roberts2022IIIPerseus}. A multitude of effects that could explain the enhanced radio luminosity of RPS galaxies are also able to influence the spectral properties of this population. 
Particle (re-)acceleration due to ISM-ICM shocks could introduce an additional source of CRe \citep{Murphy2009} with an injection spectral index that may differ from acceleration at supernova shocks associated with star-forming activity. In front of a galaxy moving at a velocity greater than the speed of sound $c_\mathrm{s}$ through the ICM, a bow shock is expected to form \citep{Stevens1999}. In principle, particle acceleration may occur at this bow shock, or at reverse shocks launched into the ISM. The bow shock should be $\sim$$1.5$\,kpc in front of the galaxy depending on it's size of the galaxy and velocity \citep{Farris1994}. Due to high speed of sound ($c_\mathrm{s}\sim500$\,km\,s$^{-1}$, e.g. \citet{Simionescu2017}) in the ICM, the shock acceleration will take place in the low Mach number regime.
That means that the injection spectral index of the shock \citep{Drury1983}:
\begin{equation}
    \alpha_\mathrm{inj}(\mathcal{M}) = -\frac{\mathcal{M}^2+ 1}{\mathcal{M}^2 - 1} + 0.5
\end{equation}
will be steeper than the $\sim$$-0.5$ in high Mach-number SN explosions. While a significant contribution of CRe accelerated at the bow shock to the total radio emission of RPS galaxies could thus explain both the radio excess and the steeper spectral indices for this population, we do not consider this scenario since we do not observe radio emission $\sim$$1-2$\,kpc in advance of the leading edges of the galaxies, the only exceptions are NGC\,4607 and NGC\,4634. Conversely, these regions are oftentimes deficient in radio emission, as already reported by \citet{Murphy2009}.
As alternative to the acceleration at the bow-shock, CRe could instead be accelerated at reverse shocks launched into the ISM. However, due to the lower temperature of the ISM and the correspondingly slow speed of sound, these shocks would then again be in the high Mach-number regime resulting in $\alpha_\mathrm{inj}\sim-0.5$, such that they would not lead to a steepening of the synchrotron spectrum. Thus, we do not consider acceleration due to ISM-ICM shocks as mechanism that can explain steeper radio spectral indices for RPS galaxies.
 
Another scenario is that of an enhanced magnetic field strength due to compression of the ISM magnetic field \citep{Boselli2006EnvironmentalClusters} or the magnetic draping mechanism \citep{Dursi2008Draping,Pfrommer2010}. This effect would cause a steepening of the synchrotron spectrum, since the  loss time $t_\mathrm{sync}$ would decrease compared to the escape time $t_\mathrm{esc}$, meaning that more spectral aging can occur before the CRe escape the galaxy. In this scenario, the enhancement of the radio emission should also be localized at the leading edge, where the  magnetic field compression takes place. This mechanism is supported by the detection of asymmetric polarized ridges in some of the Virgo cluster RPS galaxies \citep[e.g.][]{Vollmer2008NGC4501,Pfrommer2010}. If the magnetic fields are enhanced due to leading-edge compression of the ISM, the corresponding gas compression could possibly introduce relevant ionization capture losses for low-energy electrons. Those compete with the fast radiative losses at high electron energy due to the accelerated radiative aging in shaping the observed synchrotron spectral index. 

As third scenario,  \citet{Ignesti2022Gasp} suggested a rapid quenching of the SFR due to the environmental interaction as explanation for the radio excess. This would translate to a strong reduction of $\mathrm{SFR}_\mathrm{H\upalpha}$ after a delay of only few Myr but would only affect $L_{144}$ at later times due to the lifetime of the massive O-B stars and the cosmic rays that contribute to the 144\,MHz synchrotron emission. The availability of UV data presents us with the opportunity to constrain the quenching scenario, since it probes time scales in between the H$\upalpha$ and 144\,MHz continuum emission \citep{Kennicutt2012StarGalaxies,Leroy2012}. This scenario would also lead to a steeper radio spectrum for the RPS galaxies due to the declining injection of fresh CRe, resulting in an increased average radiative age of the CRe.

\subsection{Model scenarios}
\begin{figure}
    \centering
    \includegraphics[width=\linewidth]{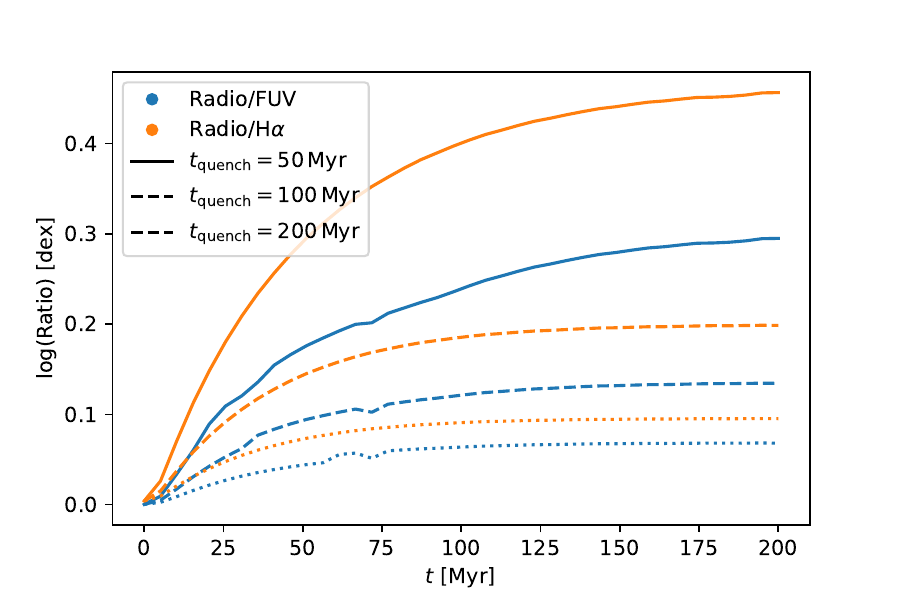}
    \includegraphics[width=\linewidth]{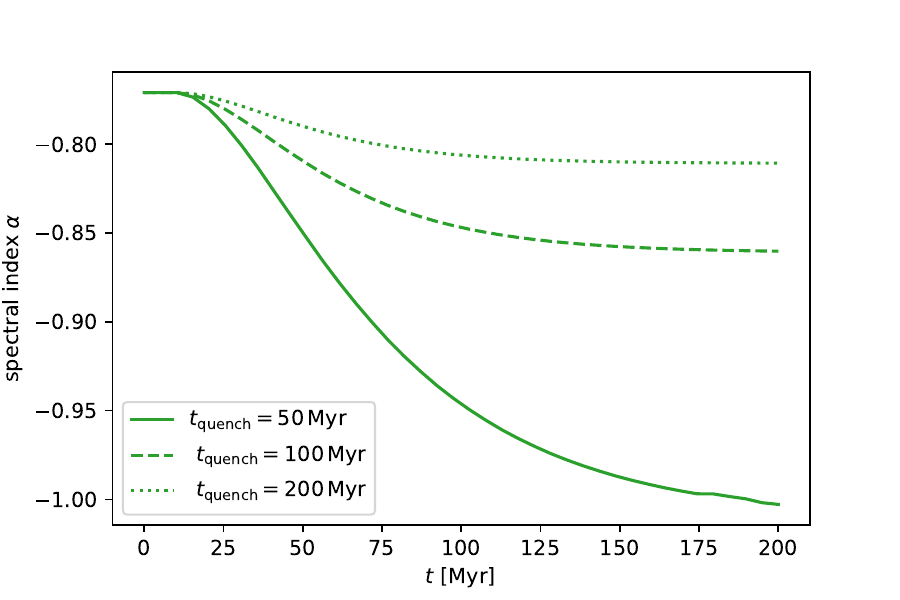}
    \caption{Scenario (1) -- quenched star formation model. \textit{Top panel:} Log-ratio of the 144\,MHz to FUV emission (blue) and H$\upalpha$ emission (orange) for a quenching time of 50\,Myr (solid lines), 100\,Myr (dashed lines) and 200\,Myr (dotted lines) as a function of time since the onset of SFR-quenching. The ratios are relative to $t=0$. \textit{Bottom panel:} Evolution of the 144\,MHz to 1.4\,GHz spectral index for the two models.}
    \label{fig:quench}
\end{figure}
\begin{figure}
    \centering
    \includegraphics[width=\linewidth]{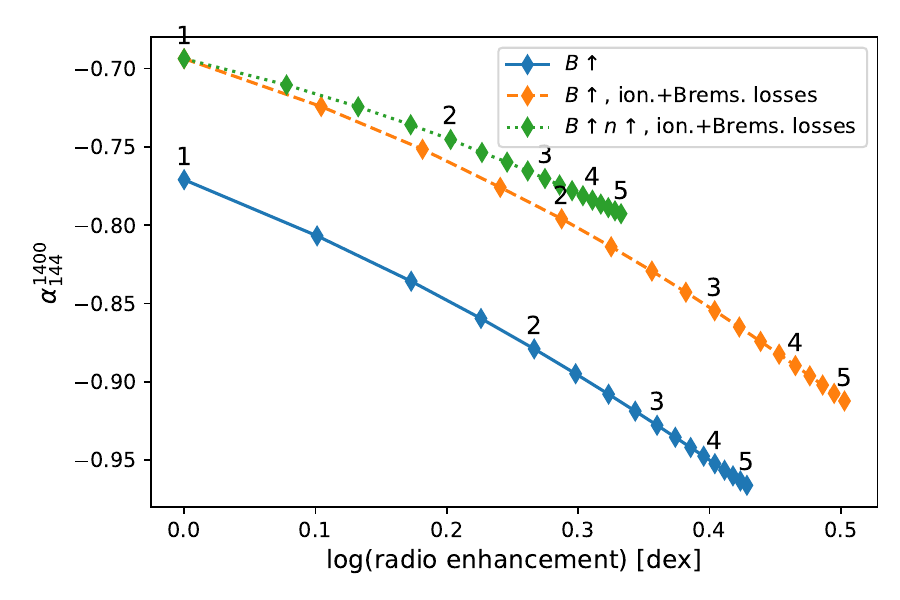}
    \caption{Scenario (2) -- compression model. Effect of an increase of the initial $10\,\mathrm{\upmu{G}}$ magnetic field on the radio excess and synchrotron spectrum (blue continuous line). Diamond symbols mark the fractional increase of the magnetic field in increments of 0.25, integer values are highlighted. The orange dashed line shows the same model when including ionization and Bremsstrahlung losses. The green dotted line shows the same model when also increasing the gas density by the same amount as the magnetic field.}
    \label{fig:comp}
\end{figure}

In the following, we will analyze the phenomenology of simple homogeneous models for the third and second scenario mentioned above.  We consider the following version of the momentum-space Fokker-Planck-equation (ignoring diffusion and re-acceleration processes) to describe the evolution of the electron spectrum $N(E,t)$ as a function of time $t$ and energy $E$:
\begin{equation}\label{eq:fokkerplanck}
    \frac{\partial N(E,t)}{\partial t} = \frac{\partial}{\partial E} \left[ N(E,t) b(E)\right] - \frac{N(E,t)}{t_\mathrm{esc}} + Q(E,t).
\end{equation} 
Here, $t_\mathrm{esc} $ is the CRe escape time for which we employ a value of 50\,Myr \citep[see e.g.][]{Dorner2023} and $b(E)$ describes the energy losses. In the case of synchrotron and inverse Compton losses,  $b(E)=b_0E^2$ with: 
\begin{equation}
b_0=3.2\times 10^{-6} \left( B^2/{2\mu_0} + u_\mathrm{rad}\right)\,\mathrm{m^3\,MeV^{-2}\,Myr^{-1}},
\end{equation}
where we assume a photon field of energy density $u_\mathrm{rad}=6\times 10^{5}$\,eV\,m$^{-3}$ and synchrotron radiation in a magnetic field $B=10\,\mathrm{\upmu{G}}$ \citep{Sarazin1999}. For the source term $Q(E,t) = q E^{-\delta}$ we consider the injection of electrons with an power-law shock-acceleration spectrum with an exponent of $\delta=-2$ and a normalization factor $q$.
The full model and numerical solution of \autoref{eq:fokkerplanck} and the calculation of the radio, FUV and H$\upalpha$ emission are described in \autoref{sec:appendix2}. 

We start from the equilibrium state electron spectrum, which corresponds to a spectral index of $\alpha=-0.77$. We then consider two scenarios: (1) the RPS causes a quenching of the SFR, modeled by an exponential decay of the source term $Q(E)$ with an e-folding time scale of $t_\mathrm{quench}$ by a factor $\exp{(-{t}/{t_\mathrm{quench}})}$; and (2) compression of the ISM magnetic field enhances the magnetic field of a galaxy by a factor of $f_B$.

For scenario (1), the exponential SFR history is of course a simplification valid only for the recent star-forming history ($\sim$ few 100\,Myr). However, this parametrization with only one free parameter allows us to describe the basic effects of a declining SFR due to RPS. In \autoref{fig:quench}, we show the time evolution of the observed radio-FUV and radio-H$\upalpha$ ratio (normalized to 1 at $t=0$) and the radio spectral index for this scenario with quenching time scales of 50, 100 and 200\,Myr. In the model with $t_\mathrm{quench}=50$\,Myr, we can reproduce the observed enhancement of $\sim 0.4$\,dex in the radio-H$\upalpha$ ratio and a significant steepening of the radio spectral index of $\Delta\alpha \approx 0.2$, in agreement with our tentative evidence of steeper spectral indices for RPS galaxies. Higher values of $t_\mathrm{quench}$ cause a less severe steepening and radio excess.
A property of this model is that, due to the longer life time of the stars responsible for the FUV emission, the radio-FUV ratio should be less enhanced compared to the radio-H$\upalpha$ ratio. Indeed for three of the objects with the highest radio excesses (NGC\,4330, NGC\,4501 and NGC\,4522) we find that the distance to the best-fitting relation is stronger relative to $\mathrm{SFR_{H\upalpha}}$ and weaker relative to $\mathrm{SFR_\mathrm{FUV}}$ (see \autoref{fig:d3d}). However, for the remaining objects, no trend can be identified and we also find no significant difference in the $\mathrm{SFR_{UV}}/\mathrm{SFR_{H\upalpha}}$ ratios for the RPS and non-RPS sample. %Furthermore, the mean of direct SFR ratios $R_\mathrm{UV/H\alpha} = \log_{10}(\mathrm{SFR_{UV}}/\mathrm{SFR_{H\alpha}})$ is $R_\mathrm{UV/H\alpha} = 0.07\pm0.09$ for the RPS sample and $R_\mathrm{UV/H\alpha} = 0.09\pm0.13$ (mean ratio of compared to the control sample.  
For the quenching scenario (1) to fully explain the radio excess and the steep radio spectra of RPS galaxies, violent quenching of the star formation activity with $t_\mathrm{quench}\leq100$\,Myr would need to be a widespread phenomenon and should also imprint in a systematic excess of UV to $H\upalpha$ emission, which we only observe for NGC\,4330, NGC\,4330 and NGC\,4522 . 
Spectral energy density fitting of Virgo cluster galaxies yields similarly short quenching time scales for NGC\,4330 and NGC\,4522\citep{Boselli2016quenching}. %Using stellar population modeling, \citet{Boselli2006NGC4569} found an best-fitting age of the RPS event in NGC\,4569 of 100\,Myr, while for NGC\,4330, \citep{Fossati2018} derived a progressive quenching ongoing since 500\,Myr. 
Less violent quenching will still yield some contribution to the radio excess and spectral steepening, but cannot fully explain the observations.
%Rapid SFR quenching was also found for galaxies outside the Virgo cluster; for the jellyfish galaxy JW100, \citet{Ignesti2022Gasp} reported that the SFR dropped by $75\%$ within 55\,Myr. These authors also analyzed six more jellyfish galaxies \citep{Ignesti2022WalkGalaxies}, finding a quenching of the RPS between $50\%$ and $90\%$ within 200\,Myr. \textbf{We conclude that the rapid quenching scenario }

In scenario (2), we model the effect of an increase of the magnetic field strength by a factor of $f_B$. This boosts the synchrotron emission and also steepens the spectral index. In the limit $t_\mathrm{syn} >> t_\mathrm{esc}$, the electron spectrum is escape-dominated and the corresponding radio slope approaches $\alpha=(\delta-1)/2=-0.5$. In the opposite case of high magnetic field strength ($t_\mathrm{esc} >> t_\mathrm{syn}$), it is loss-dominated and the radio slope is close to $\alpha=\delta/2=-1$. For the initial state with $B=10\,\upmu\mathrm{G}$, the spectral index is $\alpha=-0.77$. The blue line in \autoref{fig:comp} shows the change in radio emission and spectral index for different values of $f_B$. An increase of the magnetic field by $f_B=1.5$ steepens the spectral index to $\alpha=-0.84$ and increases the synchrotron emission by 0.17\,dex, for $f_B=2$, we find a spectral index of $\alpha=-0.88$ with a radio enhancement of 0.27\,dex. To explain the galaxies with the strongest excesses of $\geq 0.4\,$dex, we would require an increase in the magnetic field strength by $f_b>4$ (corresponding to a factor of 16 in  energy). The quoted values represent the equilibrium states, but we note that during the ramp-up phase of the magnetic field, the radio emission will be higher and the spectral index may be steeper while the `surplus' electrons (compared to the new equilibrium state) are being depleted. 
Also, less drastic increases of the magnetic field can be required to explain the enhanced radio emission if the electron distribution is strongly dominated by other processes than synchrotron radiation, i.e. in the case of strong Bremsstrahlung and ionization losses \citep[see e.g.][]{Basu2015}. If we extend our model by ionization and Bremsstrahlung losses with a gas number density of $N_\mathrm{gas}=10^6\,\mathrm{m}^{-3}$ (orange line in \autoref{fig:comp}), we still find a significant steepening ($\alpha = -0.64$ to $\alpha = -0.73$ ) and increased radio emission (by 0.3\,dex) in the case $f_B=2$. The main effect of ionization losses is that for the same magnetic field strength, the observed radio spectrum is flattened towards low frequencies (thus also explaining why the spectral indices in the model without ionization losses are mostly steeper than the observed ones). A much weaker steepening and radio enhancement is predicted by our model if we assume that the gas density $n_\mathrm{gas}$ is compressed by the same factor as the magnetic field due to the strong ionization losses in such a scenario (green dotted line in \autoref{fig:comp}). To model the relative importance of gas and magnetic field compression and cosmic ray transport, refined simulations are required. %This would also allow to probe if the overall reduced content of neutral gas in  contribution to the spectral steepening and radio enhancement in RPS galaxies might be due to their overall reduced content of neutral gas, which reduces global ionization losses compared to non-perturbed objects.

An important open question in scenario (2) is the nature of physical mechanism that is able to increase the magnetic field in RPS galaxies.
With simulations,  \citet{Farber2022} were able to reproduce a magnetic field enhancement due to compression in the range $f_B \leq 3$, however, \citet{Tonnesen2014} found that increase to only last a short time ($\sim 25$\,Myr).
An alternative to the adiabatic compression of the ISM magnetic fields is the magnetic draping mechanism, where the galaxy sweeps up ICM magnetic field lines at the leading edge \citep{Dursi2008Draping}. It was proposed to explain the asymmetric polarized radio emission for a number of Virgo cluster galaxies \citep{Pfrommer2010}. The main difficulty of both mechanisms is that they generate a magnetic field enhancement localized at the leading edge. This is only clearly the case in NGC\,4501, where the leading edge radio excess (see \autoref{fig:ratio} corresponds well with the asymmetric polarized radio emission \citep[e.g.][]{Vollmer2007TheGalaxies,Vollmer2008NGC4501}.
Many other galaxies in the RPS sample (e.g. NGC\,4330, NGC\,4396, NGC\,4402, NGC\,4522, IC\,3476) even show a deficit of radio continuum emission in that region, likely due to the fast advection of CRe or locally efficient ionization losses, amplified by the ISM compression. The compression or magnetic draping scenario can also not easily explain the enhanced radio emission across the disk that is observed for most galaxies. 
However, the limited angular resolution of our maps, the varying alignment between the galaxy disks and the velocity vector combined with projection effects (the line of sight velocity relative to M\,87 is commonly $\sim 1000$\,km/s, indicating significant radial stripping, see \autoref{tab:rpsgalaxies}) complicate the identification of such local signatures.  %In \citet{Miller2001TheClusters}, thermal compression from the ICM was instead proposed as a global mechanisms to enhance the magnetic field, but this scenario fails to explain the apparent connection between the radio excess and RPS signatures \citep{Murphy2009}.
Tuned simulations may be required to investigate if the observed disk-wide radio enhancement can be generated by magnetic field enhancement and projection effects alone.

\section{Conclusion}\label{sec:conclusion}

In this paper, we analyzed a sample of 17 Virgo cluster galaxies with a radio morphology indicative of RPS. This sample contains galaxies with masses as low as $7.8\times10^7\,\mathrm{M_\odot}$. Our LOFAR observations allowed the identification of four new RPS candidates (NGC\,4607, NGC\,4634, IC\,800, IC\,3258) and four new objects with radio continuum tails (NGC\,4302, NGC\,4424, IC\,3105, IC\,3476) in the cluster.
We compare them to a statistical sample of 120 nearby star-forming galaxies in the HRS. Using 144\,MHz observations of the LOFAR HBA Virgo Cluster Survey and LoTSS together with the multi-wavelength data of the HRS, our findings are:
\begin{itemize}
\item  The galaxies in the statistical sample without signs of environmental perturbation in LOFAR follow a super-linear radio-SFR relation with a slope of $1.39\pm0.06$ and $1.42\pm0.06$ compared to H$\upalpha$ and FUV-based SFRs, in good agreement with previous low-frequency studies of nearby galaxies.
\item Considering the stellar mass as additional parameter in the radio-SFR relation results in fits with comparable scatter ($\geq 0.25\,$dex), supporting a mass-dependent calorimetric efficiency of star-forming galaxies.
\item We find a clear radio-excess for the RPS sample, with radio luminosities that are a factor of 2-3 higher compared to the radio-SFR relation for normal star-forming galaxies. Qualitatively, we find the strongest radio excess in galaxies with pronounced radio tails.
\item We derived a radio spectral index-total mass relation of the non-RPS galaxies that is in excellent agreement with the literature. For the RPS sample, we find a relation that is shifted towards lower spectral index values by $\Delta\alpha \approx 0.15$.
\item We model the expected radio emission based on a pure diffusion scenario and hybrid NUV+100\,$\upmu\mathrm{m}$ SFR surface densities. Comparing the observed to the expected emission, we find that the radio excess is of global nature and mostly extends across the disks. In a number of cases, we can confirm leading-edge radio-deficit regions as a signature of RPS. Only for NGC\,4501 we find a leading-edge radio enhancement.
\item The radio excess and spectral steepening can be explained by variations in SF history only for NGC\,4330, NGC\,4396 and NGC\,4522 which show the highest radio excesses, which are also accompanied by FUV excesses. This explanation would require rapid quenching with e-folding times $\leq 100$\,Myr.
\item Alternatively, an increased magnetic field due to RPS is also able to generate both enhanced radio emission and a steeper synchrotron spectrum. A doubling of the magnetic field strength can explain a moderate enhancement of the radio luminosity ($\leq 0.27\,\mathrm{dex}$) and steepening ($\Delta\alpha \approx 0.1$). To explain the highest radio excesses with magnetic field increase only, a strong enhancement of more than a factor of four would be required.
\end{itemize}
This study allowed us for the first time to test different models for the radio excess, with the emerging picture that there are multiple realistic channels which can contribute to anomalous radio continuum properties of RPS galaxies. In extreme RPS galaxies, a rapid quenching of the star-forming activity may be a relevant mechanism to explain excess radio emission, however, for the broad population of objects suffering form RPS, magnetic field enhancement likely dominates.
%We considered two scenarios which can simultaneously explain the excess radio emission and the steeper spectral indices for the RPS galaxies. The first scenario, which is a reducing injection of CRe due to the environmental quenching of the SFR activity, can reproduce the radio excess and spectral steepening, but requires a rapid quenching with e-folding times $t_\mathrm{quench} \leq 100$\,Myr. Alternatively, we consider a  scenario with a globally enhanced magnetic field, this directly increases the synchrotron radiation and also causes a steepening by reducing the synchrotron lifetime $t_\mathrm{synch}$ compared to the escape lifetime $t_\mathrm{esc}$. For an magnetic field enhancement by a factor of up to two, this can explain moderately enhanced radio luminosities (by $\leq 0.27\,\mathrm{dex}$) and a moderate steepening ($\Delta\alpha \approx 0.1$). If this scenario should also fully explain the most extreme cases of enhanced radio emission, substantial increases in the magnetic field strength are required (more than a factor of five).
%We concluded that from the scenarios we considered, the objects with a strong radio excess are best explained by a rapid quenching of the SFR activity, while a magnetic field enhancement can contribute to the milder excesses and steepening for the general RPS galaxy population.
In the near future, our VIrgo Cluster multi-Telescope Observations on Radio of Interacting galaxies and AGN (ViCTORIA) project will allow to further constrain the non-thermal physical mechanisms at play in RPS galaxies with deep, polarized L-band observations by MeerKAT and ultra-low frequency measurements with the LOFAR Low-Band Antenna. Crucially, with the homogeneous multi-frequency data, we will be able to definitely confirm or reject the claim presented in this paper that RPS galaxies in the Virgo cluster show steeper radio spectral indices than normal SF galaxies. Further, we will be able to perform high-fidelity spatially resolved studies of the spectral index and to probe the magnetic field structure using polarization information.

\begin{acknowledgements}
HE acknowledges support by the Deutsche Forschungsgemeinschaft (DFG, German Research Foundation) under project number 427771150. MB acknowledges funding by the Deutsche Forschungsgemeinschaft under Germany's Excellence Strategy -- EXC 2121 ``Quantum Universe'' --  390833306.
FdG acknowledges support from the ERC Consolidator Grant ULU 101086378.
AI acknowledges funding from the European Research Council (ERC) under the European Union's Horizon 2020 research and innovation programme (grant agreement No. 833824, PI Poggianti) and the INAF founding program 'Ricerca Fondamentale 2022' (PI A. Ignesti).
LOFAR (van Haarlem et al. 2013) is the Low Frequency Array designed and constructed by
ASTRON. It has observing, data processing, and data storage facilities in several countries, which are owned by various parties (each with their own funding sources), and that are collectively operated by the ILT foundation under a joint scientific policy. The ILT resources have benefited from the following recent major funding sources: CNRS-INSU, Observatoire de Paris and Université d'Orléans, France; BMBF, MIWF-NRW, MPG, Germany; Science Foundation Ireland (SFI), Department of Business, Enterprise and Innovation (DBEI), Ireland; NWO, The Netherlands; The Science and Technology Facilities Council, UK; Ministry of Science and Higher Education, Poland; The Istituto Nazionale di Astrofisica (INAF), Italy. This research made use of the Dutch national e-infrastructure with support of the SURF Cooperative (e-infra 180169) and the LOFAR e-infra group. The Jülich LOFAR Long Term Archive and the German LOFAR network are both coordinated and operated by the Jülich Supercomputing Centre (JSC), and computing resources on the supercomputer JUWELS at JSC were provided by the Gauss Centre for Supercomputing e.V. (grant CHTB00) through the John von Neumann Institute for Computing (NIC).
This research made use of the University of Hertfordshire high-performance computing facility and the LOFAR-UK computing facility located at the University of Hertfordshire and supported by STFC [ST/P000096/1], and of the Italian LOFAR IT computing infrastructure supported and operated by INAF, and by the Physics Department of Turin university (under an agreement with Consorzio Interuniversitario per la Fisica Spaziale) at the C3S Supercomputing Centre, Italy.
\end{acknowledgements}
\bibliographystyle{aa}
\bibliography{references}

\begin{appendix}
\onecolumn
\section{LOFAR data}\label{sec:appendix0}
\begin{longtable}{ccccc}
\caption{LOFAR data.}
\label{tab:lofardata} \\
\hline\hline
Name & $S_{144}$ & $d$ & $L_{144}$ & $\alpha_{144}^{1400}$ \\
 & [Jy] & [Mpc] & [$\mathrm{W\,Hz^{-1}}$] & \\
(1) & (2) & (3) & (4) & (5) \\ 
\hline
\endfirsthead
\caption{continued.}\\
\hline
Name & $S_{144}$ & $d$ & $L_{144}$ & $\alpha_{144}^{1400}$ \\
 & [Jy] & [Mpc] & [$\mathrm{W\,Hz^{-1}}$] & \\
(1) & (2) & (3) & (4) & (5) \\
\hline
\endhead
\hline
\endfoot
HRS\,2 & (1.5$\pm$0.3)e$-2$ & 18.44 & (6.1$\pm$1.3)e+20 & $-0.47\pm0.11^\mathrm{a}$ \\
IC\,610 & (2.0$\pm$0.4)e$-2$ & 16.71 & (6.5$\pm$1.3)e+20 & $-0.58\pm0.11^\mathrm{a}$ \\
NGC\,3254 & (3.4$\pm$0.5)e$-2$ & 19.37 & (1.5$\pm$0.2)e+21 & $<-1.25^\mathrm{a}$ \\
NGC\,3277 & (1.8$\pm$0.3)e$-2$ & 20.21 & (8.8$\pm$1.4)e+20 & $-0.58\pm0.09^\mathrm{a}$ \\
HRS\,10 & (8.2$\pm$1.3)e$-3$ & 21.66 & (4.6$\pm$0.8)e+20 & $-0.56\pm0.13^\mathrm{a}$ \\
NGC\,3287 & (2.8$\pm$0.6)e$-2$ & 18.93 & (1.2$\pm$0.2)e+21 & $-0.39\pm0.10^\mathrm{a}$ \\
HRS\,12 & (3.4$\pm$0.8)e$-3$ & 19.89 & (1.6$\pm$0.4)e+20 & $<-0.27^\mathrm{a}$ \\
NGC\,3294 & (2.3$\pm$0.3)e$-1$ & 22.47 & (1.4$\pm$0.2)e+22 & $-0.70\pm0.08^\mathrm{a}$ \\
NGC\,3338 & (7.1$\pm$1.4)e$-2$ & 18.57 & (2.9$\pm$0.6)e+21 & $-0.40\pm0.10^\mathrm{a}$ \\
NGC\,3346 & (3.0$\pm$0.6)e$-2$ & 18.0 & (1.2$\pm$0.2)e+21 & $-0.34\pm0.10^\mathrm{a}$ \\
NGC\,3380 & (1.1$\pm$0.2)e$-2$ & 22.91 & (7.1$\pm$1.2)e+20 & $-0.68\pm0.12^\mathrm{a}$ \\
NGC\,3381 & (2.9$\pm$0.4)e$-2$ & 23.29 & (1.9$\pm$0.3)e+21 & $-0.70\pm0.09^\mathrm{a}$ \\
NGC\,3395 & (3.9$\pm$0.6)e$-1$ & 23.1 & (2.5$\pm$0.4)e+22 & $-0.64\pm0.08^\mathrm{a}$ \\
NGC\,3424 & (3.0$\pm$0.4)e$-1$ & 21.44 & (1.6$\pm$0.2)e+22 & $-0.75\pm0.08^\mathrm{a}$ \\
NGC\,3430 & (1.4$\pm$0.2)e$-1$ & 22.64 & (8.7$\pm$1.3)e+21 & $-0.68\pm0.08^\mathrm{a}$ \\
NGC\,3437 & (2.6$\pm$0.5)e$-1$ & 18.24 & (1.0$\pm$0.2)e+22 & $-0.59\pm0.10^\mathrm{a}$ \\
HRS\,26 & (4.8$\pm$0.8)e$-3$ & 22.41 & (2.9$\pm$0.5)e+20 & $<-0.42^\mathrm{a}$ \\
NGC\,3442 & (4.6$\pm$0.7)e$-2$ & 24.77 & (3.4$\pm$0.5)e+21 & $-0.56\pm0.08^\mathrm{a}$ \\
NGC\,3451 & (3.0$\pm$0.6)e$-2$ & 19.03 & (1.3$\pm$0.3)e+21 & $-0.43\pm0.10^\mathrm{a}$ \\
NGC\,3448 & (1.9$\pm$0.3)e$-1$ & 19.63 & (8.6$\pm$1.3)e+21 & $-0.54\pm0.08^\mathrm{a}$ \\
NGC\,3504 & (8.8$\pm$1.3)e$-1$ & 21.94 & (5.1$\pm$0.8)e+22 & $-0.51\pm0.08^\mathrm{a}$ \\
NGC\,3512 & (3.8$\pm$0.6)e$-2$ & 19.61 & (1.7$\pm$0.3)e+21 & $-0.56\pm0.08^\mathrm{a}$ \\
NGC\,3596 & (5.8$\pm$1.2)e$-2$ & 17.04 & (2.0$\pm$0.4)e+21 & $-0.51\pm0.10^\mathrm{a}$ \\
NGC\,3629 & (2.1$\pm$0.4)e$-2$ & 21.53 & (1.2$\pm$0.2)e+21 & $-0.46\pm0.10^\mathrm{a}$ \\
NGC\,3631 & (4.3$\pm$0.6)e$-1$ & 16.5 & (1.4$\pm$0.2)e+22 & $-0.72\pm0.08^\mathrm{a}$ \\
NGC\,3655 & (1.6$\pm$0.3)e$-1$ & 21.43 & (8.6$\pm$1.7)e+21 & $-0.43\pm0.10^\mathrm{a}$ \\
NGC\,3657 & (9.1$\pm$1.4)e$-3$ & 17.2 & (3.2$\pm$0.5)e+20 & $-0.39\pm0.10^\mathrm{a}$ \\
NGC\,3666 & (5.1$\pm$1.0)e$-2$ & 15.14 & (1.4$\pm$0.3)e+21 & $-0.45\pm0.10^\mathrm{a}$ \\
NGC\,3684 & (6.4$\pm$1.6)e$-2$ & 16.54 & (2.1$\pm$0.5)e+21 & $-0.64\pm0.11^\mathrm{a}$ \\
NGC\,3683 & (4.9$\pm$0.7)e$-1$ & 24.4 & (3.5$\pm$0.5)e+22 & $-0.69\pm0.08^\mathrm{a}$ \\
NGC\,3686 & (8.7$\pm$1.8)e$-2$ & 16.51 & (2.8$\pm$0.6)e+21 & $-0.74\pm0.10^\mathrm{a}$ \\
NGC\,3729 & (6.8$\pm$1.0)e$-2$ & 15.14 & (1.9$\pm$0.3)e+21 & $-0.51\pm0.08^\mathrm{a}$ \\
HRS\,61 & (4.3$\pm$0.7)e$-3$ & 17.39 & (1.6$\pm$0.3)e+20 & $<-0.37^\mathrm{a}$ \\
NGC\,3755 & (3.2$\pm$0.5)e$-2$ & 22.44 & (2.0$\pm$0.3)e+21 & $-0.57\pm0.08^\mathrm{a}$ \\
NGC\,3756 & (2.8$\pm$0.4)e$-2$ & 18.41 & (1.1$\pm$0.2)e+21 & $-0.81\pm0.09^\mathrm{a}$ \\
NGC\,3795 & (3.7$\pm$0.6)e$-3$ & 17.33 & (1.3$\pm$0.2)e+20 & $<-0.31^\mathrm{a}$ \\
NGC\,3794 & (1.0$\pm$0.2)e$-2$ & 19.76 & (4.8$\pm$0.8)e+20 & $<-0.74^\mathrm{a}$ \\
NGC\,3813 & (3.4$\pm$0.5)e$-1$ & 20.97 & (1.8$\pm$0.3)e+22 & $-0.57\pm0.08^\mathrm{a}$ \\
HRS\,67 & (6.2$\pm$1.0)e$-3$ & 20.51 & (3.1$\pm$0.5)e+20 & $-0.34\pm0.11^\mathrm{a}$ \\
HRS\,68 & (1.1$\pm$0.2)e$-2$ & 20.17 & (5.6$\pm$0.8)e+20 & $-0.49\pm0.10^\mathrm{a}$ \\
NGC\,3898 & (3.9$\pm$0.6)e$-2$ & 16.73 & (1.3$\pm$0.2)e+21 & $-0.78\pm0.08^\mathrm{a}$ \\
NGC\,3953 & (2.4$\pm$0.4)e$-1$ & 15.0 & (6.4$\pm$1.0)e+21 & $-0.75\pm0.08^\mathrm{a}$ \\
NGC\,3982 & (2.5$\pm$0.4)e$-1$ & 15.83 & (7.6$\pm$1.1)e+21 & $-0.61\pm0.08^\mathrm{a}$ \\
HRS\,76 & (4.7$\pm$0.8)e$-3$ & 15.27 & (1.3$\pm$0.2)e+20 & $-0.18\pm0.11^\mathrm{a}$ \\
NGC\,4100 & (2.1$\pm$0.3)e$-1$ & 15.31 & (6.0$\pm$0.9)e+21 & $-0.60\pm0.08^\mathrm{a}$ \\
NGC\,4178 & (5.6$\pm$1.2)e$-2$ & 16.5 & (1.8$\pm$0.4)e+21 & $-0.38\pm0.10^\mathrm{a}$ \\
IC\,3061 & (1.2$\pm$0.3)e$-2$ & 16.5 & (3.8$\pm$0.9)e+20 & $<-0.80^\mathrm{a}$ \\
NGC\,4207 & (3.4$\pm$0.7)e$-2$ & 16.5 & (1.1$\pm$0.2)e+21 & $-0.25\pm0.10^\mathrm{a}$ \\
NGC\,4212 & (1.3$\pm$0.3)e$-1$ & 16.5 & (4.1$\pm$0.8)e+21 & $-0.72\pm0.10^\mathrm{a}$ \\
NGC\,4216 & (6.4$\pm$1.3)e$-2$ & 16.5 & (2.1$\pm$0.4)e+21 & $-0.72\pm0.10^\mathrm{b}$ \\
NGC\,4222 & (1.3$\pm$0.3)e$-2$ & 16.5 & (4.4$\pm$1.0)e+20 & $-0.58\pm0.12^\mathrm{a}$ \\
NGC\,4237 & (2.3$\pm$0.5)e$-2$ & 16.5 & (7.5$\pm$1.7)e+20 & $-0.58\pm0.11^\mathrm{a}$ \\
IC\,3105$^\dag$ & (4.0$\pm$1.1)e$-3$ & 16.5 & (1.3$\pm$0.4)e+20 & $<-0.36^\mathrm{c}$ \\
NGC\,4254$^\dag$ & (2.5$\pm$0.5)e$+0$ & 16.5 & (8.3$\pm$1.7)e+22 & $-0.77\pm0.10^\mathrm{a}$ \\
NGC\,4289 & (1.5$\pm$0.4)e$-2$ & 16.5 & (4.8$\pm$1.2)e+20 & $-0.53\pm0.12^\mathrm{b}$ \\
NGC\,4294 & (8.6$\pm$1.7)e$-2$ & 16.5 & (2.8$\pm$0.6)e+21 & $-0.51\pm0.10^\mathrm{b}$ \\
NGC\,4298 & (1.3$\pm$0.3)e$-1$ & 16.5 & (4.2$\pm$0.8)e+21 & $-0.88\pm0.10^\mathrm{d}$ \\
NGC\,4302$^\dag$ & (2.5$\pm$0.5)e$-1$ & 16.5 & (8.0$\pm$1.6)e+21 & $-0.77\pm0.10^\mathrm{a}$ \\
NGC\,4303 & (2.0$\pm$0.4)e$+0$ & 16.5 & (6.5$\pm$1.3)e+22 & $-0.67\pm0.10^\mathrm{a}$ \\
NGC\,4307 & (9.5$\pm$2.1)e$-3$ & 23.0 & (6.0$\pm$1.4)e+20 & $-0.57\pm0.14^\mathrm{a}$ \\
NGC\,4312 & (3.6$\pm$0.7)e$-2$ & 16.5 & (1.2$\pm$0.2)e+21 & $-0.50\pm0.10^\mathrm{a}$ \\
NGC\,4316 & (2.1$\pm$0.4)e$-2$ & 23.0 & (1.3$\pm$0.3)e+21 & $-0.54\pm0.10^\mathrm{c}$ \\
NGC\,4321 & (1.5$\pm$0.3)e$+0$ & 16.5 & (4.9$\pm$1.0)e+22 & $-0.74\pm0.10^\mathrm{d}$ \\
NGC\,4330$^\dag$ & (5.6$\pm$1.1)e$-2$ & 16.5 & (1.8$\pm$0.4)e+21 & $-0.63\pm0.11^\mathrm{a}$ \\
NGC\,4343 & (4.2$\pm$0.9)e$-2$ & 23.0 & (2.7$\pm$0.5)e+21 & $-1.01\pm0.11^\mathrm{b}$ \\
IC\,3258$^\dag$ & (9.8$\pm$2.1)e$-3$ & 16.5 & (3.2$\pm$0.7)e+20 & $<-0.74^\mathrm{a}$ \\
NGC\,4351 & (2.0$\pm$0.4)e$-2$ & 16.5 & (6.5$\pm$1.3)e+20 & $<-1.03^\mathrm{a}$ \\
IC\,3268 & (8.0$\pm$1.9)e$-3$ & 23.0 & (5.1$\pm$1.2)e+20 & $-0.26\pm0.12^\mathrm{a}$ \\
NGC\,4359 & (1.1$\pm$0.2)e$-2$ & 17.9 & (4.2$\pm$0.7)e+20 & $<-0.77^\mathrm{a}$ \\
NGC\,4370 & (1.1$\pm$0.2)e$-2$ & 23.0 & (6.8$\pm$1.5)e+20 & $<-0.77^\mathrm{a}$ \\
NGC\,4376 & (5.0$\pm$1.4)e$-3$ & 23.0 & (3.1$\pm$0.9)e+20 & $<-0.43^\mathrm{b}$ \\
NGC\,4380 & (2.0$\pm$0.4)e$-2$ & 23.0 & (1.2$\pm$0.3)e+21 & $<-0.68^\mathrm{a}$ \\
NGC\,4383 & (1.2$\pm$0.2)e$-1$ & 16.5 & (3.9$\pm$0.8)e+21 & $-0.45\pm0.10^\mathrm{a}$ \\
VCC\,827 & (5.1$\pm$1.1)e$-2$ & 23.0 & (3.2$\pm$0.7)e+21 & $-0.59\pm0.10^\mathrm{a}$ \\
NGC\,4390 & (1.8$\pm$0.4)e$-2$ & 23.0 & (1.1$\pm$0.2)e+21 & $<-0.98^\mathrm{d}$ \\
IC\,3322 & (1.9$\pm$0.4)e$-2$ & 23.0 & (1.2$\pm$0.3)e+21 & $-0.79\pm0.12^\mathrm{d}$ \\
NGC\,4396$^\dag$ & (9.6$\pm$1.9)e$-2$ & 16.5 & (3.1$\pm$0.6)e+21 & $-0.79\pm0.10^\mathrm{a}$ \\
NGC\,4402$^\dag$ & (4.3$\pm$0.9)e$-1$ & 16.5 & (1.4$\pm$0.3)e+22 & $-0.81\pm0.10^\mathrm{a}$ \\
NGC\,4413 & (8.8$\pm$1.9)e$-3$ & 16.5 & (2.9$\pm$0.6)e+20 & $<-0.68^\mathrm{a}$ \\
NGC\,4412 & (4.0$\pm$0.9)e$-2$ & 16.5 & (1.3$\pm$0.3)e+21 & $-0.44\pm0.11^\mathrm{d}$ \\
NGC\,4416 & (1.8$\pm$0.4)e$-2$ & 16.5 & (6.0$\pm$1.3)e+20 & $-0.57\pm0.11^\mathrm{a}$ \\
NGC\,4424$^\dag$ & (3.0$\pm$0.6)e$-2$ & 23.0 & (1.9$\pm$0.4)e+21 & $-0.67\pm0.11^\mathrm{a}$ \\
NGC\,4430 & (2.5$\pm$0.5)e$-2$ & 23.0 & (1.6$\pm$0.3)e+21 & $-0.57\pm0.10^\mathrm{b}$ \\
VCC\,1091 & (6.1$\pm$1.4)e$-3$ & 23.0 & (3.9$\pm$0.9)e+20 & $-0.25\pm0.12^\mathrm{a}$ \\
NGC\,4450 & (4.3$\pm$0.9)e$-2$ & 16.5 & (1.4$\pm$0.3)e+21 & $-0.71\pm0.10^\mathrm{a}$ \\
NGC\,4451 & (2.1$\pm$0.4)e$-2$ & 23.0 & (1.3$\pm$0.3)e+21 & $-0.52\pm0.10^\mathrm{b}$ \\
IC\,3392 & (8.2$\pm$1.8)e$-3$ & 16.5 & (2.7$\pm$0.6)e+20 & $-0.44\pm0.13^\mathrm{a}$ \\
NGC\,4457 & (8.0$\pm$1.7)e$-2$ & 16.5 & (2.6$\pm$0.5)e+21 & $-0.45\pm0.10^\mathrm{a}$ \\
NGC\,4470 & (2.1$\pm$0.4)e$-2$ & 16.5 & (7.0$\pm$1.4)e+20 & $-0.14\pm0.10^\mathrm{a}$ \\
NGC\,4480 & (2.2$\pm$0.5)e$-2$ & 16.5 & (7.1$\pm$1.6)e+20 & $-0.86\pm0.13^\mathrm{a}$ \\
NGC\,4491 & (7.5$\pm$2.0)e$-3$ & 16.5 & (2.4$\pm$0.7)e+20 & $<-0.61^\mathrm{a}$ \\
NGC\,4505 & (2.7$\pm$0.6)e$-2$ & 16.5 & (8.7$\pm$2.0)e+20 & $<-1.15^\mathrm{d}$ \\
IC\,797 & (1.1$\pm$0.2)e$-2$ & 16.5 & (3.6$\pm$0.8)e+20 & $<-0.78^\mathrm{a}$ \\
NGC\,4501$^\dag$ & (1.8$\pm$0.4)e$+0$ & 16.5 & (6.0$\pm$1.2)e+22 & $-0.75\pm0.10^\mathrm{a}$ \\
IC\,3476$^\dag$ & (5.0$\pm$1.0)e$-2$ & 16.5 & (1.6$\pm$0.3)e+21 & $-0.81\pm0.10^\mathrm{c}$ \\
NGC\,4519 & (6.4$\pm$1.3)e$-2$ & 16.5 & (2.1$\pm$0.4)e+21 & $-0.83\pm0.10^\mathrm{a}$ \\
NGC\,4522$^\dag$ & (8.1$\pm$1.6)e$-2$ & 16.5 & (2.6$\pm$0.5)e+21 & $-0.48\pm0.10^\mathrm{a}$ \\
NGC\,4525 & (2.1$\pm$0.4)e$-3$ & 16.77 & (7.2$\pm$1.4)e+19 & $<-0.07^\mathrm{d}$ \\
IC\,800$^\dag$ & (1.1$\pm$0.2)e$-2$ & 16.5 & (3.5$\pm$0.8)e+20 & $<-0.76^\mathrm{b}$ \\
NGC\,4532 & (5.0$\pm$1.0)e$-1$ & 16.5 & (1.6$\pm$0.3)e+22 & $-0.63\pm0.10^\mathrm{a}$ \\
NGC\,4535 & (6.4$\pm$1.3)e$-1$ & 16.5 & (2.1$\pm$0.4)e+22 & $-0.96\pm0.10^\mathrm{a}$ \\
IC\,3521 & (2.6$\pm$0.5)e$-2$ & 16.5 & (8.3$\pm$1.7)e+20 & $-0.84\pm0.12^\mathrm{b}$ \\
NGC\,4540 & (1.4$\pm$0.3)e$-2$ & 16.5 & (4.5$\pm$1.0)e+20 & $-0.64\pm0.12^\mathrm{a}$ \\
NGC\,4548$^\dag$ & (1.1$\pm$0.2)e$-1$ & 16.5 & (3.7$\pm$0.8)e+21 & $-1.32\pm0.10^\mathrm{b}$ \\
NGC\,4565 & (8.6$\pm$1.7)e$-1$ & 17.61 & (3.2$\pm$0.6)e+22 & $-0.80\pm0.10^\mathrm{b}$ \\
NGC\,4567 & (7.1$\pm$1.4)e$-2$ & 16.5 & (2.3$\pm$0.5)e+21 & $-0.76\pm0.10^\mathrm{c}$ \\
NGC\,4568 & (5.8$\pm$1.2)e$-1$ & 16.5 & (1.9$\pm$0.4)e+22 & $-0.60\pm0.10^\mathrm{c}$ \\
NGC\,4569$^\dag$ & (5.5$\pm$1.1)e$-1$ & 16.5 & (1.8$\pm$0.4)e+22 & $-0.55\pm0.10^\mathrm{c}$ \\
NGC\,4579 & (6.5$\pm$1.3)e$-1$ & 16.5 & (2.1$\pm$0.4)e+22 & $-0.59\pm0.10^\mathrm{a}$ \\
NGC\,4580 & (1.2$\pm$0.3)e$-2$ & 16.5 & (3.8$\pm$0.9)e+20 & $<-0.79^\mathrm{b}$ \\
NGC\,4595 & (9.6$\pm$2.2)e$-3$ & 16.5 & (3.1$\pm$0.7)e+20 & $-0.31\pm0.12^\mathrm{b}$ \\
NGC\,4606 & (8.9$\pm$2.1)e$-3$ & 16.5 & (2.9$\pm$0.7)e+20 & $<-0.68^\mathrm{a}$ \\
NGC\,4607$^\dag$ & (9.8$\pm$2.0)e$-2$ & 16.5 & (3.2$\pm$0.6)e+21 & $-0.70\pm0.10^\mathrm{a}$ \\
NGC\,4630 & (2.4$\pm$0.5)e$-2$ & 16.5 & (7.9$\pm$1.7)e+20 & $-0.28\pm0.10^\mathrm{a}$ \\
NGC\,4634$^\dag$ & (1.5$\pm$0.3)e$-1$ & 16.5 & (5.0$\pm$1.0)e+21 & $-0.66\pm0.10^\mathrm{a}$ \\
NGC\,4639 & (3.7$\pm$0.8)e$-2$ & 16.5 & (1.2$\pm$0.3)e+21 & $-0.68\pm0.10^\mathrm{b}$ \\
NGC\,4647 & (3.0$\pm$0.6)e$-1$ & 16.5 & (9.8$\pm$2.0)e+21 & $-0.92\pm0.10^\mathrm{a}$ \\
NGC\,4654$^\dag$ & (8.5$\pm$1.7)e$-1$ & 16.5 & (2.8$\pm$0.6)e+22 & $-0.87\pm0.10^\mathrm{a}$ \\
NGC\,4689 & (1.4$\pm$0.3)e$-1$ & 16.5 & (4.6$\pm$0.9)e+21 & $-1.15\pm0.10^\mathrm{a}$ \\
NGC\,4725 & (1.8$\pm$0.3)e$-1$ & 17.27 & (6.4$\pm$1.0)e+21 & $-0.79\pm0.08^\mathrm{a}$ \\
NGC\,4747 & (1.6$\pm$0.3)e$-2$ & 16.84 & (5.4$\pm$0.9)e+20 & $-0.33\pm0.08^\mathrm{a}$ \\
NGC\,4746 & (1.8$\pm$0.4)e$-1$ & 16.5 & (5.9$\pm$1.2)e+21 & $-0.52\pm0.10^\mathrm{a}$ \\
NGC\,4758 & (6.7$\pm$1.5)e$-3$ & 16.5 & (2.2$\pm$0.5)e+20 & $-0.34\pm0.13^\mathrm{a}$ \\
NGC\,4779 & (2.6$\pm$0.5)e$-2$ & 16.5 & (8.4$\pm$1.8)e+20 & $-0.49\pm0.10^\mathrm{a}$ \\
NGC\,5014 & (3.2$\pm$0.5)e$-2$ & 16.23 & (10.0$\pm$1.5)e+20 & $-0.44\pm0.08^\mathrm{a}$ \\
NGC\,5145 & (1.3$\pm$0.2)e$-1$ & 17.5 & (4.7$\pm$0.7)e+21 & $-0.58\pm0.08^\mathrm{a}$ \\
IC\,902 & (2.1$\pm$0.3)e$-2$ & 22.97 & (1.3$\pm$0.2)e+21 & $-0.46\pm0.08^\mathrm{a}$ \\
NGC\,5248 & (5.4$\pm$1.1)e$-1$ & 16.46 & (1.8$\pm$0.4)e+22 & $-0.58\pm0.10^\mathrm{a}$ \\
NGC\,5301 & (7.2$\pm$1.1)e$-2$ & 21.54 & (4.0$\pm$0.6)e+21 & $-0.63\pm0.08^\mathrm{a}$ \\
NGC\,5303 & (7.4$\pm$1.1)e$-2$ & 20.27 & (3.6$\pm$0.5)e+21 & $-0.46\pm0.08^\mathrm{a}$ \\
HRS\,300 & (3.0$\pm$0.6)e$-3$ & 19.34 & (1.4$\pm$0.3)e+20 & $<-0.22^\mathrm{a}$ \\
NGC\,5372 & (5.2$\pm$0.8)e$-2$ & 24.53 & (3.7$\pm$0.6)e+21 & $-0.51\pm0.08^\mathrm{a}$ \\
NGC\,5486 & (9.5$\pm$1.5)e$-3$ & 19.76 & (4.4$\pm$0.7)e+20 & $<-0.71^\mathrm{a}$ \\
HRS\,315 & (3.2$\pm$0.7)e$-3$ & 20.57 & (1.6$\pm$0.4)e+20 & $<-0.24^\mathrm{a}$ \\
NGC\,5645 & (4.1$\pm$0.9)e$-2$ & 19.57 & (1.9$\pm$0.4)e+21 & $-0.44\pm0.10^\mathrm{a}$ \\
NGC\,5669 & (2.6$\pm$0.5)e$-2$ & 19.54 & (1.2$\pm$0.2)e+21 & $-0.38\pm0.10^\mathrm{a}$ \\
NGC\,5692 & (3.0$\pm$0.6)e$-2$ & 22.59 & (1.8$\pm$0.4)e+21 & $-0.68\pm0.10^\mathrm{a}$ \\ \hline
\end{longtable}
\noindent\makebox[\textwidth][c]{%
    \begin{minipage}{0.55\linewidth}
    {\small\textbf{Notes.} Column (1) lists the galaxy names, objects that are marked with a dagger symbol are in the RPS sample. Column (2) shows the LOFAR flux density measured at $20''$ resolution, column (3) the distance (taken form \citet{Boselli2010TheSurvey} for objects outside the Virgo cluster, otherwise assumed to be 16.5\,Mpc), column (4) the 144\,MHz  luminosity and column (5) the spectral index between 144\,MHz and 1.4\,GHz. The 1.4\,GHz flux densities (or upper limits) are taken from  $^\mathrm{a}$ \citet{Boselli2015HGalaxies}; $^\mathrm{b}$ measured from the maps of \citet{Chung2009}; $^\mathrm{c}$ \citet{Murphy2009} and $^\mathrm{d}$ \citet{Vollmer2010}.} 
    \end{minipage}}

\FloatBarrier
\newpage
\section{Radio-SFR maps}\label{sec:appendix1}
\begin{figure*}[ht!]
\centering
    \includegraphics[width=0.14\textwidth]{figures/legend.png}
    \hspace{0.10\textwidth}
    \includegraphics[width=0.5\textwidth]{figures/cbar.png}
    \includegraphics[width=0.24\textwidth]{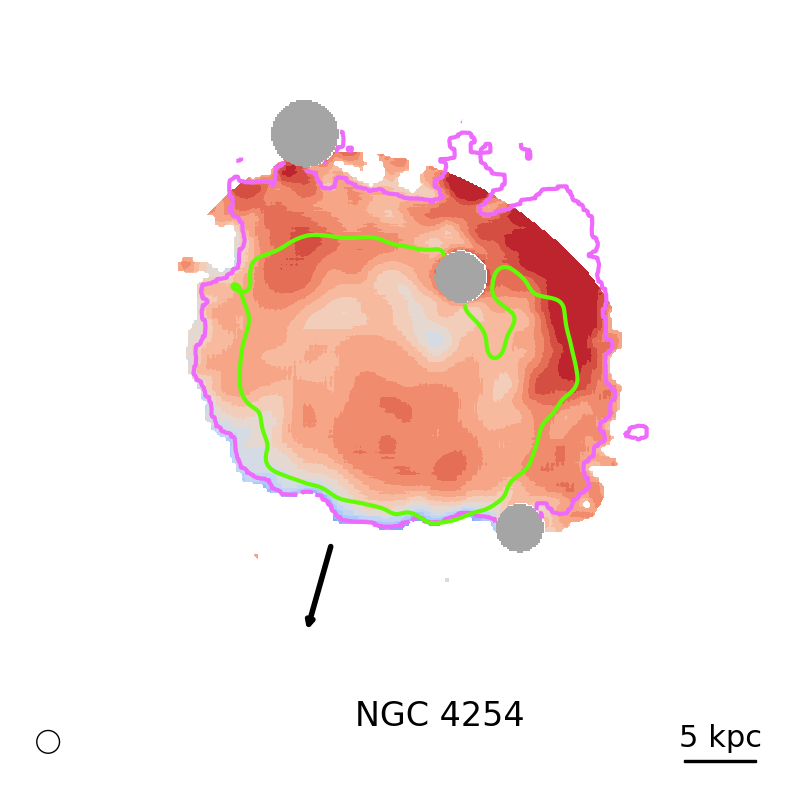}
    \includegraphics[width=0.24\textwidth]{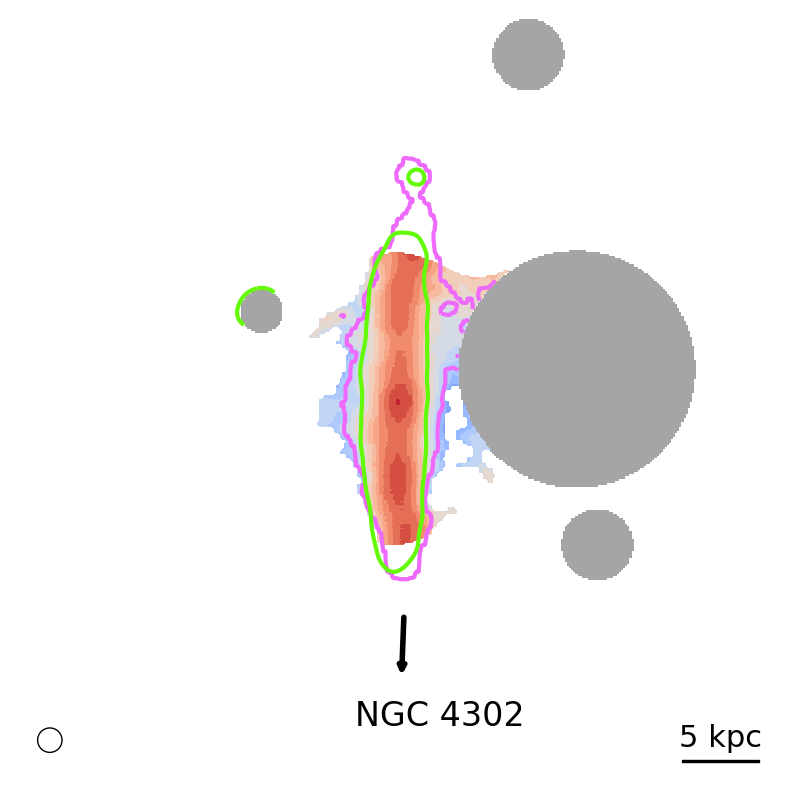}
    \includegraphics[width=0.24\textwidth]{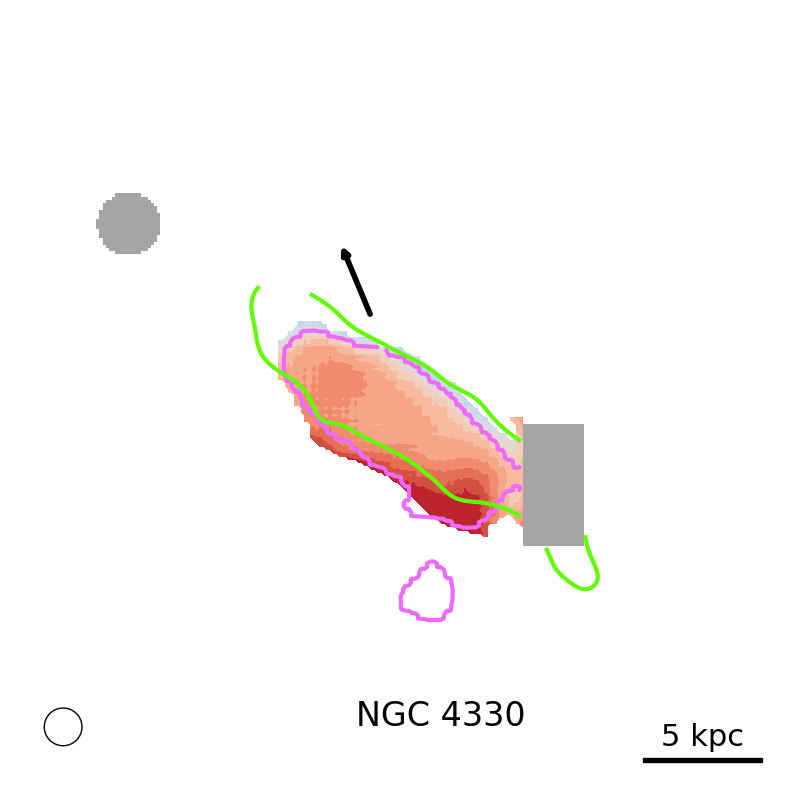}
    \includegraphics[width=0.24\textwidth]{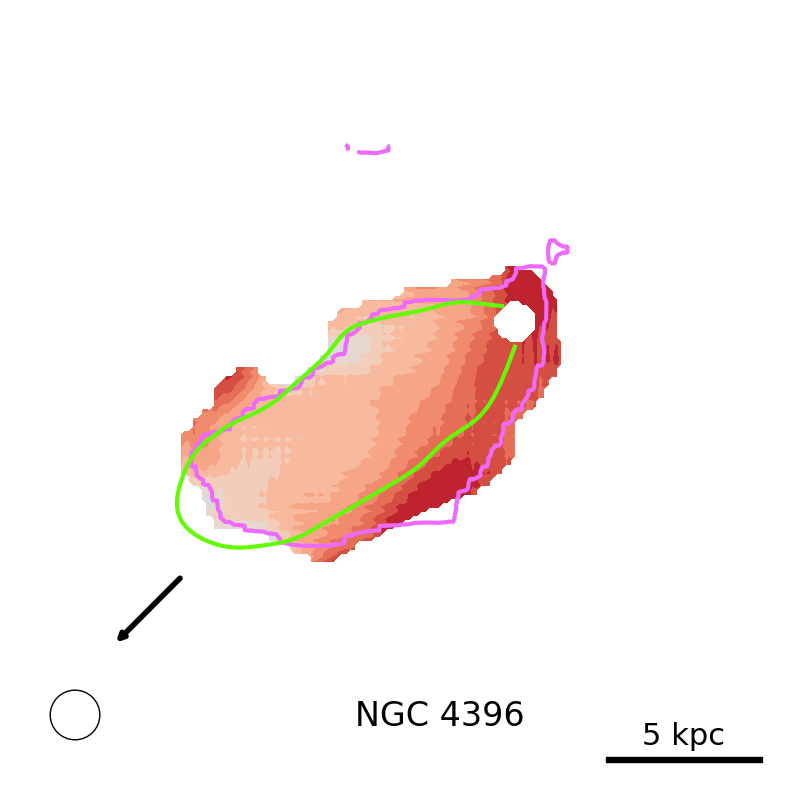}
    \includegraphics[width=0.24\textwidth]{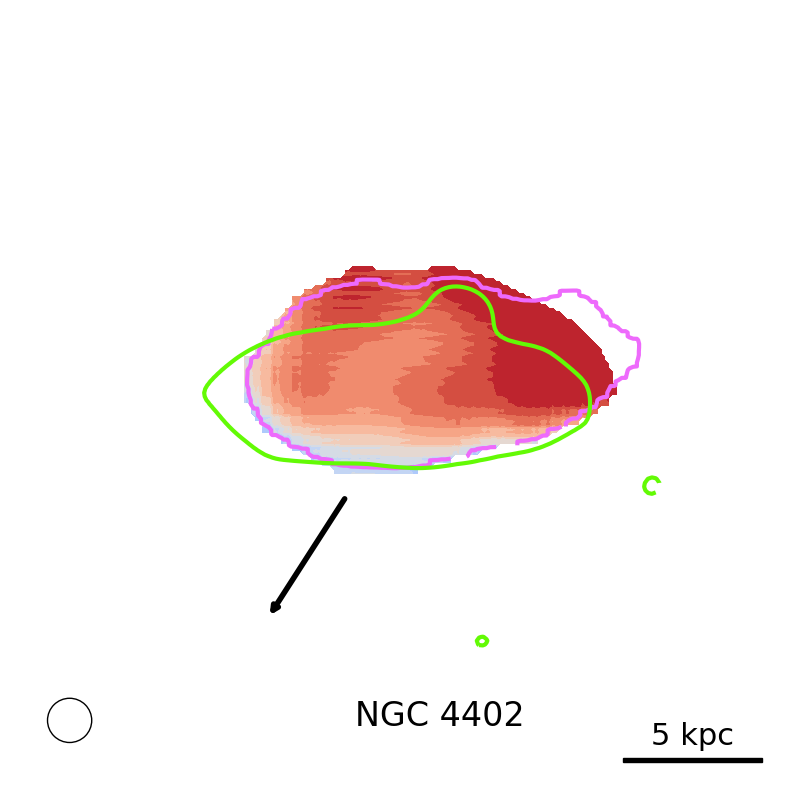}
    \includegraphics[width=0.24\textwidth]{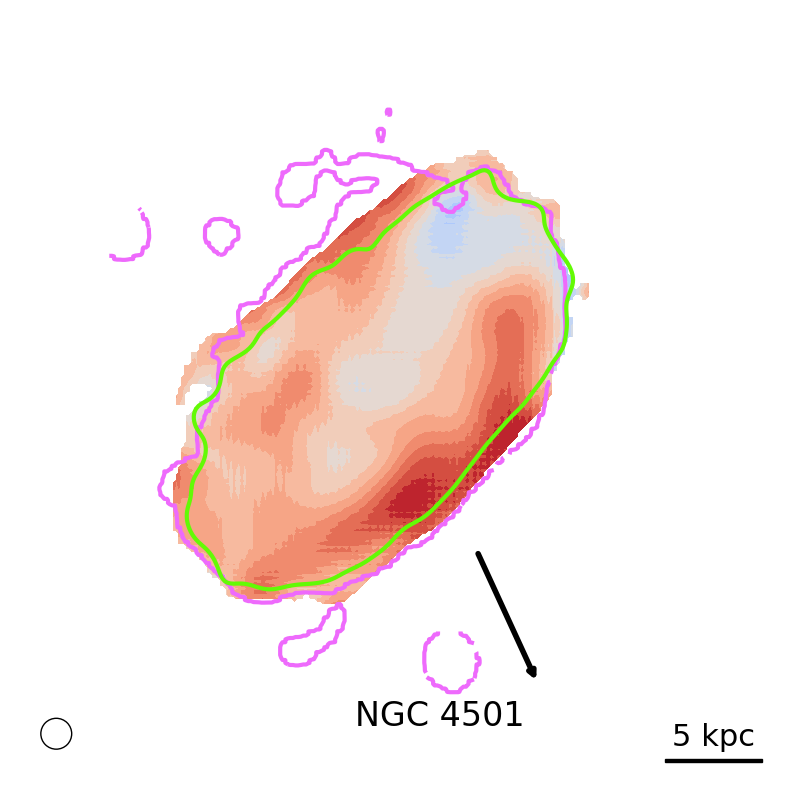}
    \includegraphics[width=0.24\textwidth]{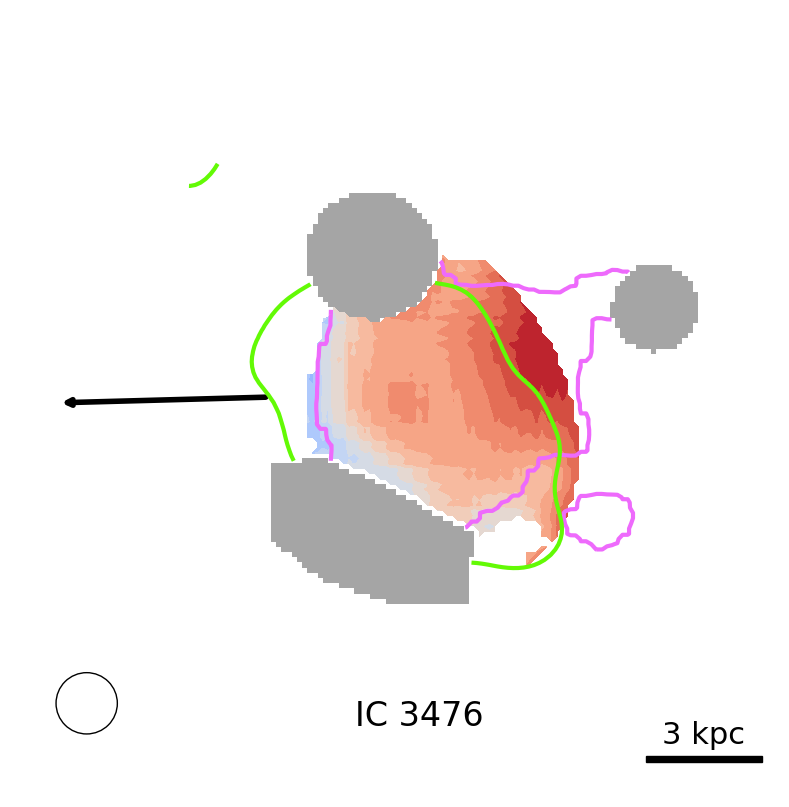}
    \includegraphics[width=0.24\textwidth]{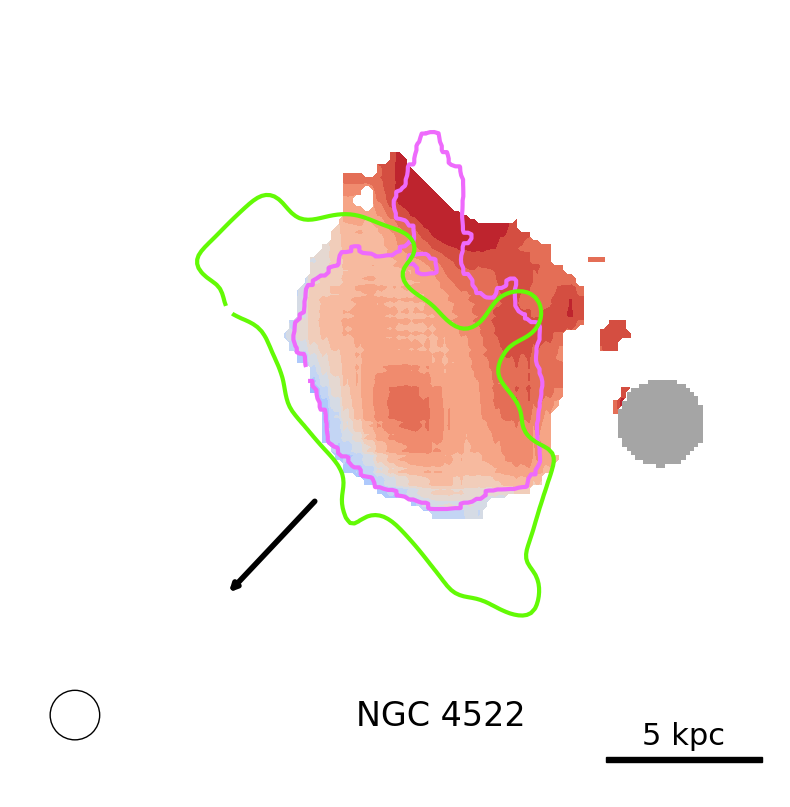}
    \includegraphics[width=0.24\textwidth]{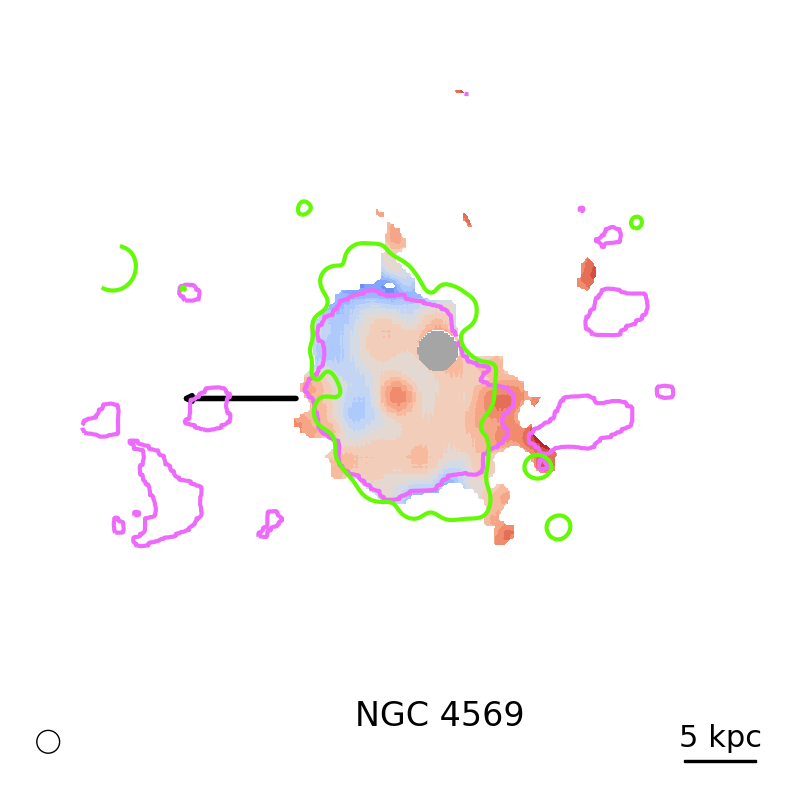}
    \includegraphics[width=0.24\textwidth]{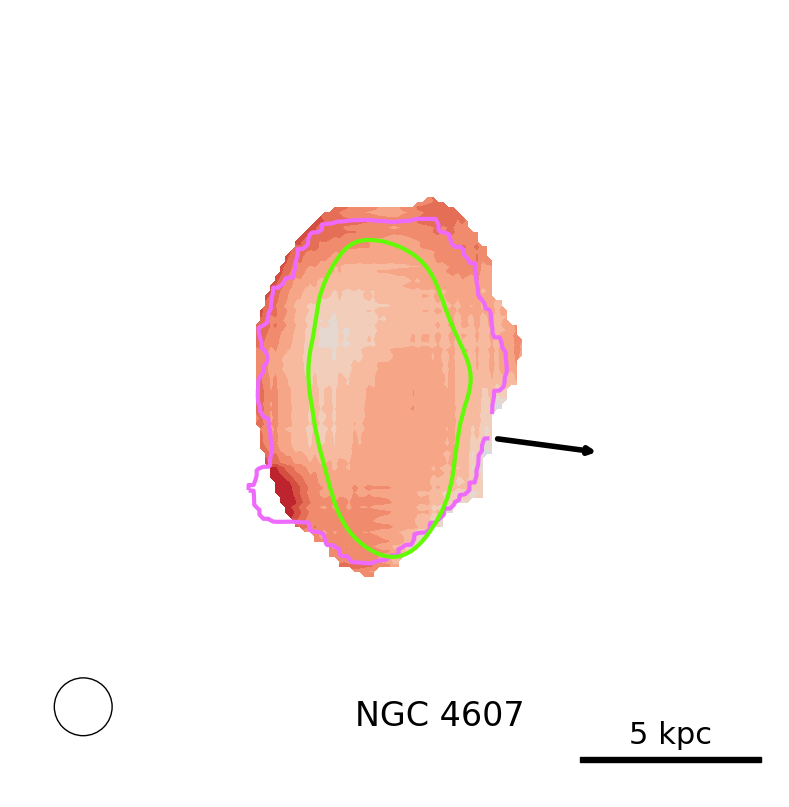}
    \includegraphics[width=0.24\textwidth]{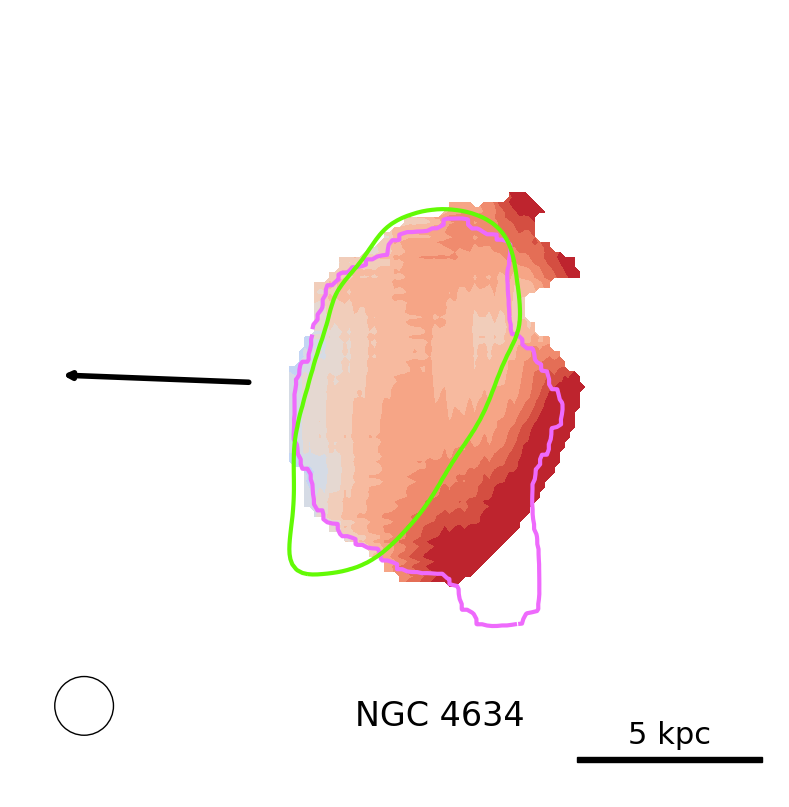}
    \includegraphics[width=0.24\textwidth]{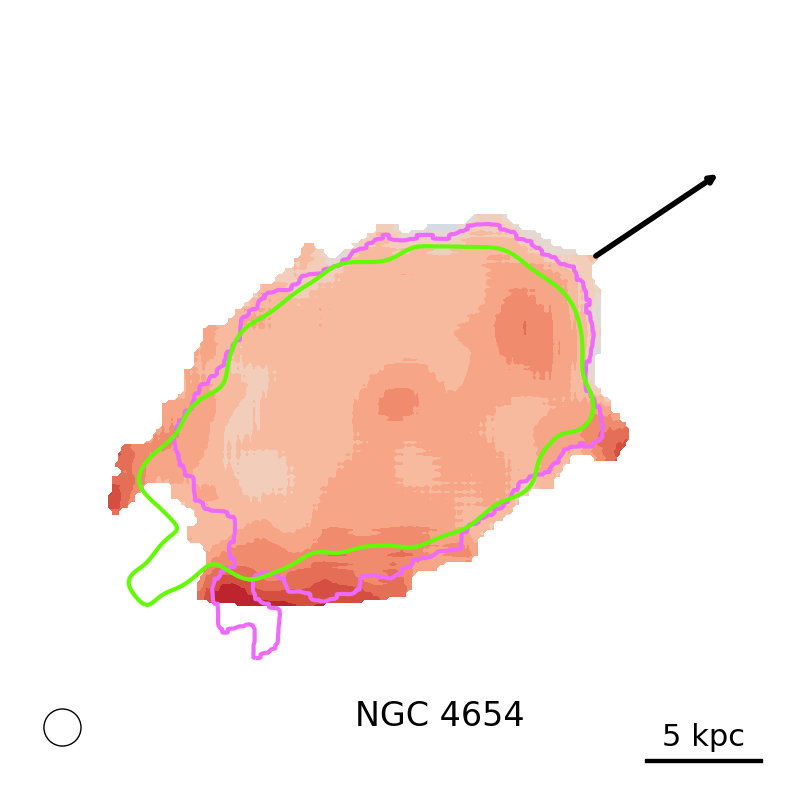}
    \caption{Logarithmic ratio between the observed and modeled radio emission. The figure is identical to \autoref{fig:ratio}, except that we only show the images based on the $20''$ radio maps which were not displayed previously.}\label{fig:ratioapp}
\end{figure*}
\FloatBarrier
\twocolumn

\section{Synchrotron emission model}\label{sec:appendix2}
To numerically solve \autoref{eq:fokkerplanck}, we use the Chang-Cooper discretization scheme \citep{chang1970practical} implemented in the \texttt{pychangcooper}\footnote{\url{https://github.com/grburgess/pychangcooper}} library. We extend this code to include an escape term $N(E,t)/t_\mathrm{esc}$, a time-dependent injection term $Q(E,t)=qE^{-2}\exp{(-t/t_\mathrm{quench})}$ and the relevant loss processes, i.e. Synchrotron radiation, inverse Compton scattering, ionization and Bremsstrahlung. Those are described by \citep{longair2010high}:        
%ionization_factor = 7.64e-15*(3*np.log(self._half_grid/0.51) + 19.8)*1e-6*(3600*24*365*1e6) # MeV / Myr, everything but Ngas
%        brems_factor = 3.66e-22*(3600*24*365*1e6)*self._half_grid
%        self._heating_term = self._C0 * self._half_grid**2  + self._N_gas*(ionization_factor+brems_factor)
\begin{equation}
    b(E) = b_0 E^2 + N_\mathrm{gas}\left( b_1 \left(3\log\frac{E}{m_ec^2} + 19.8\right)+ b_2E \right),
\end{equation}
with $b_0=3.2\times 10^{-6} \left( B^2/{2\mu_0} + u_\mathrm{rad}\right)\,\mathrm{m^3\,MeV^{-2}\,Myr^{-1}}$, $b_1=2.4\times10^{-7}$\,MeV\,m$^3$\,Myr$^{-1}$, and $b_2 = 1.2\times10^{-8}$\,m$^{3}$\,Myr$^{-1}$. Here, $m_e$ the is electron rest mass, $c$ the speed of light and $N_\mathrm{gas}$ the gas density. For the latter, we assume that only the \hi{} content contributes and approximate a typical mean density $N_\mathrm{gas}=4.4\times10^5$\,$\mathrm{m}^{-3}$ assuming the \hi{} mass of NGC\,4396 $M_\hi{}=8.6\times10^8\,\mathrm{M_\odot}$ \citep{Chung2009} is distributed in a cylinder with a radius of 5\,kpc and a height of 1\,kpc.

We split the energy domain into 100 logarithmically spaced points ranging from 1\,MeV to 100\,GeV. The time domain is discretized in steps of 0.1\,Myr. For the initial electron spectrum, we use the source term $Q(E)$ for constant injection and let the spectrum evolve for $t>>t_\mathrm{esc}$ until it reaches a steady state. This spectrum is then either subject to a time-dependent decrease of the source term on a scale $t_\mathrm{quench}$ or to an increase of the magnetic field by a factor of $f_B$ and evolved in time.

The flux density of the corresponding synchrotron emission at a frequency $\nu$ and after a time $t$ is given by \citep{Harwood2013}:
\begin{equation}
    S(\nu,B,t) = S_0 B \int_0^\pi \dif{\delta} \sin(\delta)^2 \int \dif{E} F(x) N(E,B,t),  
\end{equation}
where $\delta$ is the pitch angle, $S_0$ a normalization constant and $F(x)$ the following integral of the Bessel function of order 5/3:
\begin{equation}
F(x) = x \int_x^\infty K_{5/3}(z)\dif{z}.    
\end{equation}
The variable $x=\nu/\nu_\mathrm{c}$ is the ratio of the frequency $\nu$ and the critical frequency $\nu_\mathrm{c}$ , which is given by:
\begin{equation}
    \nu_\mathrm{c}= \frac{3 E^2 e B\sin(\delta)}{4\pi m^3_\mathrm{e} c^4}.
\end{equation} 
Here, $e$ is the electron charge.

For the electron spectrum an the corresponding synchrotron emission, we include a 10\,Myr delay between the star formation and the CRe injection due to the lifetime of the massive O and B stars. The FUV and H$\upalpha$ emission will also be delayed compared to the star formation, although less severely. We define $I_\mathrm{burst}(t)$ as the time-dependent intensity after an instantaneous burst of star formation, where we use the distribution provided in \citet{Leroy2012}. To take into account the contribution of the prior SFR history to the intensity $I(t)$, we use the convolution:
\begin{equation}
    I(t) = \int \mathrm{SFR}(\tau) I_\mathrm{burst}(t-\tau) \dif{\tau}.
\end{equation}
In the scenario of reducing SFR due to RPS, the SFR is given by:
\begin{equation}
    \mathrm{SFR}(t) =\begin{cases}
    \mathrm{SFR}_0,& \text{if } t\leq 0\\
    \mathrm{SFR}_0 \mathrm{e}^{-\frac{t}{t_\mathrm{quench}}},       & \text{otherwise.}
\end{cases}
\end{equation}
% from 100 MeV has impact at 10 MHz
% from 300 MeV has impact at 100 MHz
% below 20 GeV has impact at 1.4 Ghz
\end{appendix}
\end{document}